\documentstyle[emulateapj,rotate,times]{article} 
 
\newcommand{\etal}{et~al.\ }
\newcommand{\PVdblt}{{\rm P}\kern 0.1em{\sc v}~$\lambda\lambda 1117, 1128$}
\newcommand{\CaIIdblt}{{\rm Ca}\kern 0.1em{\sc ii}~$\lambda\lambda 3934, 3969$}
\newcommand{\AlIIIdblt}{{\rm Al}\kern 0.1em{\sc iii}~$\lambda\lambda 1854, 1862$}
\newcommand{\CIVdblt}{{\rm C}\kern 0.1em{\sc iv}~$\lambda\lambda 1548, 1550$}
\newcommand{\MgIIdblt}{{\rm Mg}\kern 0.1em{\sc ii}~$\lambda\lambda 2796, 2803$}
\newcommand{\NVdblt}{{\rm N}\kern 0.1em{\sc v}~$\lambda\lambda 1238, 1242$}  
\newcommand{\SVIdblt}{{\rm S}\kern 0.1em{\sc vi}~$\lambda\lambda 933, 944$} 
\newcommand{\OVIdblt}{{\rm O}\kern 0.1em{\sc vi}~$\lambda\lambda 1031, 1037$} 
\newcommand{\SiIIdblt}{{\rm Si}\kern 0.1em{\sc ii}~$\lambda\lambda 1190, 1193$} 
\newcommand{\SiIVdblt}{{\rm Si}\kern 0.1em{\sc iv}~$\lambda\lambda 1393, 1402$} 
\newcommand{\PV}{\hbox{{\rm P}\kern 0.1em{\sc v}}}
\newcommand{\AlI}{\hbox{{\rm Al}\kern 0.1em{\sc i}}}
\newcommand{\AlII}{\hbox{{\rm Al}\kern 0.1em{\sc ii}}}
\newcommand{\AlIII}{{\hbox{\rm Al}\kern 0.1em{\sc iii}}}
\newcommand{\CaII}{\hbox{{\rm Ca}\kern 0.1em{\sc ii}}}
\newcommand{\CII}{\hbox{{\rm C}\kern 0.1em{\sc ii}}}
\newcommand{\CIIe}{\hbox{{\rm C$^{\ast}$}\kern 0.1em{\sc ii}}}
\newcommand{\CIII}{\hbox{{\rm C}\kern 0.1em{\sc iii}}}
\newcommand{\CIV}{\hbox{{\rm C}\kern 0.1em{\sc iv}}}
\newcommand{\CV}{\hbox{{\rm C}\kern 0.1em{\sc v}}}
\newcommand{\HI}{\hbox{{\rm H}\kern 0.1em{\sc i}}}
\newcommand{\HII}{\hbox{{\rm H}\kern 0.1em{\sc ii}}}
\newcommand{\Lya}{\hbox{{\rm Ly}\kern 0.1em$\alpha$}}
\newcommand{\Lyb}{\hbox{{\rm Ly}\kern 0.1em$\beta$}}
\newcommand{\Lyg}{\hbox{{\rm Ly}\kern 0.1em$\gamma$}}
\newcommand{\Lyd}{\hbox{{\rm Ly}\kern 0.1em$\delta$}}
\newcommand{\Lye}{\hbox{{\rm Ly}\kern 0.1em$\epsilon$}}
\newcommand{\Lyphi}{\hbox{{\rm Ly}\kern 0.1em$\phi$}}
\newcommand{\Lyfive}{\hbox{{\rm Ly}\kern 0.1em$5$}}
\newcommand{\Lysix}{\hbox{{\rm Ly}\kern 0.1em$6$}}
\newcommand{\Lyseven}{\hbox{{\rm Ly}\kern 0.1em$7$}}
\newcommand{\Lyeight}{\hbox{{\rm Ly}\kern 0.1em$8$}}
\newcommand{\Lynine}{\hbox{{\rm Ly}\kern 0.1em$9$}}
\newcommand{\Lyten}{\hbox{{\rm Ly}\kern 0.1em$10$}}
\newcommand{\Lyeleven}{\hbox{{\rm Ly}\kern 0.1em$11$}}
\newcommand{\HeI}{\hbox{{\rm He}\kern 0.1em{\sc i}}}
\newcommand{\HeII}{\hbox{{\rm He}\kern 0.1em{\sc ii}}}
\newcommand{\FeI}{\hbox{{\rm Fe}\kern 0.1em{\sc i}}}
\newcommand{\FeII}{\hbox{{\rm Fe}\kern 0.1em{\sc ii}}}
\newcommand{\FeIII}{\hbox{{\rm Fe}\kern 0.1em{\sc iii}}}
\newcommand{\MnII}{\hbox{{\rm Mn}\kern 0.1em{\sc ii}}}
\newcommand{\MgI}{\hbox{{\rm Mg}\kern 0.1em{\sc i}}}
\newcommand{\MgII}{\hbox{{\rm Mg}\kern 0.1em{\sc ii}}}
\newcommand{\MgIII}{\hbox{{\rm Mg}\kern 0.1em{\sc iii}}}
\newcommand{\NI}{\hbox{{\rm N}\kern 0.1em{\sc i}}}
\newcommand{\NII}{\hbox{{\rm N}\kern 0.1em{\sc ii}}}
\newcommand{\NIII}{\hbox{{\rm N}\kern 0.1em{\sc iii}}}
\newcommand{\NV}{\hbox{{\rm N}\kern 0.1em{\sc v}}}
\newcommand{\OVI}{\hbox{{\rm O}\kern 0.1em{\sc vi}}}
\newcommand{\OI}{\hbox{{\rm O}\kern 0.1em{\sc i}}}
\newcommand{\OII}{\hbox{[{\rm O}\kern 0.1em{\sc ii}]}}
\newcommand{\OIV}{\hbox{{\rm O}\kern 0.1em{\sc iv}]}}
\newcommand{\SI}{{\rm S}\kern 0.1em{\sc i}}
\newcommand{\SIV}{{\rm S}\kern 0.1em{\sc iv}}
\newcommand{\SVI}{{\rm S}\kern 0.1em{\sc vi}}
\newcommand{\SiI}{\hbox{{\rm Si}\kern 0.1em{\sc i}}}
\newcommand{\SiII}{\hbox{{\rm Si}\kern 0.1em{\sc ii}}}
\newcommand{\SiIII}{\hbox{{\rm Si}\kern 0.1em{\sc iii}}}
\newcommand{\SiIV}{\hbox{{\rm Si}\kern 0.1em{\sc iv}}}
\newcommand{\SII}{\hbox{{\rm S}\kern 0.1em{\sc ii}}}
\newcommand{\SIII}{\hbox{{\rm S}\kern 0.1em{\sc iii}}}
\newcommand{\NaI}{\hbox{{\rm Na}\kern 0.1em{\sc i}}}
\newcommand{\TiII}{\hbox{{\rm Ti}\kern 0.1em{\sc ii}}}
\newcommand{\kms}{\hbox{km~s$^{-1}$}}
\newcommand{\cmsq}{\hbox{cm$^{-2}$}}

\tighten
\begin{document}
 
\slugcomment{The Astrophysical Journal, {\rm in press}}
 
\lefthead{CHURCHILL ET~AL.}
\righthead{{\MgII} ABSORPTION SELECTED GALAXIES}


\title{Low and High Ionization Absorption Properties of {\MgII}
Absorption--Selected Galaxies at Intermediate Redshifts. I. General
Properties\altaffilmark{1,2}}

\author{Christopher~W.~Churchill\altaffilmark{3}, Richard~R.~Mellon,
Jane~C.~Charlton\altaffilmark{4}}
\affil{The Pennsylvania State University, University Park, PA 16802}

\author{Buell~T.~Jannuzi}
\affil{National Optical Astronomy Observatories, Tucson, AZ 85719}

\author{Sofia~Kirhakos}
\affil{Institute for Advanced Study, Princeton, NJ 08544}

\author{Charles~C.~Steidel\altaffilmark{5}}
\affil{California Institute of Technology, Palomar Observatories,
Pasadena, CA 91125}

\and

\author{Donald~P.~Schneider}
\affil{The Pennsylvania State University, University Park, PA 16802}

\altaffiltext{1}{Based in part on observations obtained at the
W.~M. Keck Observatory, which is operated as a scientific partnership
among Caltech, the University of California, and NASA. The Observatory
was made possible by the generous financial support of the W. M. Keck
Foundation.}
\altaffiltext{2}{Based in part on observations obtained with the
NASA/ESA {\it Hubble Space Telescope}, which is operated by the STScI
for the Association of Universities for Research in Astronomy, Inc.,
under NASA contract NAS5--26555.}
\altaffiltext{3}{Visiting Astronomer at the W.~M. Keck Observatory}
\altaffiltext{4}{Center for Gravitational Physics and Geometry}
\altaffiltext{5}{NSF Young Investigator}

\begin{abstract}
We present extensive metal--line absorption properties for 45
absorption systems that were selected by their {\MgII} absorption at
redshifts between 0.4 and 1.4.  
For each system the properties of several chemical species are
determined, including a wide range of ionization conditions. 
In the optical, the absorption systems have been observed at $\sim
6$~{\kms} resolution with HIRES/Keck, which covered {\MgII}, several
{\FeII} transitions, {\MgI}, and in some cases (depending upon
redshift), {\CaII}, {\TiII}, {\MnII}, and {\AlIII}.
Ultraviolet, lower resolution ($\sim 230$~{\kms}) Faint Object
Spectrograph data (1600~{\AA}--3275~{\AA}) were obtained from the {\it
Hubble Space Telescope\/} archive.
These spectra covered {\AlII}, {\AlIII}, {\SiII}, {\SiIII}, {\SiIV},
{\CII}, {\CIII}, {\CIV}, {\NV}, {\OVI}, and several Lyman series
transitions, with coverage dependent upon the absorption system
redshift.
From these data, we infer that {\MgII} absorbing galaxies
at intermediate redshifts have multiphase gaseous structures.
\end{abstract}


\section{Introduction}
\label{sec:intro}

During the last decade a great deal of progress has been made toward
understanding the physical properties of intervening metal--line
absorption systems measured in the spectra of high redshift quasars.
This is particularly true at intermediate redshifts, $0.5\leq z \leq
1.5$, for absorbers selected by the presence of the resonant
{\MgIIdblt} doublet (e.g.\ \cite{ltw87}; \cite{tytler87};
\cite{sbs88}; \cite{pb90}, \cite{ss92}).
One of the most notable achievements was the demonstration that {\MgII}
absorbers with $W_{r}(2796)\geq0.3$~{\AA} are almost always associated
with galaxies (\cite{bb91}; \cite{steidel95}).
Those works substantiated the $\sim 30$--year standing hypothesis by
Bahcall \& Spitzer (1969\nocite{bs69}) that metal--line absorption in
quasar spectra arises in extended gaseous envelopes surrounding
intervening galaxies.

The general picture today is that a wide variety of morphological
types (from ellipticals to irregulars) have gaseous ``halos'' that
extend to roughly 40~kpc, with the most common being Sbc--Scd types
(Steidel, Dickinson, \& Persson 1994\nocite{sdp94}; \cite{guillemin}).
The line--of--sight gas kinematics of the absorbers is
consistent with that expected for material bound in galactic potential
wells (\cite{pb90}; Churchill, Steidel, \& Vogt 1996\nocite{csv96};
\cite{kinematicpaper}).
This picture, however, is not without its counter examples or
ambiguities.
In some cases there is evidence that compact star forming
objects spread out over $\sim 200$ kpc are seen at the {\MgII}
absorption redshift and there is no directly associable bright galaxy
(\cite{yanny92}; \cite{yannyyork92}).
It is also not yet established whether the more numerous ``weak''
{\MgII} absorbers, those with $W_{r}(2796)<0.3$~{\AA}, are related 
to galaxies similar in type to those associated with ``strong''
{\MgII} absorption.
There is mounting evidence that a fair number of the weak systems do not
arise within $\sim 40$~kpc of normal, bright galaxies
(\cite{cwc-lb98}; \cite{weak}).

What are the typical low to high ionization absorption conditions in 
intermediate redshift {\MgII} absorption--selected galaxies?
Do the majority of {\MgII} systems have an associated high ionization
phase as seen in {\SiIV}, {\CIV}, {\NV}, and {\OVI} absorption?
Are there any trends between the high ionization and low ionization
absorption strengths?
Are there other relationships (or lack of relationships!) that provide
clues to the physical nature of galactic gas at intermediate
redshifts?

Motivated by these and similar questions, we have undertaken a program
to measure the absorption properties of a wide variety of chemical
and ionization species associated with {\MgII} absorbers.
Unique to our study is that the {\MgII} systems have been observed at
high resolution ($\sim 6$~{\kms}) with HIRES/Keck~I (\cite{thesis}).
These spectra also provide a population of weak systems, which are
significantly more numerous in their redshift path density
(\cite{weak}).
The HIRES spectra cover {\MgII}, several {\FeII} transitions,
{\MgI}, and depending upon redshift coverage, {\CaII}, {\TiII},
{\MnII}, and {\AlIII}.
The remaining absorption properties, including neutral hydrogen and
higher ionization species, have been measured in lower resolution
($\sim 230$~{\kms}) spectra obtained from the {\it Hubble Space
Telescope\/} archive of the Faint Object Spectrograph.

In this paper, we present the measurements of the absorption lines
found in the FOS spectra.
Additional analysis focused on the above motivational questions is
presented in a parallel companion paper (\cite{paper2}, hereafter
Paper II).
In \S~\ref{sec:sample}, we outline our sample selection.
Details of the data analysis are presented in \S~\ref{sec:data}.
In \S~\ref{sec:systems}, we provide a brief description of each system.
The general absorption properties are presented in
\S~\ref{sec:discussion} and a brief synopsis is given in 
\S~\ref{sec:conclusions}.

\section{Sample Selection}
\label{sec:sample}

The {\MgII} systems for our study were selected from the samples of
Steidel \& Sargent (1992\nocite{ss92}, hereafter SS92) and Sargent,
Boksenberg, \& Steidel (1988\nocite{sbs88}, hereafter SBS).
We obtained HIRES spectra (\cite{vogt94}) on the Keck I telescope of
many of the brightest quasars from the SS92 and SBS database.
The optical wavelength coverage for our HIRES spectra translates to a
redshift interval of $0.4 \leq z \leq 1.4$ for 28 {\MgII} systems.
An additional 23 systems with equivalent widths below the detection
threshold in the SS92 and SBS quasar spectra were discovered in the
HIRES spectra, which have a detection threshold of $0.02$~{\AA}
(\cite{weak}).
In order to study the wealth of ionization species with transitions
further into the ultraviolet, we then searched the {\it Hubble Space
Telescope\/} archive for Faint Object Spectrograph (FOS) observations
of the quasars observed with HIRES/Keck.
A number of the FOS spectra were obtained fully reduced from the {\it
HST\/} QSO Absorption Line Key Project (hereafter KP, \cite{cat1};
\cite{cat2}; \cite{cat3}). 
The remaining FOS spectra were obtained from the {\it HST\/} archive
and reduced using the same methodology as for the KP spectra.

The journal of the HIRES/Keck observations is given in
Table~\ref{tab:hiresobsjournal}.
The spectral resolution was $R=45,000$, which corresponds to
$6.6$~{\kms}.
{}From left to right, the columns are the quasar name, the visual
magnitude, the emission redshift, the observation date, the total
exposure time, and the observed wavelength range.
In the continuum near the observed {\MgII} features, the spectra have
signal--to--noise ratios ranging from 15 to 50, with the majority
being around 30.
The spectra do not have continuous wavelength coverage; there are
small gaps redward of 5100~{\AA}, where the free spectral range of the
projected echelle format exceeds the width of the $2048 \times 2048$
Tektronix CCD.

There are several FOS observational modes, including grating,
polarimetry, and slit settings.
We have chosen to limit our survey to the highest resolution spectra
($R=1300$), obtained with the G130H (1150--1600~{\AA}), G190H
(1600--2300~{\AA}) and G270H (2225--3275~{\AA}) gratings.
We chose to exclude spectra obtained at lower resolution in order to
maintain uniformity and because they are not very useful for
narrow--line absorption work.
However, we did use a lower resolution G160L spectrum (in one case) to
search for a Lyman limit break.
We also opted to exclude spectropolarimetry data because they
required additional reduction and calibration steps that did not yield
spectra of comparable quality to those obtained with
non--spectropolarimetry mode.
However, when available, we did use $R=1300$ spectropolarimetry data
to search for Lyman limit breaks and damped {\Lya} lines at the
absorber redshifts.
Finally, we excluded spectra with slit widths greater than
$1${\arcsec}.
Some of the FOS spectra in our sample were obtained before the
COSTAR refurbishing mission.
The instrumental spread function of pre--COSTAR spectra introduces  
broad wings in proportion to the aperture width of the observing mode
(\cite{jannuzi-hartig}).
We studied only those FOS spectra obtained with fairly narrow
apertures in order to mitigate systematics in the absorption line
strengths between the pre--COSTAR and COSTAR spectra.

The final selection of FOS spectra are listed in
Table~\ref{tab:fosobsjournal}.
From left to right the columns are the quasar name, the alias useful
for searching the {\it HST\/} archive, the program identification
numbers,  and the names of the principal investigators for each of the
gratings.
The observing modes of the KP spectra are described in Jannuzi \etal
(1998\nocite{cat3}); the slit widths are all less than
$0.26${\arcsec}.
The observing modes of the non--KP spectra vary and are noted in
\S~\ref{sec:systems}, where each spectrum is individually discussed.
We also excluded quasars with complicated broad absorption line
features.
We make no corrections to these narrow aperture pre--COSTAR spectra
and assume a Gaussian instrumental spread function (e.g.\
\cite{dpsKP}) for measuring equivalent widths in all FOS spectra.

\begin{figure*}[t]
\plotone{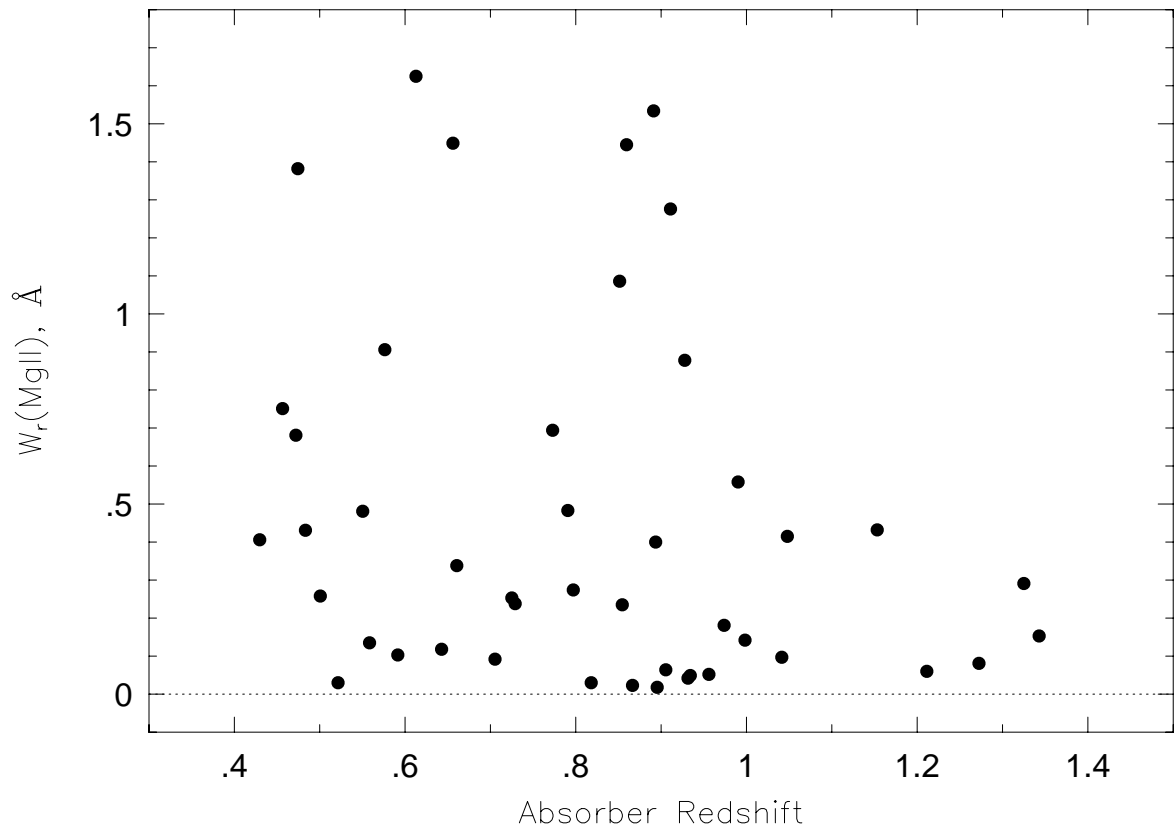}
\caption[fig1.ps]
{The rest--frame equivalent width of {\MgII} $\lambda 2796$ vs.\
absorption redshift for the sample. \label{fig:Wvsz}}
\end{figure*}

These selection criteria resulted in a sample of 45 {\MgII} absorbing
systems.
These systems and their rest--frame equivalent widths are listed in
Table~\ref{tab:hiresresults}.
The first three columns are the quasar name, the absorber redshift,
and the {\MgII} $\lambda 2796$ rest--frame equivalent width,
$W_{r}({\MgII})$.
The remaining columns are explained in \S~\ref{sec:systems}.
In Figure~\ref{fig:Wvsz}, we show a plot of $W_{r}({\MgII})$ 
vs.\ absorber redshift, $z_{\rm abs}$.
Note that the sample is devoid of $W_{r}({\MgII}) \geq 0.5$~{\AA} systems
for $z\geq 1$.
This is not surprising because the total redshift path above $z \simeq 1$
covered by the HIRES spectra is very small compared to the coverage 
from $0.4 \leq z \leq 1.0$ and because small equivalent width systems
are very common.

\begin{figure*}[t]
\plotone{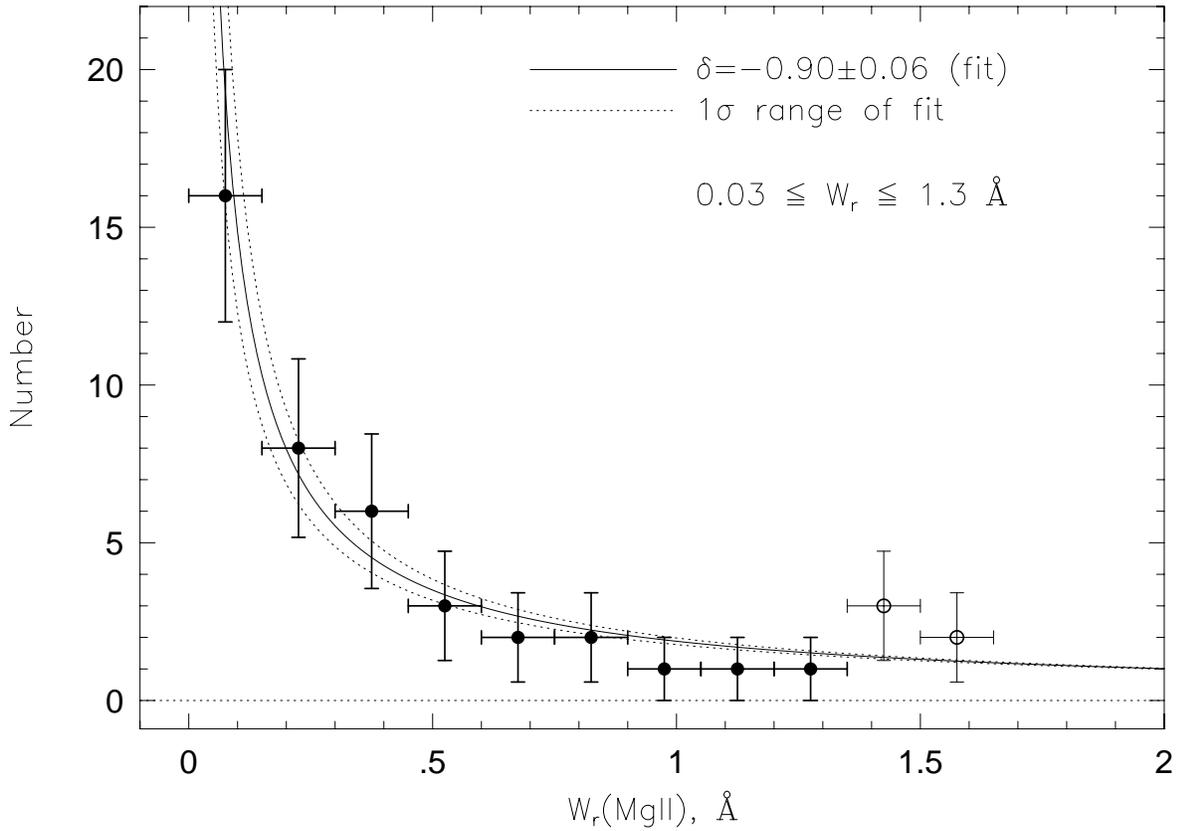}
\caption[fig2.ps]
{The rest--frame equivalent width distribution of {\MgII} $\lambda
2796$ for the sample. The data are binned at 0.15~{\AA} for
presentation and given Poissonian uncertainties.  The solid line is a
maximum likelihood power--law fit (to the unbinned data) over the
range $0.03 \leq W_{r} \leq 1.3$~{\AA}.  The dotted lines shows the
$1\sigma$ range in the distribution function parameters.  The unfilled
data points were not included in the fit (see text). \label{fig:fW}}
\end{figure*}

The distribution of $W_{r}({\MgII})$ is shown in Figure~\ref{fig:fW}.
Using a maximum likelihood fit (e.g.\ \cite{ltw87}), we find that
the distribution of our sample follows a power--law with $f(W) \propto
W^{-0.9\pm0.6}$ for $0.03 \leq W_{r}({\MgII}) < 1.3$~{\AA}, which is
consistent with that of an unbiased sample, $f(W) \propto
W^{-1.0\pm0.1}$, over this equivalent width range (\cite{weak}).
Between $1.0$--$1.3$~{\AA} the data are systematically below the fit,
but are consistent with the maximum likelihood result to $1~\sigma$ in
all bins.
Above $1.3$~{\AA}, there is a slight overabundance of systems (not
included in the maximum likelihood analysis), due to a selection bias
toward larger equivalent width absorbers in the HIRES/Keck survey
(\cite{thesis}).
These are all damped {\Lya} absorbers.
Overall, the sample studied here is consistent with an unbiased sample
in its equivalent width distribution for $W_{r}({\MgII}) < 1.3$~{\AA}
and for $z\leq 1$.

\newpage
\section{Data Analysis}
\label{sec:data}

The HIRES data were processed in the standard manner using the
IRAF\footnote{IRAF is distributed by the National Optical Astronomy
Observatories, which are operated by AURA, Inc., under contract to the
NSF.} {\it Apextract\/} package for echelle data.  
The details of the HIRES data reduction are given in Churchill
(1995\nocite{lotr}, 1997\nocite{thesis}).
The KP FOS spectra were reduced by the KP collaboration, as described
in Schneider \etal (1993\nocite{dpsKP}),  Bahcall \etal
(1996\nocite{cat2}), and  Jannuzi \etal (1998\nocite{cat3}).
The archival FOS spectra were reduced and calibrated using KP methods.

As a consistency check, we performed two separate, full analyses of
the FOS data.
The primary analysis invoked {\it a priori\/} knowledge of the
{\MgII} redshifts for searching, identifying, and measuring the
absorption lines (described in greater detail below).
This analysis also used {\it a priori\/} knowledge of other
metal--line systems along the line of sight (from an exhaustive search
of the literature), in order to assess the possibility of
misidentifications and blends from these other systems.
We developed our own set of automated and graphically interactive
software routines for objective feature finding, identifying blending
due to other systems, and measuring equivalent widths (including
Gaussian deblending).
Many of our algorithms are based upon those described by 
Schneider \etal (1993\nocite{dpsKP}) and Bahcall \etal
(1996\nocite{cat2}).
The secondary analysis, which we call the ``KP analysis'', was
performed as a consistency check.
This was an unbiased search for metal--line systems using the KP
methodologies (\cite{cat1}; \cite{cat2}; \cite{cat3}) and software,
i.e.\ the ZSEARCH and JASON programs.
The two full analyses were compared for each FOS absorption line
measurement.
Though the ``KP analysis'' yielded different results in detail from
those of the primary analysis (i.e.\  small differences in equivalent
widths and their errors), there were very few inconsistencies.
In the few cases of differing results, both analyses were studied and
a consensus was reached.

\subsection{Continuum Fitting}

Since the continuum fitting is the most subjective aspect of the data
reduction and probably has the most significant impact on the
analysis, we briefly elaborate.
For the HIRES data, continuum fitting is fairly straight forward
because the continuum is well sampled and the signal--to--noise ratio
is quite high. 
We used the technique of Sembach \& Savage (1992\nocite{kenspaper}),
which employs Legendre polynomials across a limited regions of
spectrum around each of the transitions.  
The HIRES spectra were not flux calibrated, so the normalized flux is
simply relative to the flux levels resulting from the HIRES
sensitivity function.
Following flux calibration, the FOS spectra were continuum fit as 
described in Schneider \etal (1993\nocite{dpsKP}); this entails
several iterations of interactive cubic spline fitting to the flux
calibrated spectra.

\subsubsection{Line Finding}
\label{sec:detect}

For both the HIRES and the FOS spectra, absorption features were
detected using a method slightly modified from that of Schneider \etal
(1993\nocite{dpsKP}), which is optimal in the case of unresolved
lines.
One constructs a discrete model of the instrumental spread function
(ISF) consisting of $M=2J_{0} + 1$ elements, where $J_{0}$ is a
non--negative integer and $\sum_{1}^{M} P_{j} = 1$.
The term $P_{j}$ is the value at pixel $j$ of a symmetric ISF having
a peak value at $j=J_{0}+1$.
For both HIRES and FOS, the ISF is modeled as a Gaussian with
$R = 45,000$ and $1300$, respectively.
We use $J_{0} = 6$, which gives a model ISF over 13 pixels.

The equivalent width of an unresolved feature centered in pixel $i$ is
calculated by centering the ISF on the pixel and then computing
\begin{equation}
w(\lambda_{i}) =  \Delta \lambda _{i}
\frac
{      \sum _{j_{1}} ^{j_{2}} P_{j} F(\lambda_{k})
}
{      \sum _{j_{1}} ^{j_{2}} P^{2}_{j},
}
\label{eq:donsew}
\end{equation}
where $F(\lambda_{k}) =  1-I(\lambda _{k})/I_{c}(\lambda _{k}) $, 
$I(\lambda_{k})$ is the flux in pixel $k$, $I_{c}(\lambda_{k})$ is its
fitted continuum level, $\Delta \lambda_{i} =  (\lambda_{i-1} -
\lambda_{i+1})/2$ is the wavelength interval spanned by the pixel,
$j_{1}$ and $j_{2}$ are the minimum and the maximum points in the ISF,
and the index\footnote{The index $k$, as written here, provides a
minor correction to a previously published version of this formula
in Schneider \etal (1993).} $k = i+(j-1)-J_{0}$, and where
$w(\lambda_{i}) < 0$ for an absorption  feature.
At the spectra ends the continuum is extrapolated over $J_0$
pixels (similar to a wrap around convolution).
The uncertainty in $w(\lambda_{i})$, is given by
\begin{equation}
\sigma_{w}(\lambda_{i}) = \Delta \lambda _{i}  
\frac
{     \left\{ 
      \sum _{j_{1}} ^{j_{2}} 
      P^{2}_{j} \sigma ^{2}_{F}(\lambda_{k})
      \right\} ^{1/2} 
}
{    \sum _{j_{1}} ^{j_{2}} P^{2}_{j},
}
\label{eq:donssigma}
\end{equation}
where $\sigma _{F}(\lambda_{k}) = \sigma_{I}(\lambda_{k}) /
I_{c}(\lambda_{k}) $,  and where
$\sigma_{I}(\lambda_{k})$ is the $1\sigma$ uncertainty in
$I(\lambda_{k})$ resulting from data reduction and calibration
sources and Poisson noise in the quasar flux and sky.
This uncertainty ``spectrum'' serves as the $1\sigma$ observed
equivalent width detection threshold.

In general terms, an unresolved {\it absorption\/} line at pixel $i$
is defined when the pixel equivalent width, $w(\lambda_{i})$, is less 
than $-N\sigma _{w}(\lambda_{i})$, where $N$ is an arbitrarily defined
number giving the number of $\sigma$ beyond the equivalent width
detection threshold.
For resolved features, the detection is defined over a spectral {\it
region}.
The region extremes are defined at the pixels where the
$w(\lambda_{i})$, which are smooth over the scale length of the ISF,
become greater then zero.
 
\begin{figure*}[t]
\plotone{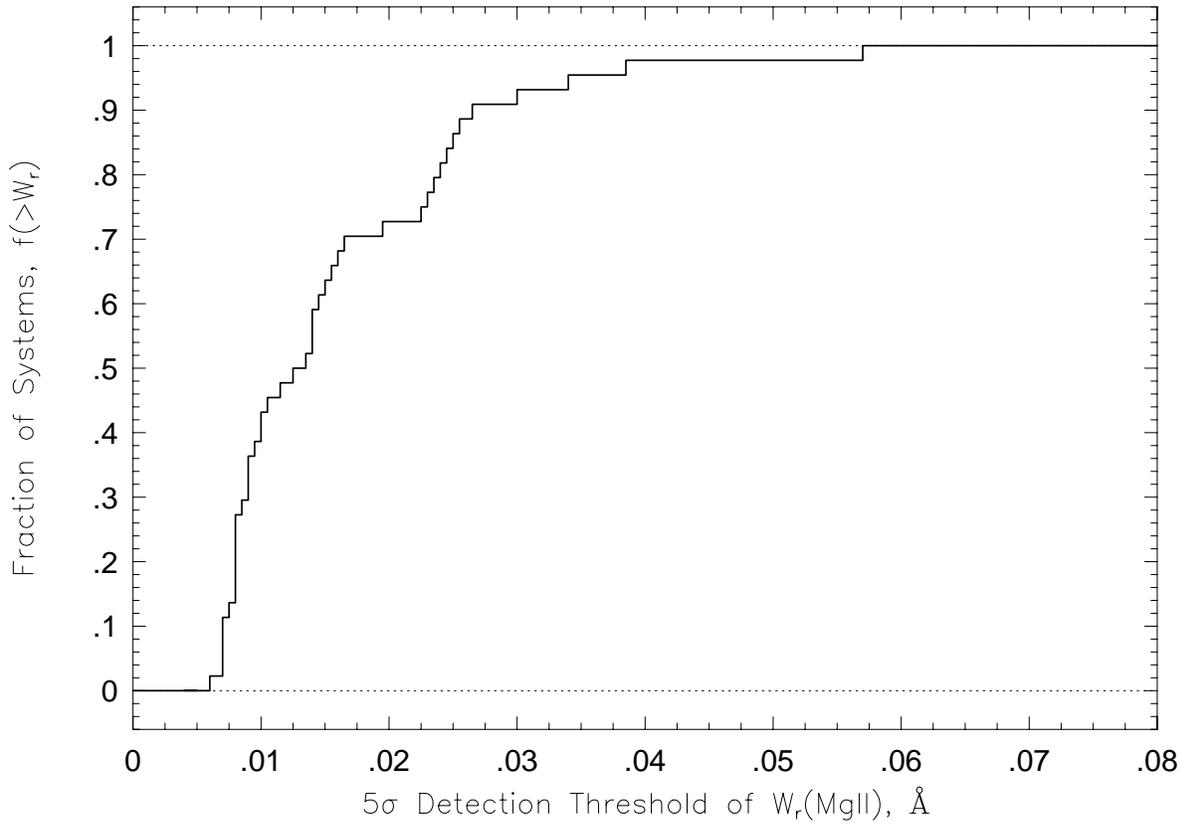}
\caption[fig3.ps]
{The cumulative distribution of the $5\sigma$ rest--frame equivalent
width detection threshold of {\MgII} $\lambda 2796$ for the overall
sample \label{fig:ewlim}}
\end{figure*}

For the HIRES spectra, we enforce a $5\sigma$ detection threshold.
In Figure~\ref{fig:ewlim}, we show the cumulative distribution of the 
$5\sigma$ rest--frame equivalent width detection threshold of {\MgII}
$\lambda 2796$.
Our sample is 100\% 
complete to a $5\sigma$ threshold of $0.06$~{\AA}, 93\% 
complete to $0.03$~{\AA}, and 73\%
complete to $0.02$~{\AA}.
For the FOS spectra, we require only a $3~\sigma$ detection threshold
(whereas the ``KP analysis'' enforced a $4.5~\sigma$ threshold).
We applied a less stringent threshold because we used {\it a priori\/}
knowledge of the expected location of the absorption lines, whereas
the ``KP analysis'' was an unbiased search for absorption lines, and
was therefore more conservative.

\subsubsection{Equivalent Widths}
\label{sec:equivalentwidths}

For the HIRES spectra, the equivalent widths, $W$, are measured
directly by summing the quantity $[1-I(\lambda _{i})/I_{c}(\lambda
_{i})] \Delta \lambda _{i}$ across the profiles.
The measurement uncertainties are obtained from quadrature summing the
quantity $\sigma _{I}(\lambda _{i}) [\partial W / \partial I(\lambda
_{i})]$.
The equivalent width uncertainties do not account for subjectivity in
the continuum placement.

For the FOS spectra, the equivalent widths and their uncertainties
were measured by fitting Gaussians to the absorption features.
The quoted equivalent widths and uncertainties were taken from our own
measurements (i.e.\ not from the ``KP analysis'').
We have used an interactive $\chi ^{2}$ minimization scheme, where the
minimization is performed by the NETLIB--{\sc slatec}
routine\footnote{NETLIB is a collection of mathematical software,
papers, and databases maintained by AT\&T Bell Laboratories, the
University of Tennessee, and Oak Ridge National Laboratory ({\it
www.netlib.org}).} {\sc dnls1} (\cite{more78}).
The uncertainties in the fitted parameters, the Gaussian amplitudes,
widths, and centers, are computed using a modified version of the
routine {\sc dfridr} (\cite{recipes}).
The equivalent width uncertainties are based upon standard error
propagation, including correlated terms.
For a given line, the minimum allowed Gaussian width is set by the
instrumental resolution.

\begin{figure*}[t]
\plotone{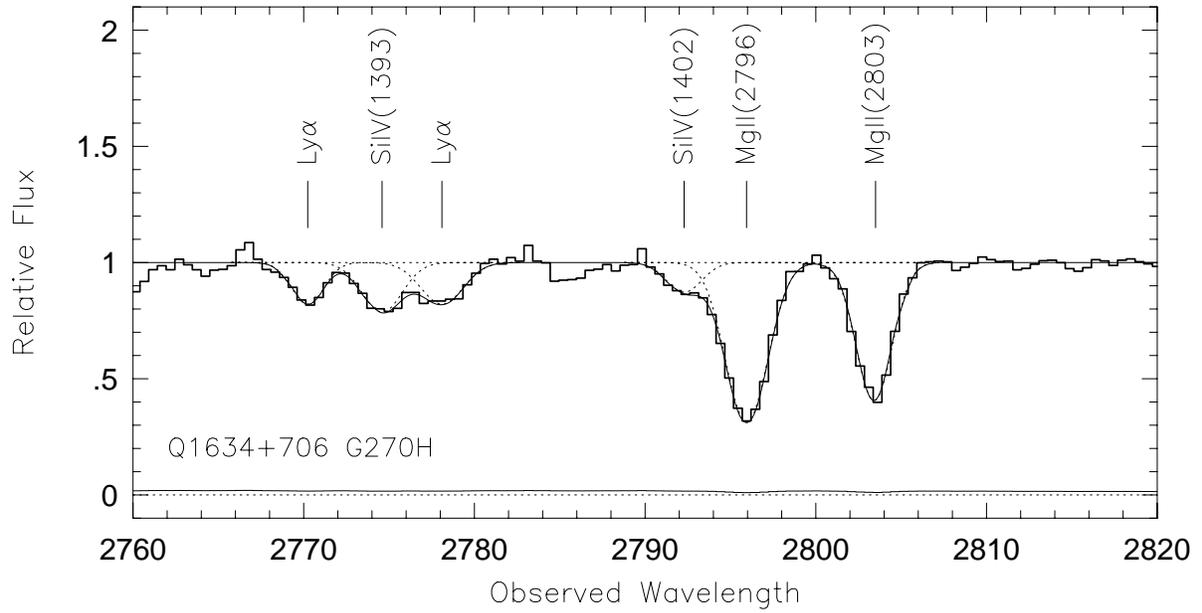}
\caption[fig4.ps]
{Example of Gaussian deblending to obtain equivalent widths from FOS
spectra.   Shown is the {\SiIV} doublet at $z=0.9902$ in the G270H
spectrum of PG~$1634+706$; $\lambda 1393$ lies between two {\Lya}
lines and $\lambda 1402$ resides in the wing of Galactic {\MgII}
$\lambda 2796$. \label{fig:deblending}}
\end{figure*}

Due to the number density of absorption features and the limited
resolution of the FOS spectra, Gaussian deblending was sometimes
required.
Deblending was used only in cases where the individual line centroids 
were clearly separated or where a weak blend in the wing of a stronger
line showed clear ``double'' structure.
Otherwise, a blend was quoted.
Examples of the deblending cases are illustrated in
Figure~\ref{fig:deblending}, which shows the {\SiIV} doublet at
$z=0.9902$ in PG~$1634+706$; $\lambda 1393$ lies between two {\Lya}
lines and $\lambda 1402$ resides in the wing of Galactic {\MgII}
$\lambda 2796$.

\subsubsection{Wavelength Zero Point Shifting of FOS Spectra}

The velocity zero points of the KP FOS spectra were defined by setting
the mean of the singly ionized transitions from Galactic clouds to
redshift zero (see \cite{savageKP}).
The archival FOS spectra, on the other hand, were not zero point
shifted immediately following their reduction.
In both the KP and archival spectra, analysis of the
different absorption lines associated with the {\MgII} systems often
revealed a systematic velocity shift with respect to the HIRES
spectra, which yielded very precise ($\sigma_{z}/z \simeq 10^{-6}$),
heliocentric, vacuum wavelength, absorption redshifts from {\MgII}.

We adjusted the zero points of the FOS spectra using a constant
velocity shift, which was often determined from the mean velocity
offsets of two or more {\MgII} systems along the line of sight.
Since {\SiII} and {\CII} have ionization states similar to {\MgII},
they were used to derive the shift.
If neither {\SiII} nor {\CII} was available, then we used {\Lya}
and/or {\CIV}.
The largest shift we applied was no larger than $\sim 70$\% of a FOS
spectral resolution element.

\subsubsection{Line Identification Policy in FOS Spectra}
\label{sec:fosids}

Our philosophy for the detection and identification of lines in the
crowded FOS spectra is as follows [also see \S~4 of Bahcall \etal
(1996\nocite{cat2})].
We generated an objective list of lines using a $3\sigma$ detection
criterion (see \S~\ref{sec:detect}).
This relatively ``liberal'' threshold was chosen because we are not
producing a formal catalog of all absorbers, but have precise
redshifts from {\MgII} in the HIRES spectra.
In addition, the continuum fits to the FOS spectra were conservative
in that they are probably systematically low in regions of dense line
blending.

We examined, case by case, those lines that happen to be coincident
with the predicted positions of the transitions from {\SI}, {\OI},
{\NI}, {\FeII}, {\SiII}, {\AlII}, {\SII}, {\CII}, {\AlIII}, {\NII},
{\FeIII}, {\SiIII}, {\SIII}, {\SiIV},  {\SIV},  {\NIII}, {\CIII},
{\CIV}, {\SVI}, {\NV}, {\OVI}, and the Lyman series.
For several of the above species, multiple transitions were covered,
allowing either examination of a ``clean'' region of spectrum, or 
data that corroborated or ruled out identifications in ``confused''
regions.
Some of these transitions were rarely detected at the $3 \sigma$
level.  
We did not search regions of the spectra having an observed equivalent
width detection threshold greater than $3$~{\AA}.

A fair fraction of the time, line blending with transitions from other
absorbing systems (some from our sample) was problematic.
We attempted to identify all lines involved with blends as follows:
We constructed a list of 100 transitions with accurate
rest--frame wavelengths (\cite{morton91}) and a list of all known
absorption redshifts, including {\Lya} ``systems'' for which {\Lyb}
could be confirmed, from an exhaustive literature search (optical
and ultraviolet observations) for known systems.
In most all cases, a viable candidate for a blend could be determined.
 
For a given system, a species/transition is identified under the
following conditions:
1) there is at least one 3$\sigma$ detection that can be identified
with a transition of the species;
2) all other covered transitions of the species are not inconsistent
with the identification\footnote{Additional transitions can be
consistent if they: a) are not covered, b) have equivalent widths
between the scaled $f\lambda$ value and the equivalent width of
the strongest transitions, c) are in a blend for which the
contribution of the transition of interest could be consistent.}; 
3) at least one transition is not blended with a possible transition 
from another system, or, if it is, can be unambiguously deblended using
Gaussian fitting (meaning that it is on the wing of a line or has its
own clear line center, see \S~\ref{sec:equivalentwidths}).
In the cases of doublets, if the weaker member of a doublet is not
inconsistent with the first,  then we quote it as a detection.  
This can occur when the second member is in a blend or is not covered
by the spectrum. 
It is always possible that a doublet is a chance match with two
``random'' lines, but this type of false match is expected to have
small probability in our spectra, based upon the simulations by
Bahcall \etal (1996\nocite{cat2}) and by Jannuzi \etal
(1998\nocite{cat3}).

If a species was detected in fewer than 10\% of the systems, it was
eliminated from our overall presentation.
Each species has what we call a ``flag transition''.
The flag transitions are those with the strongest $f\lambda$ for each
chemical/ionization species. 
For the non--doublet transitions, they are {\FeII} $\lambda 2600$,
{\MgI} $\lambda 2853$, {\AlII} $\lambda 1670$, {\CII} $\lambda 1334$,
{\SiII} $\lambda 1260$, and {\SiIII} $\lambda 1206$.
If the flag transition for the species is not identifiable using the
above criteria, but a weaker transition is, the equivalent width of
the weaker transition is quoted (in Table~\ref{tab:fosresults}).
We list these for completeness, but do not include them in our
analysis.

In spectral regions with no detected absorption lines, the
$3\sigma$ equivalent width limits were computed from
Equation~\ref{eq:donssigma}, assuming unresolved features.
For any given undetected species, the transition with the most
stringent limit was adopted.  
If there was a blend at the expected position of the transition,
then an ``upper limit'' was obtained by quoting the
equivalent width of the blended absorption line(s).
Due to flat fielding uncertainties, the minimum limit quoted (observer
frame) is $0.13$~{\AA} (see Bahcall \etal 1996\nocite{cat2}).

\subsubsection{Measuring Lyman Limit Breaks}

Measurements of the Lyman limit breaks were made using KP techniques,
as discussed in Schneider \etal (1993\nocite{dpsKP}) and Jannuzi \etal
(1998\nocite{cat3}).
In the cases of multiple or double breaks, we modeled the data
employing the same technique applied by Churchill \& Charlton
(1999\nocite{q1206}; see their Figure 3$a$).
The redshifts of the Lyman limit break models were fixed at the
redshifts of the {\MgII} absorbers.
We varied the {\HI} column densities and superimposed synthetic
spectra on the data.
The quoted optical depths were obtained from the best matching
model spectrum.
Because the signal--to--noise ratio was often low, we did not use any
fitting statistics, but simply attempted to make the depth of the
break consistent with the model.
In the cases of possible double or multiple breaks (due to redshift
proximity), this modeling technique was successful at either singling
out which {\MgII} system actually gives rise to the break, or placing
constraints on the relative strengths of the double break.
When a Lyman break is present, we quote a rough estimate of its
optical depth in \S~\ref{sec:discusssystems}, where each absorber is
discussed individually.
We also place a ``$+$'' in column six of Table~\ref{tab:fosresults}
(described in \S~\ref{sec:systems}).
If a break was not present, we tabulated a ``$-$'' and assign an upper
limit on $\log N({\HI})$ of $16.8$~{\cmsq}.
When the location of the break was not covered in the spectra, we
placed a ``$\cdots$''.



\section{Presentation of System Properties}
\label{sec:systems}

Line identifications of absorption features in FOS spectra
is often plagued by line blending and other sources of confusion.
Thus, we believe that a fairly complete description of the issues
encountered with each system is warranted, especially in view of the 
possibility that any single system may be the subject of future
detailed study as higher quality data become available.

\begin{figure*}[t]
\plotfiddle{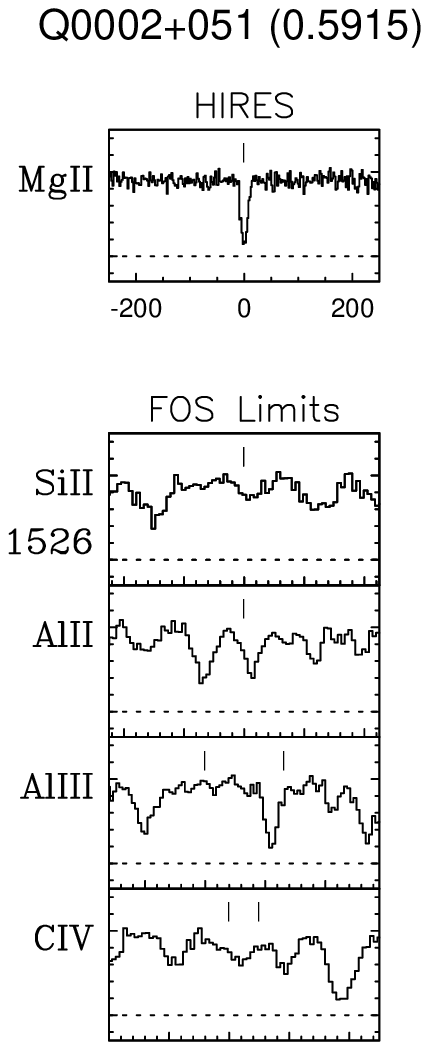}{8.in}{0}{100}{100}{-338}{-135}
\plotfiddle{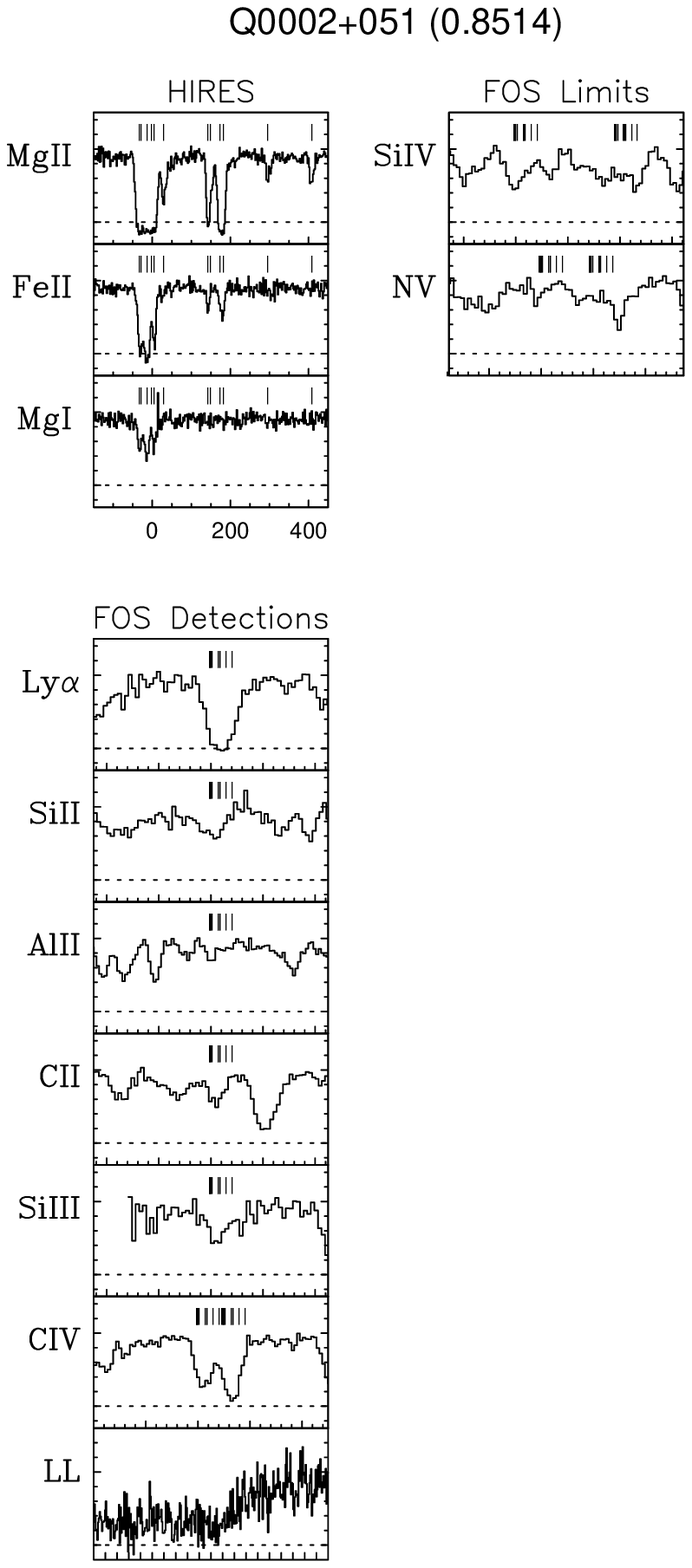}{0.in}{0.}{100.}{100.}{-212}{-113}
\plotfiddle{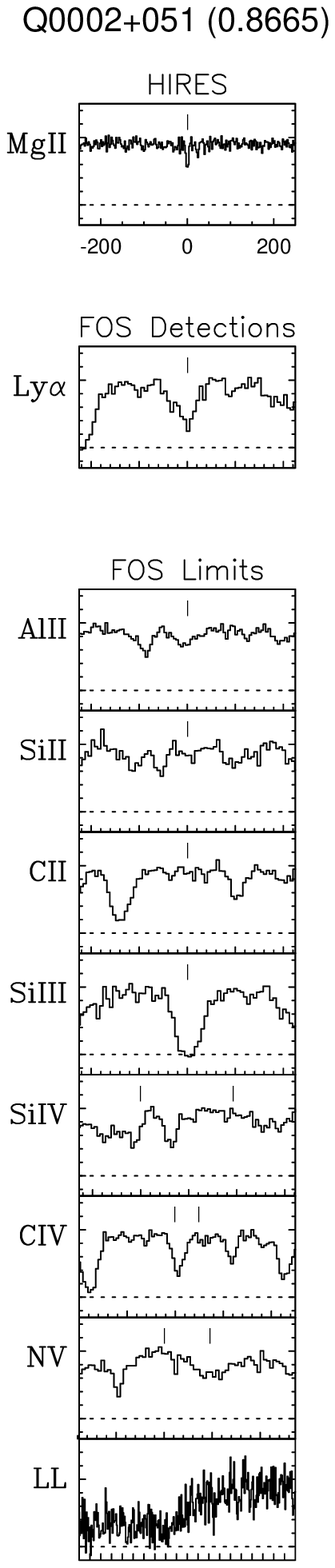}{0.in}{0.}{100.}{100.}{40}{-91}
\figcaption[fig5a.ps, fig5b.ps, fig5c.ps, 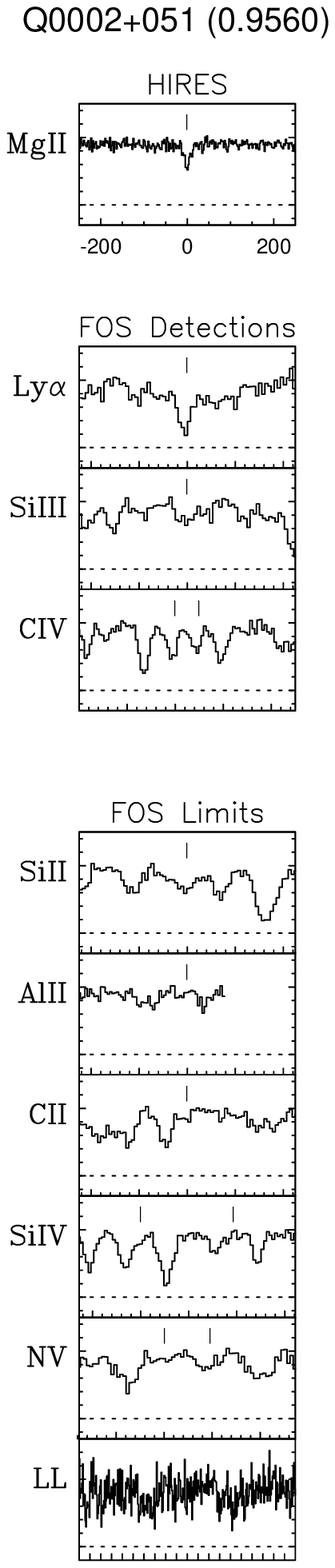, 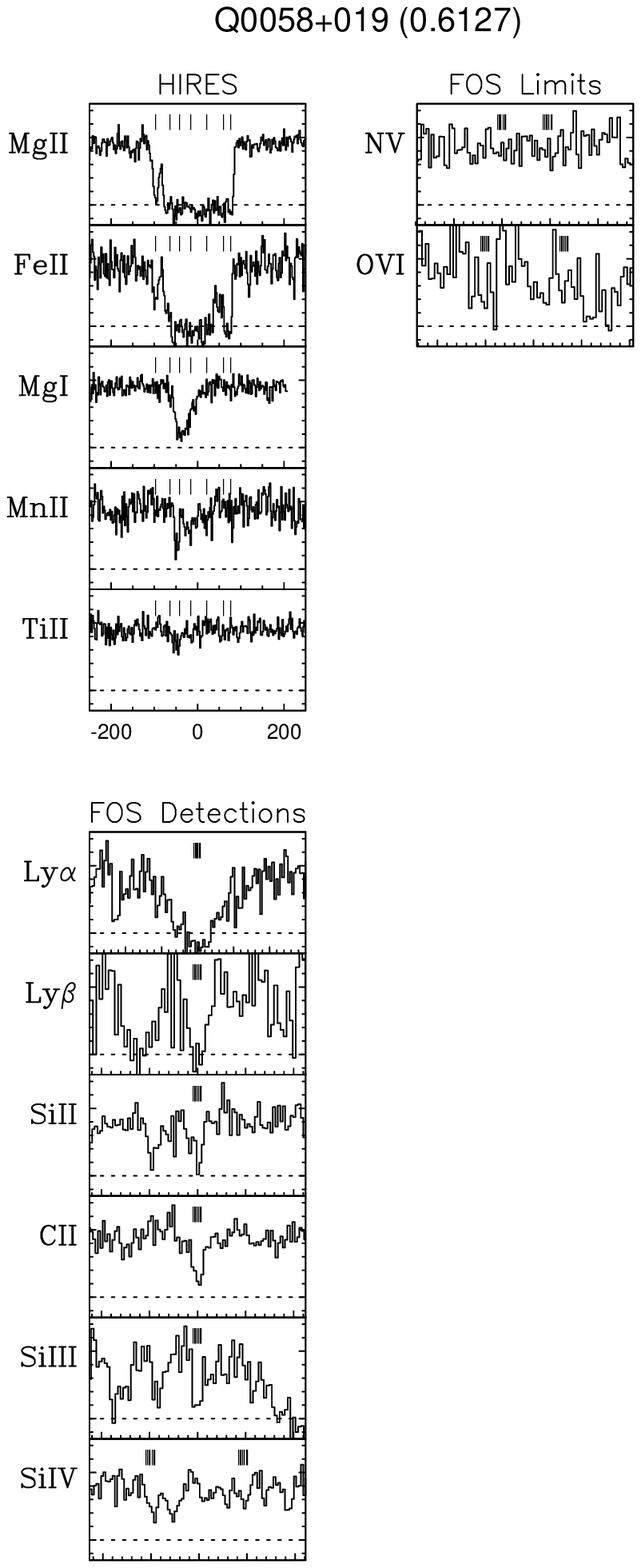, 
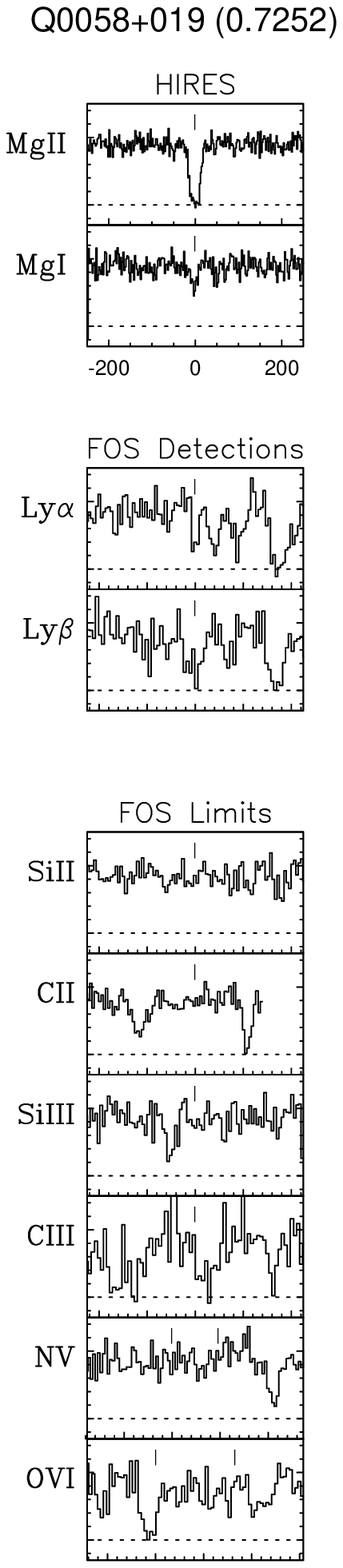, 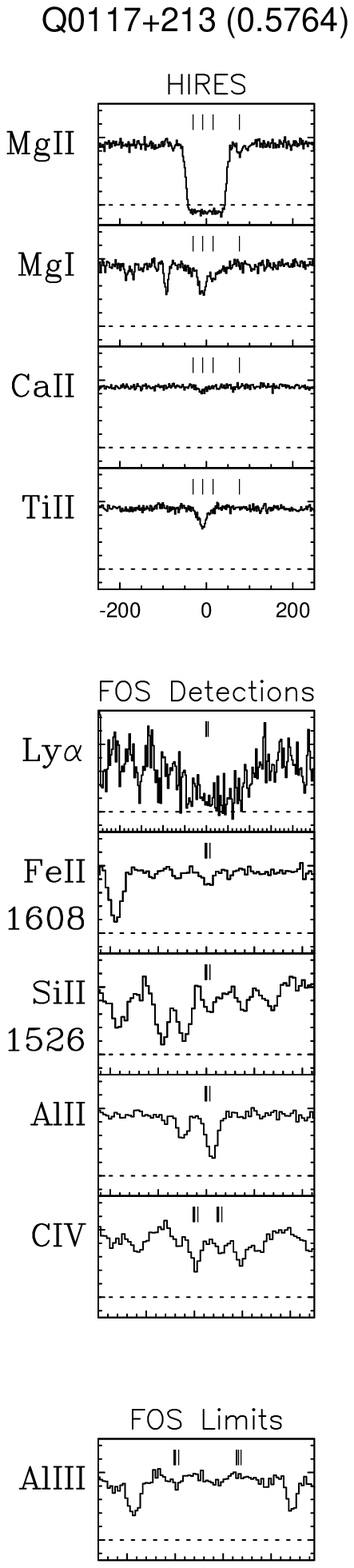, 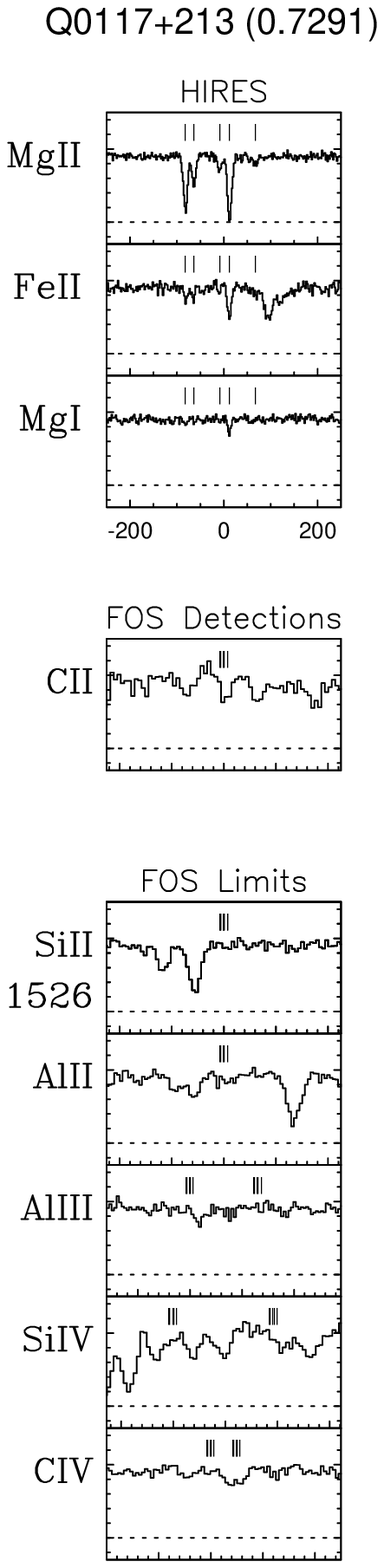, 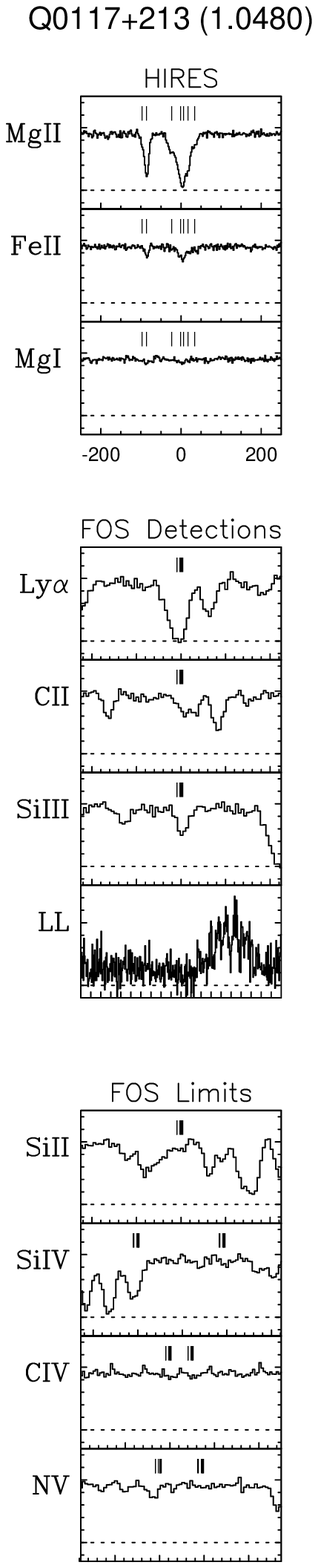, 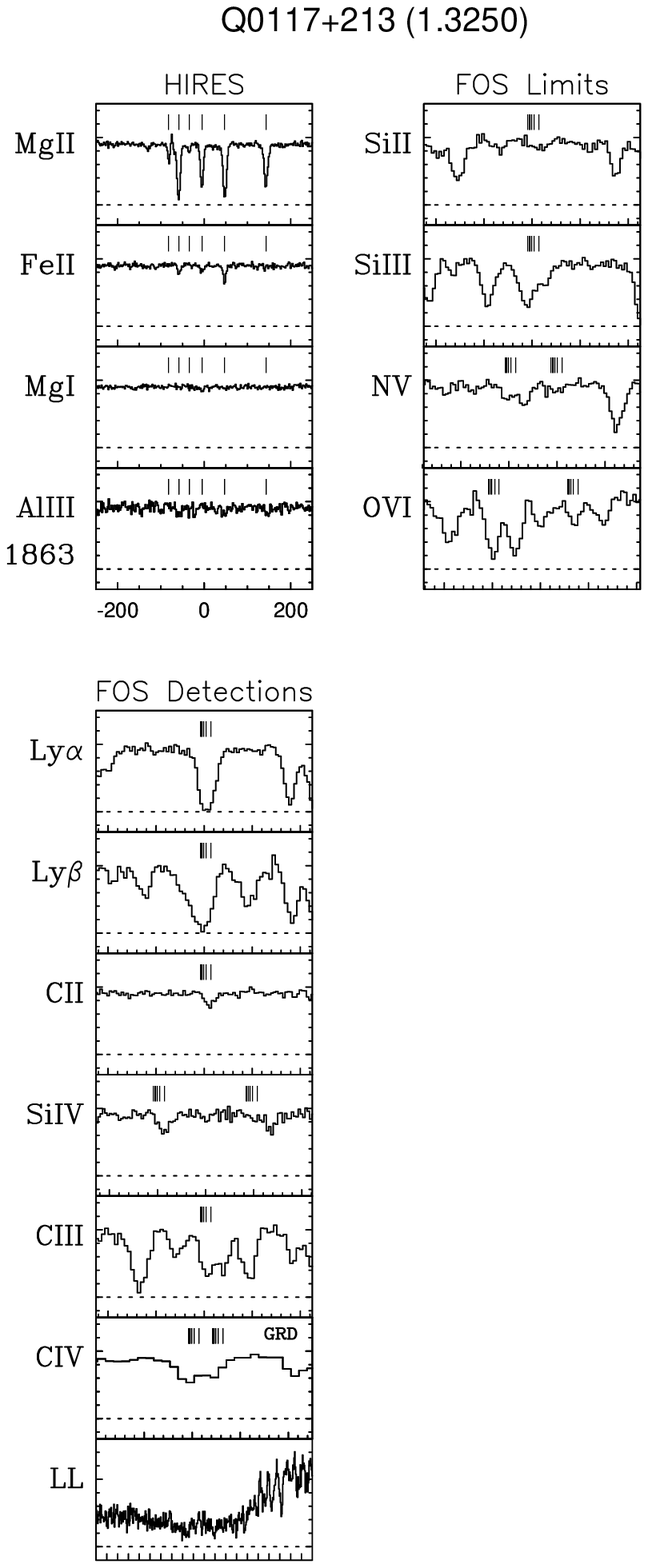, 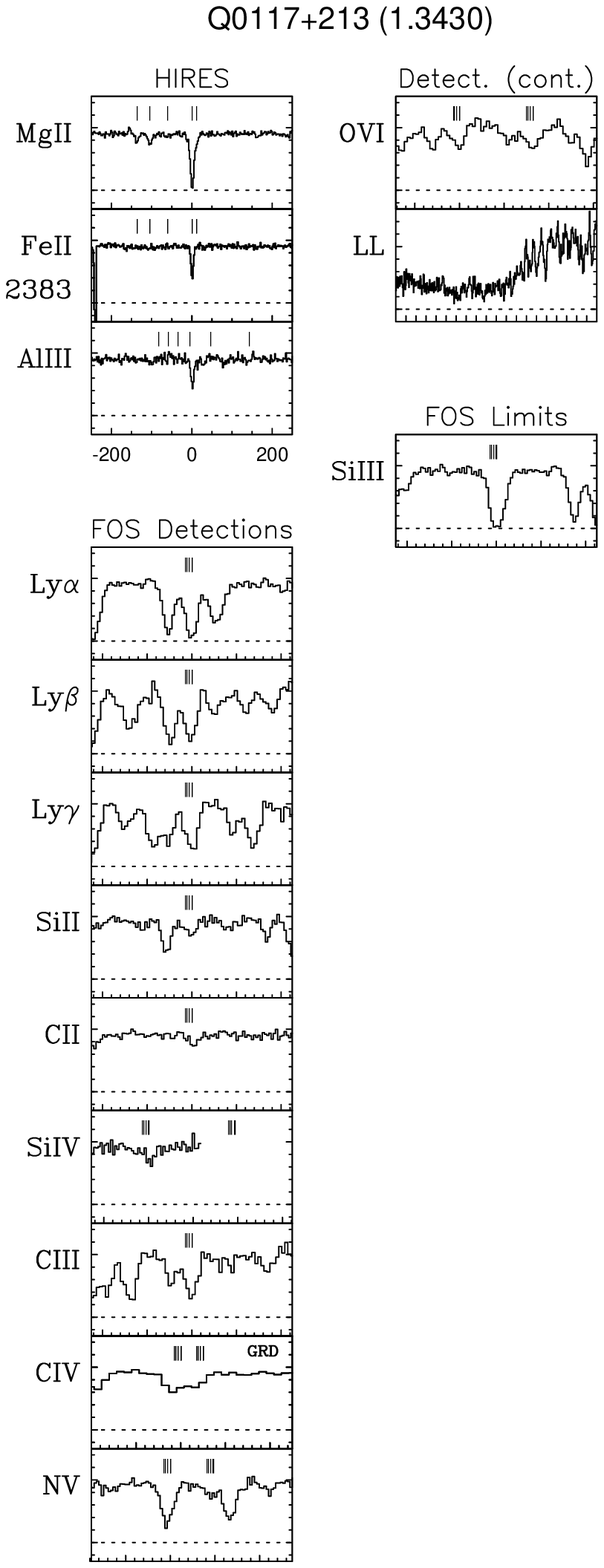, 
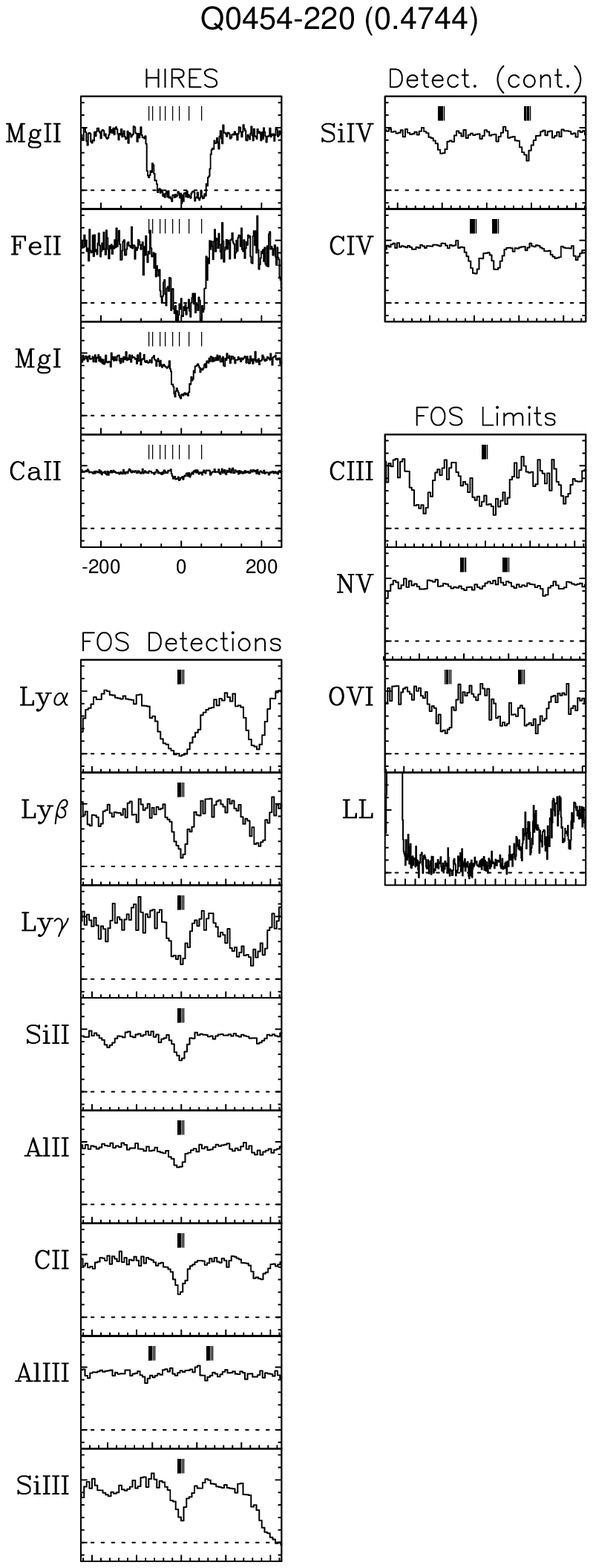, 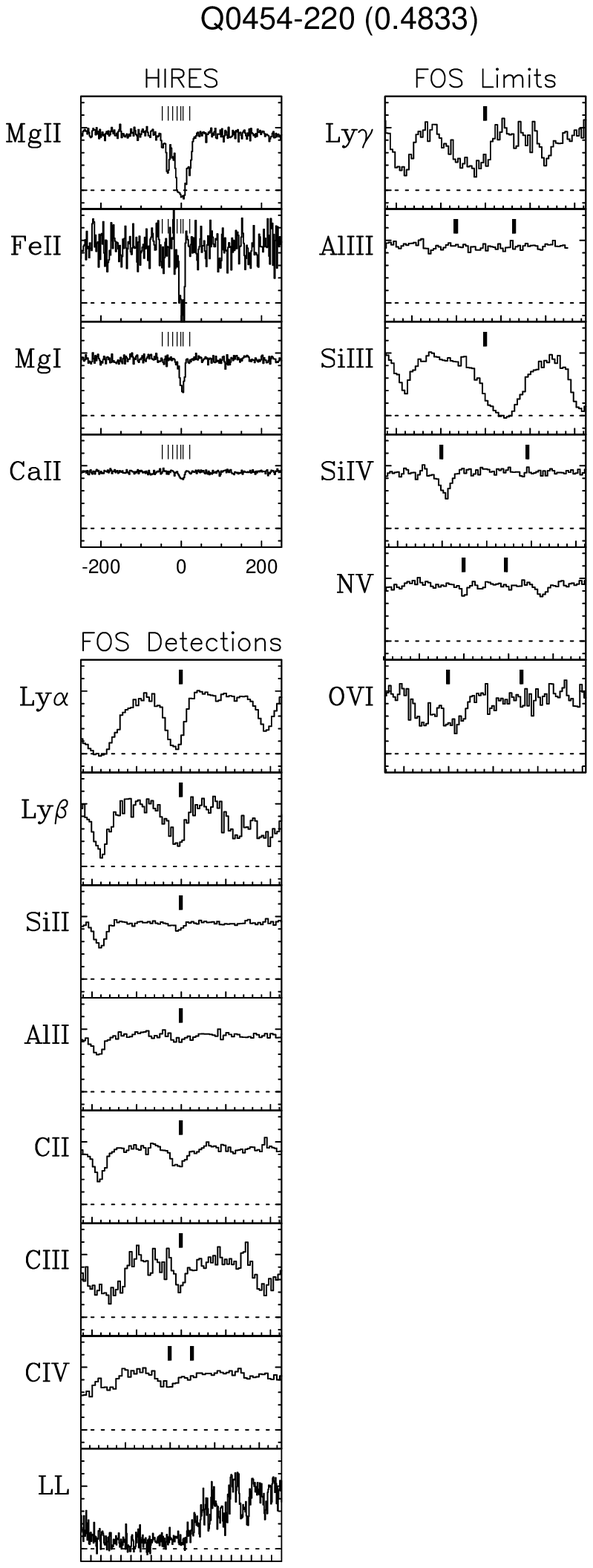, 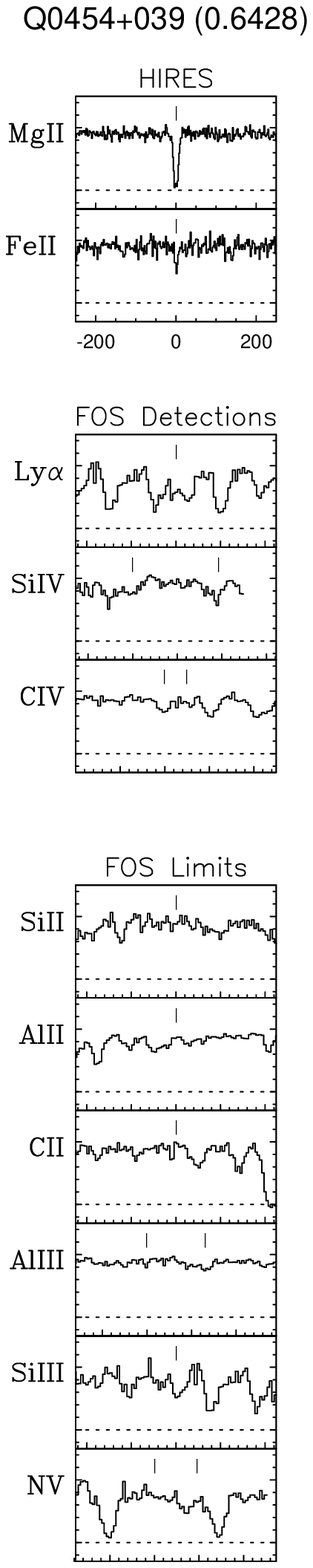, 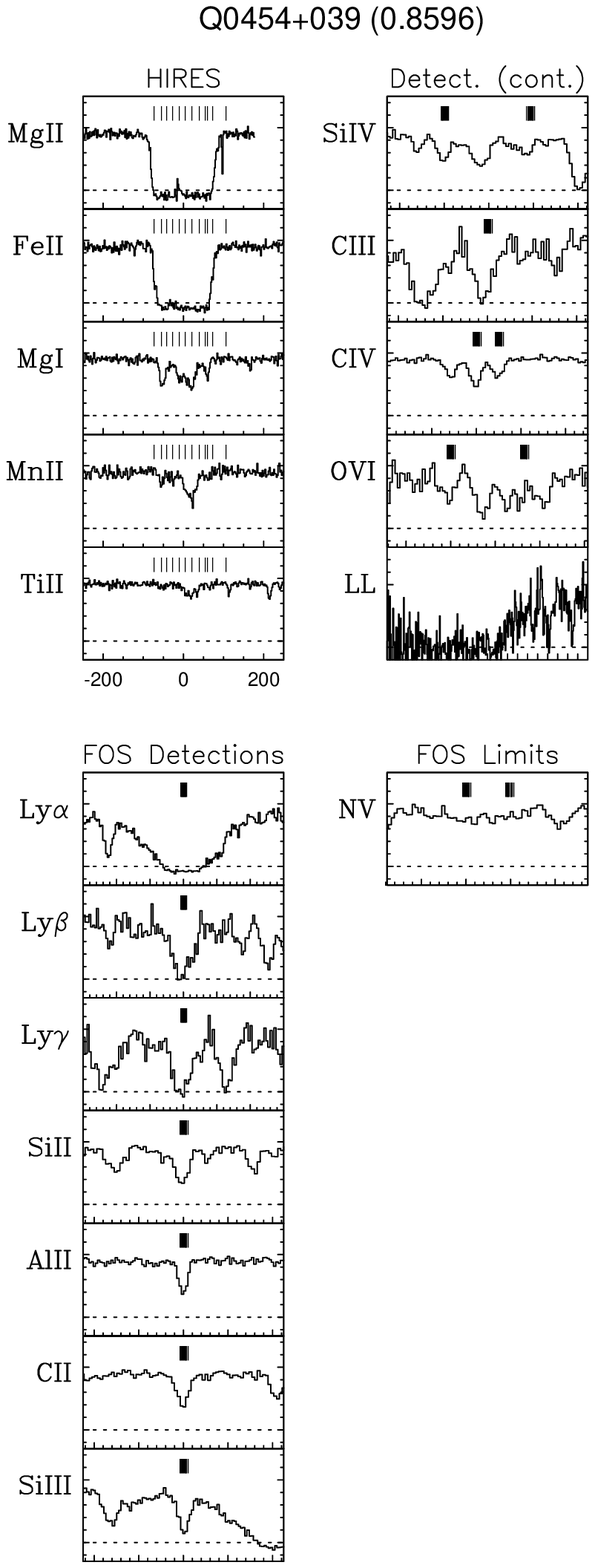, 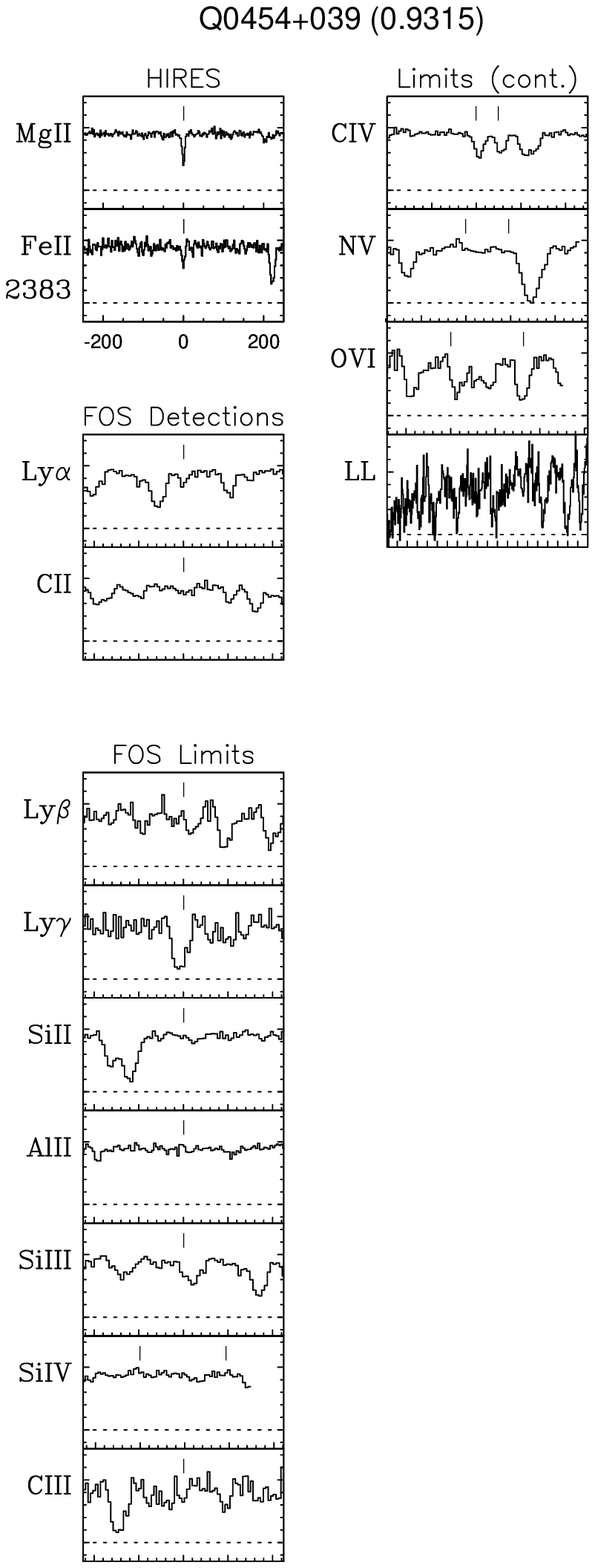, 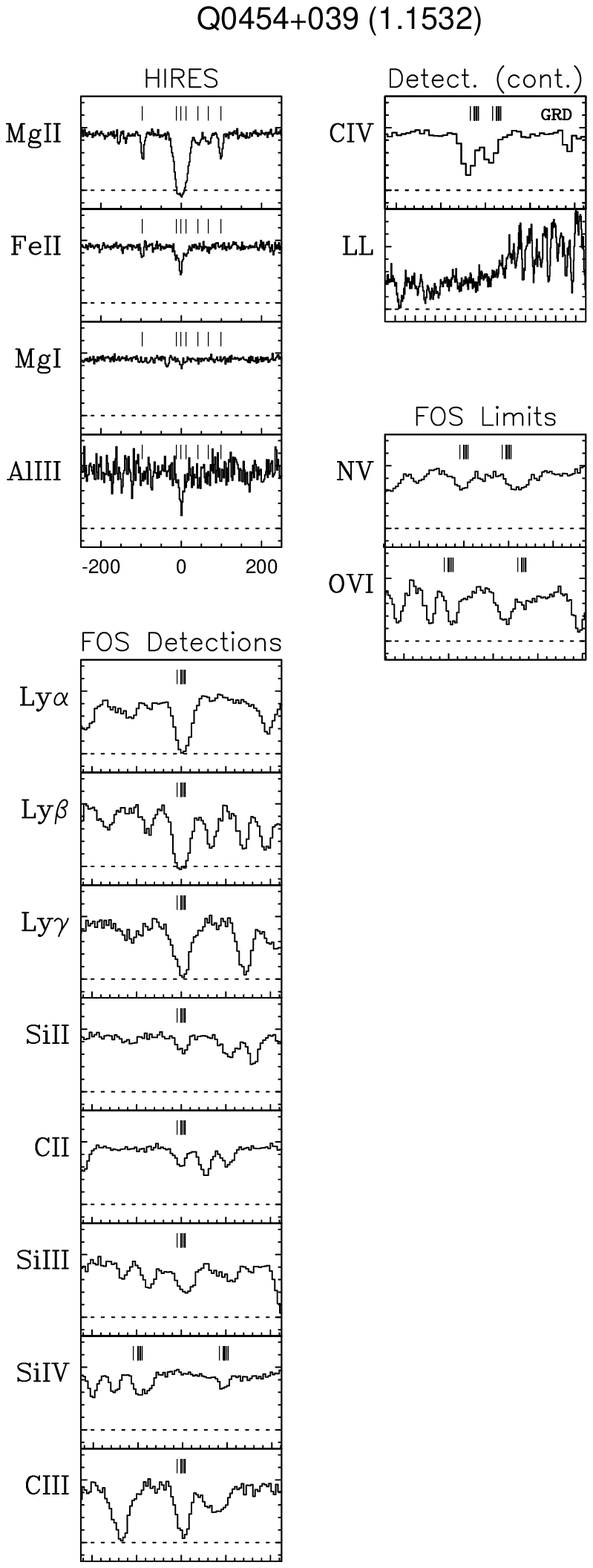, 
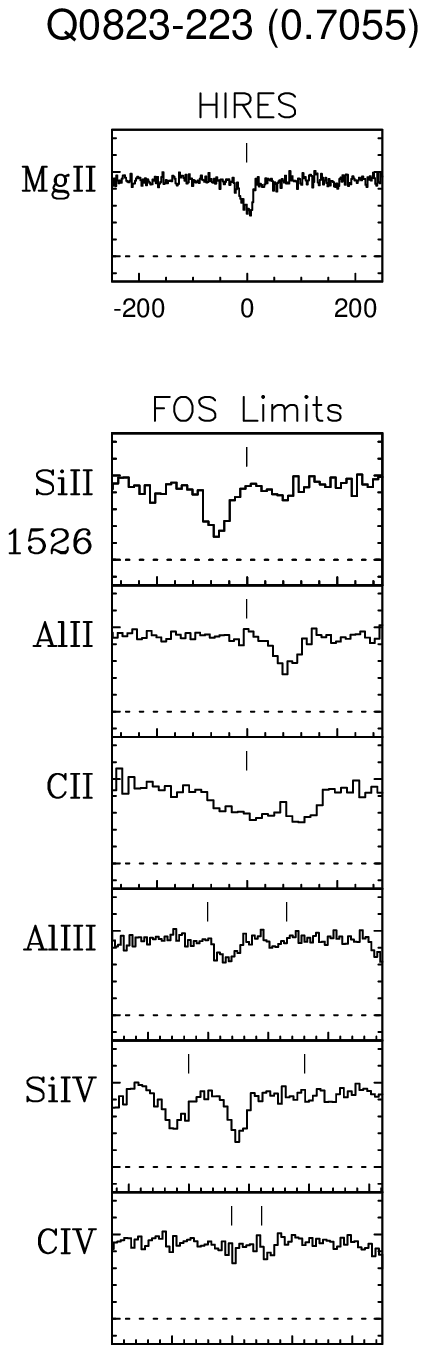, 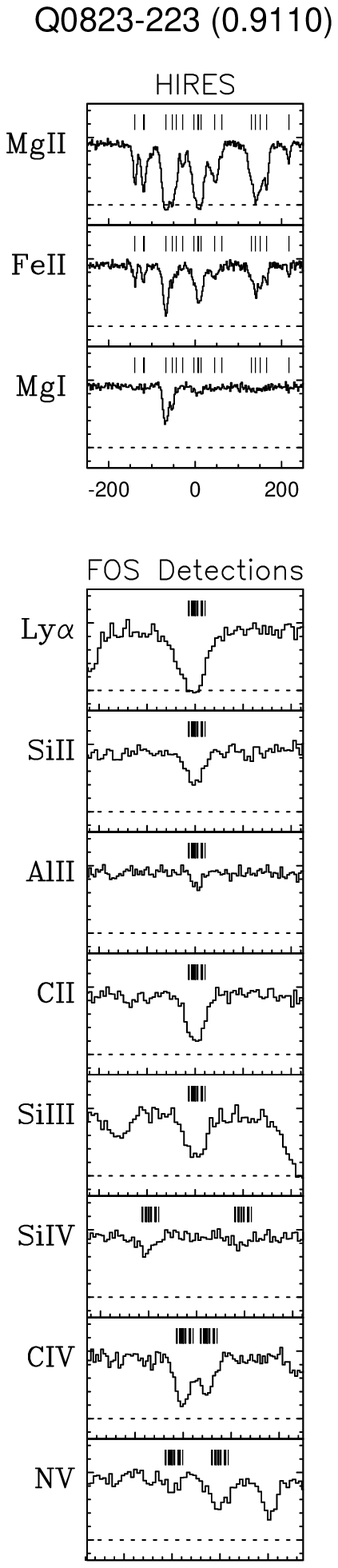, 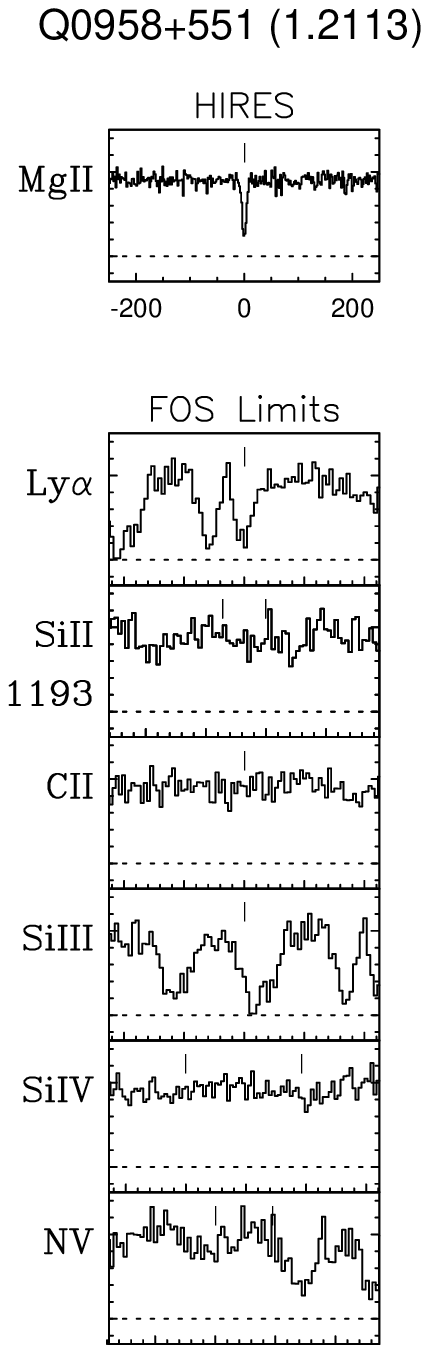, 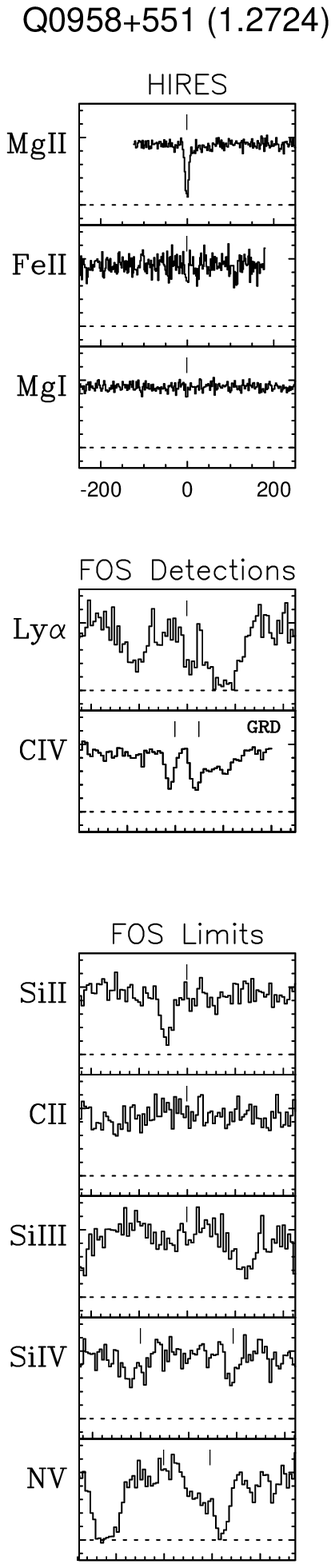, 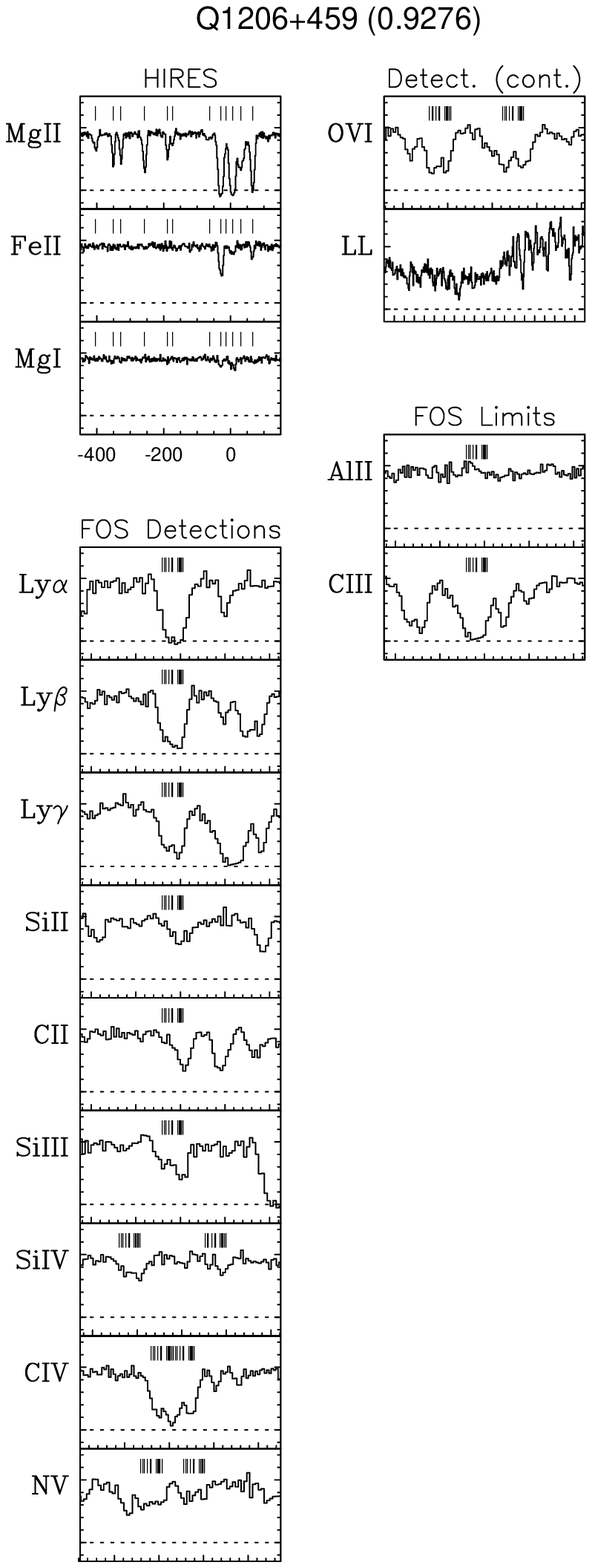, 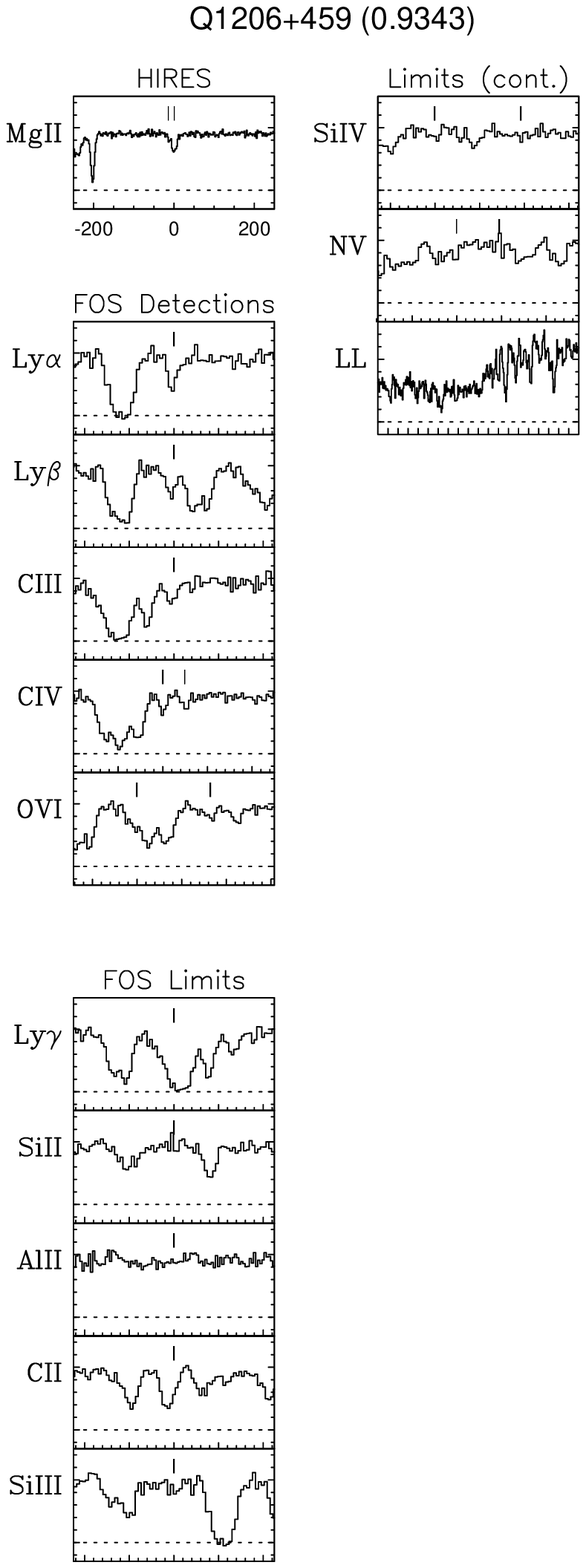, 
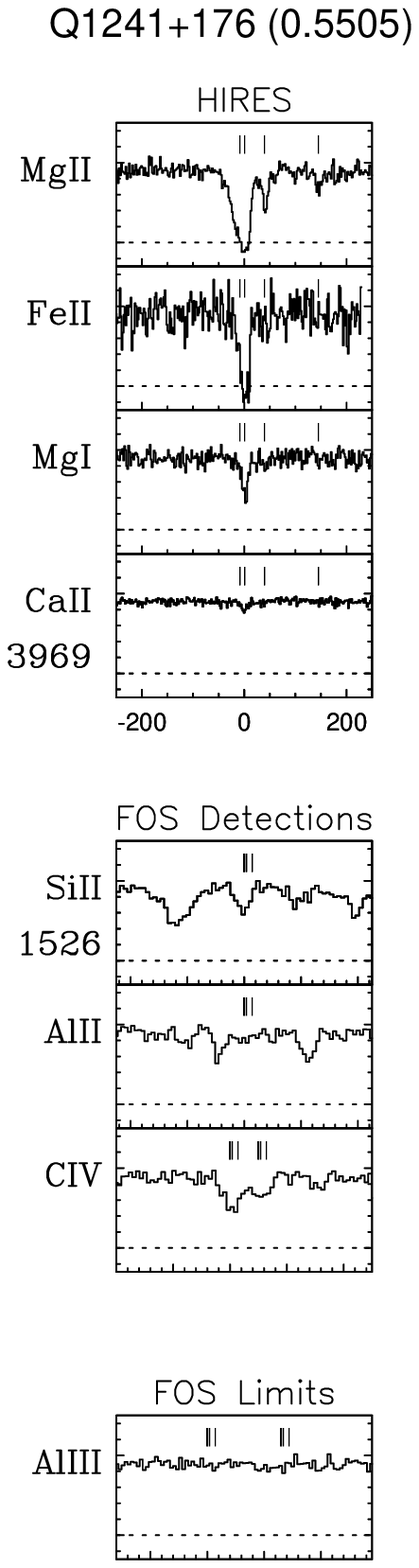, 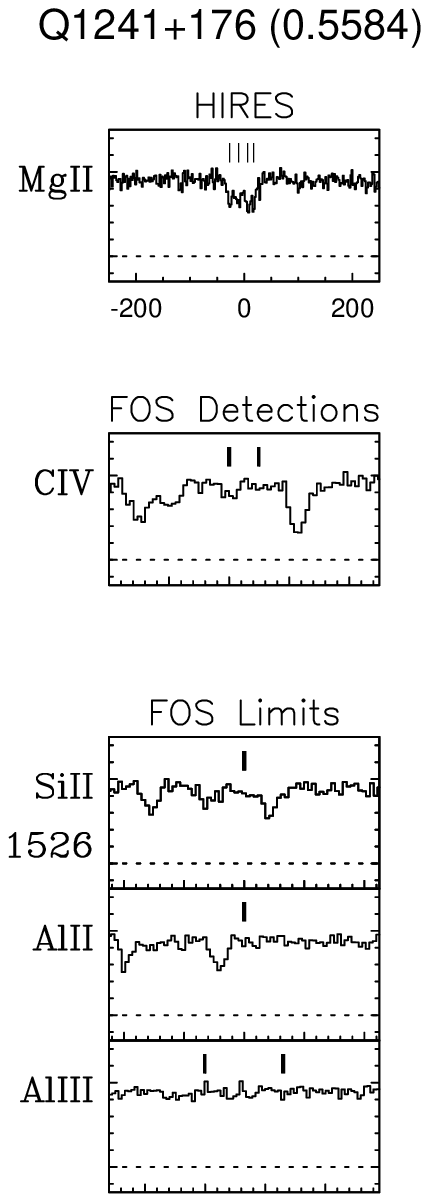, 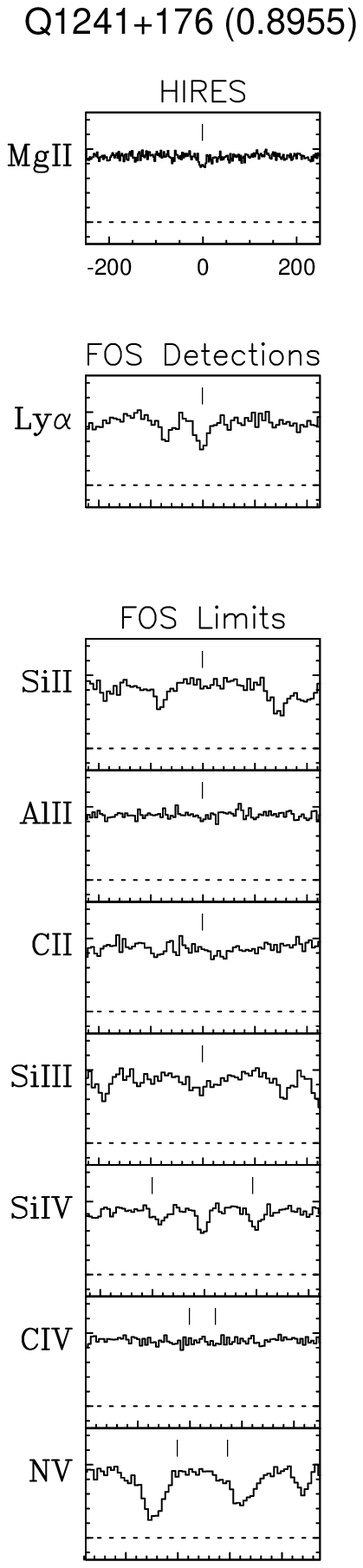, 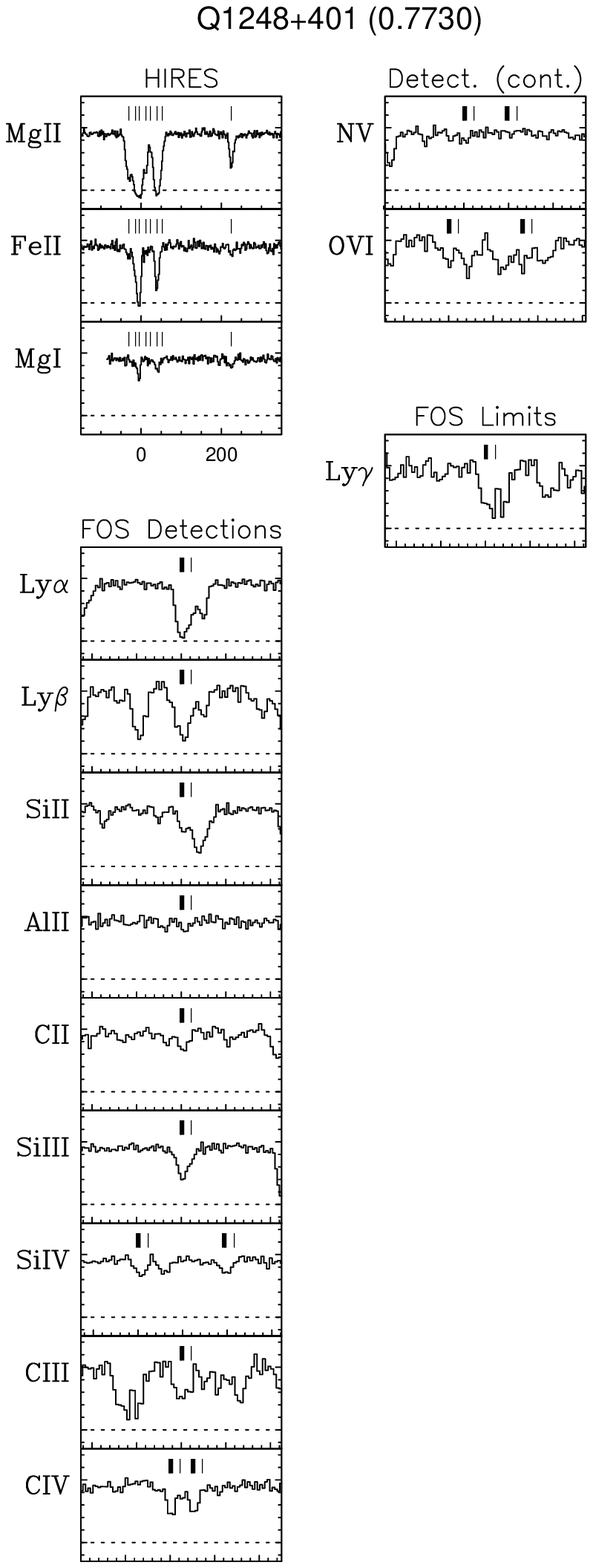, 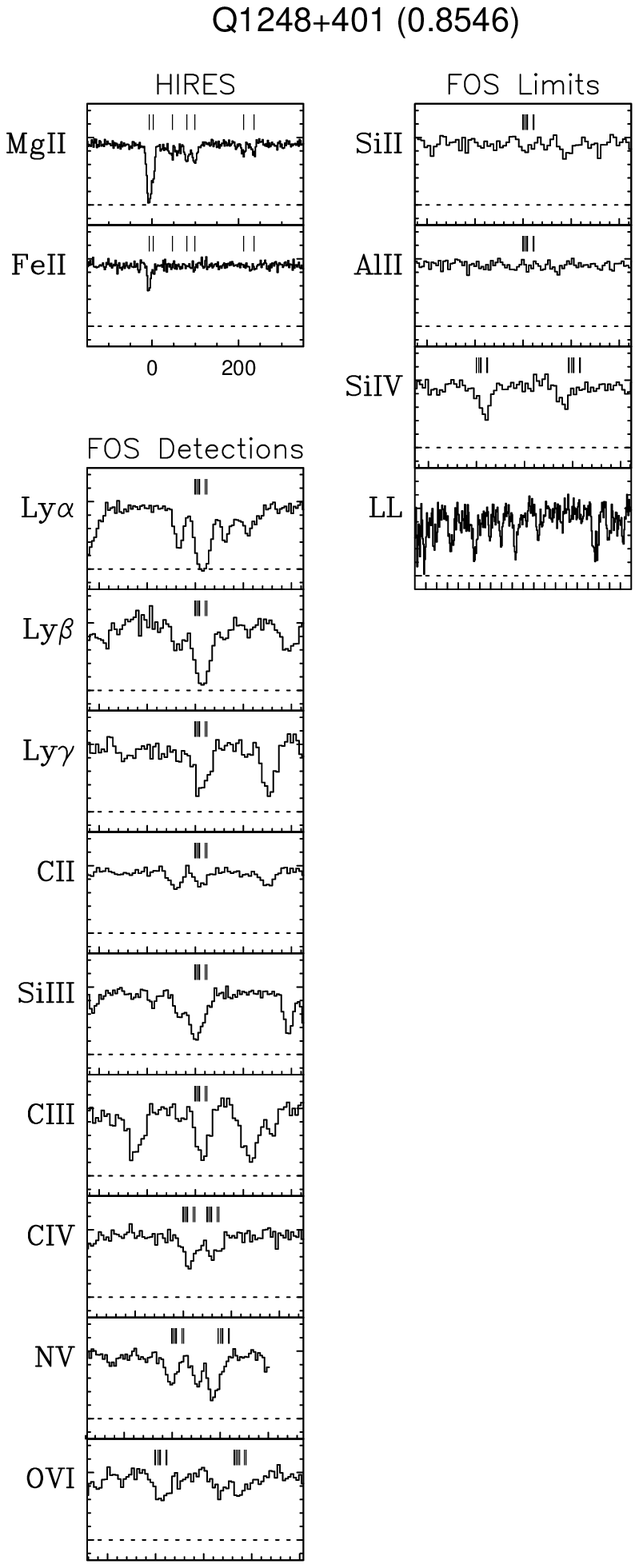, 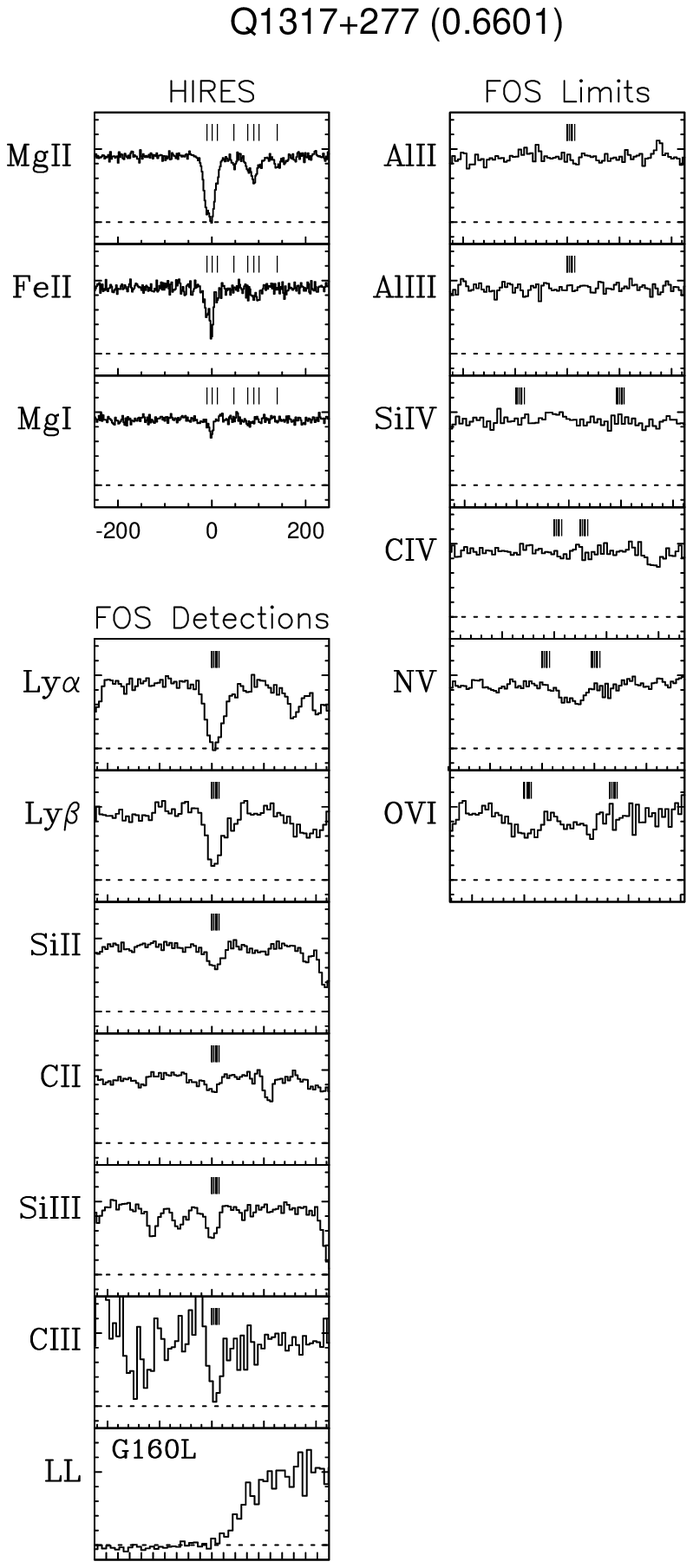, 
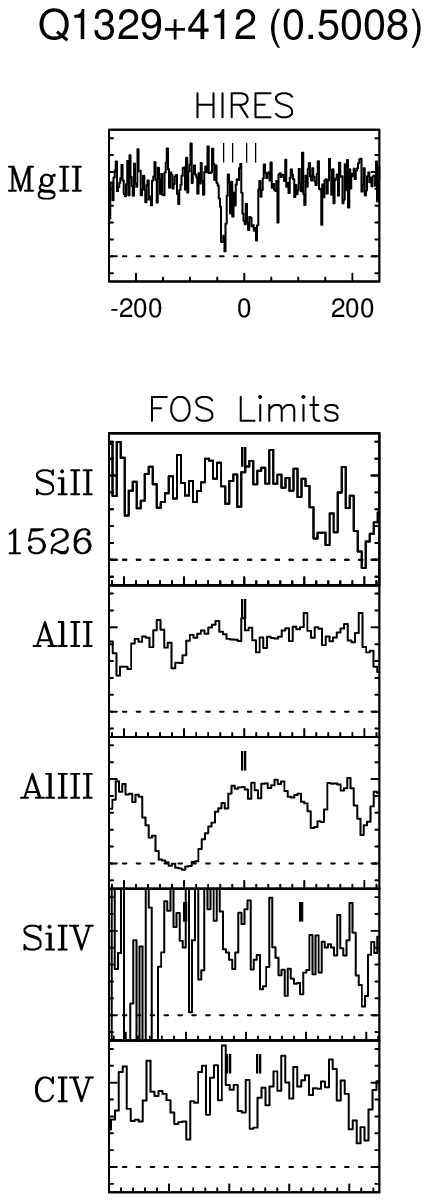, 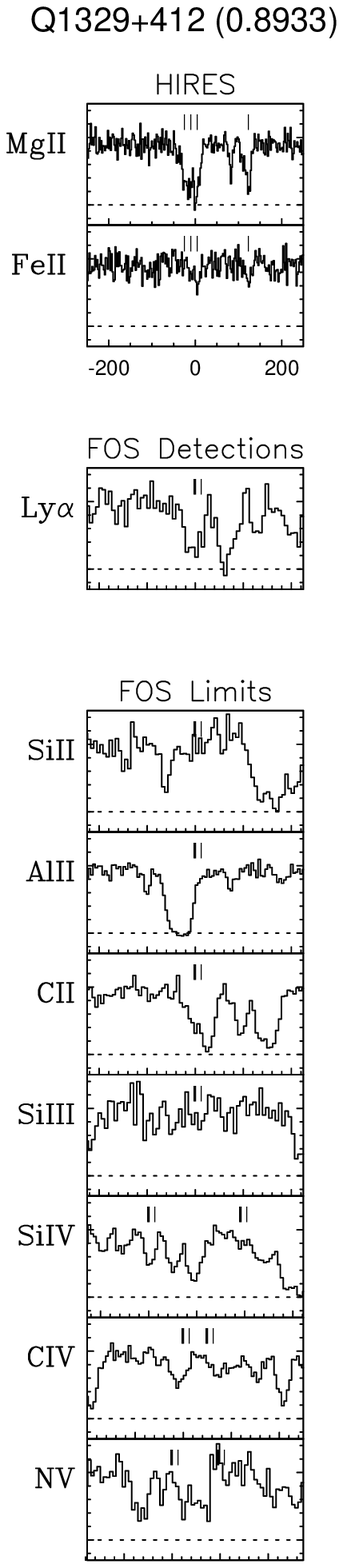, 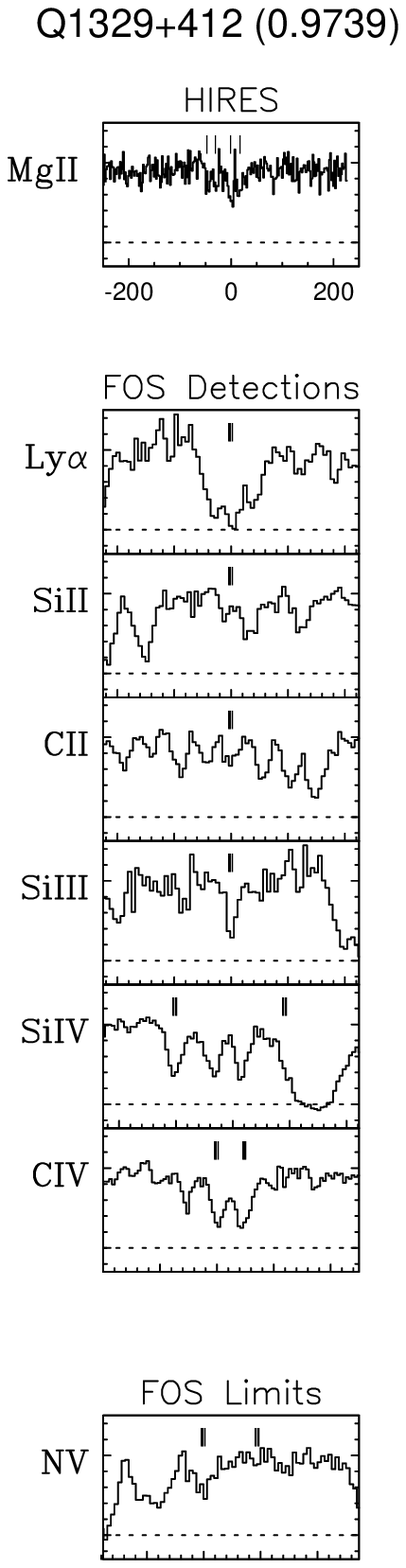, 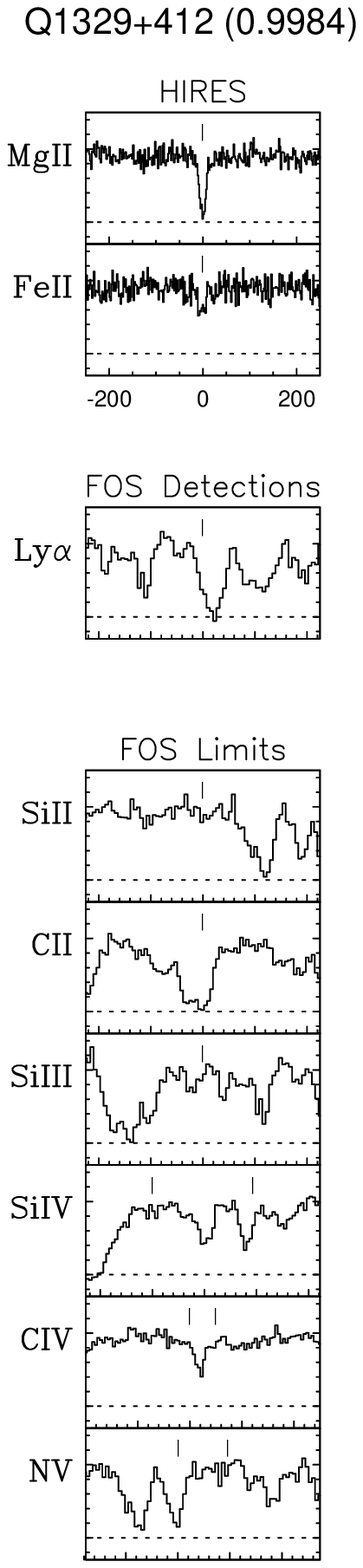, 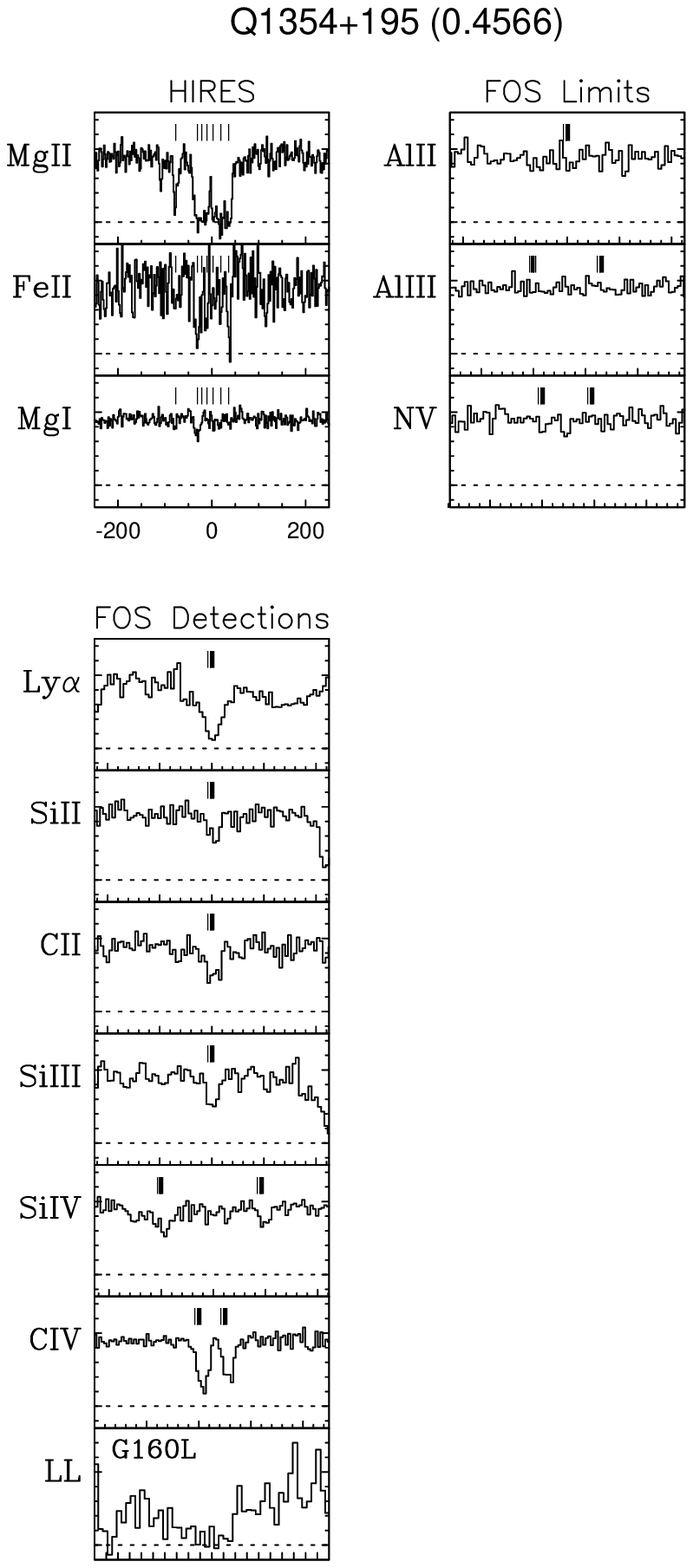, 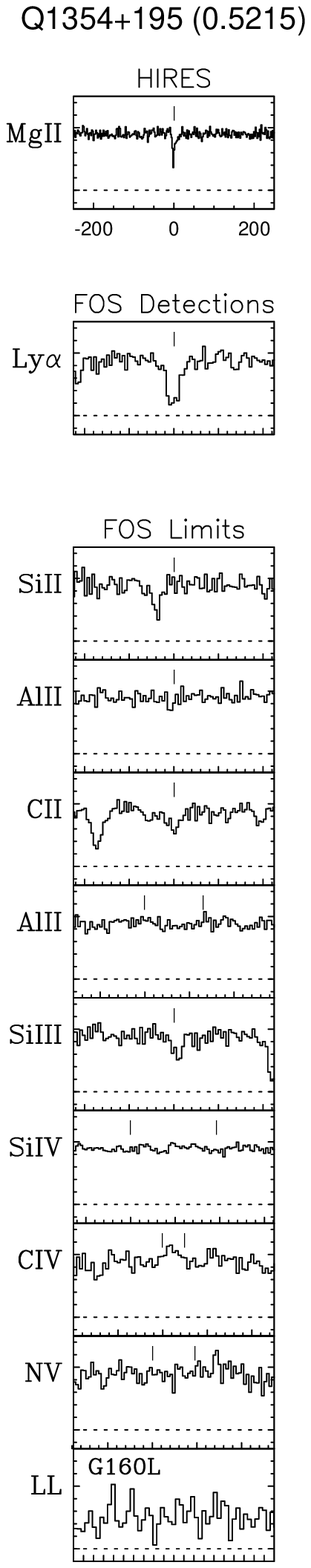, 
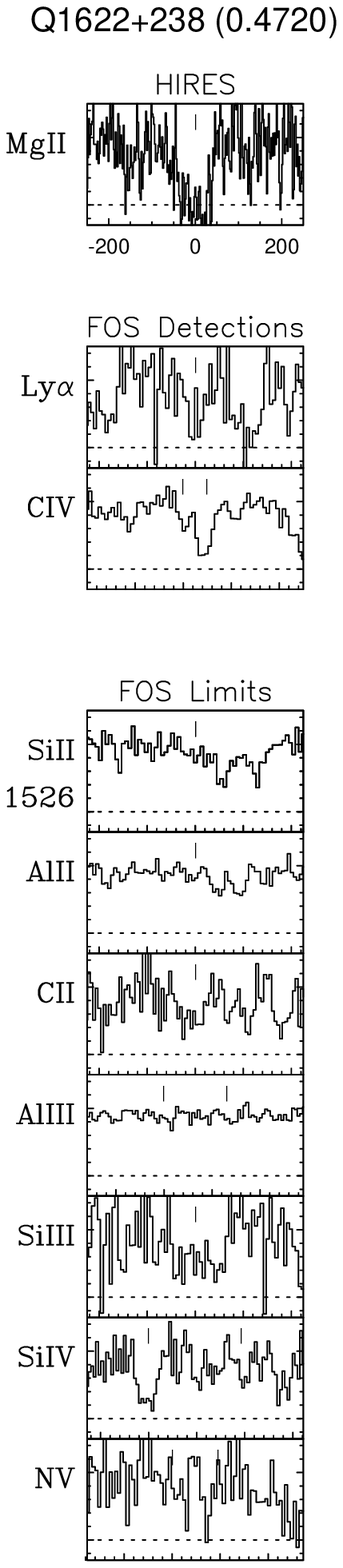, 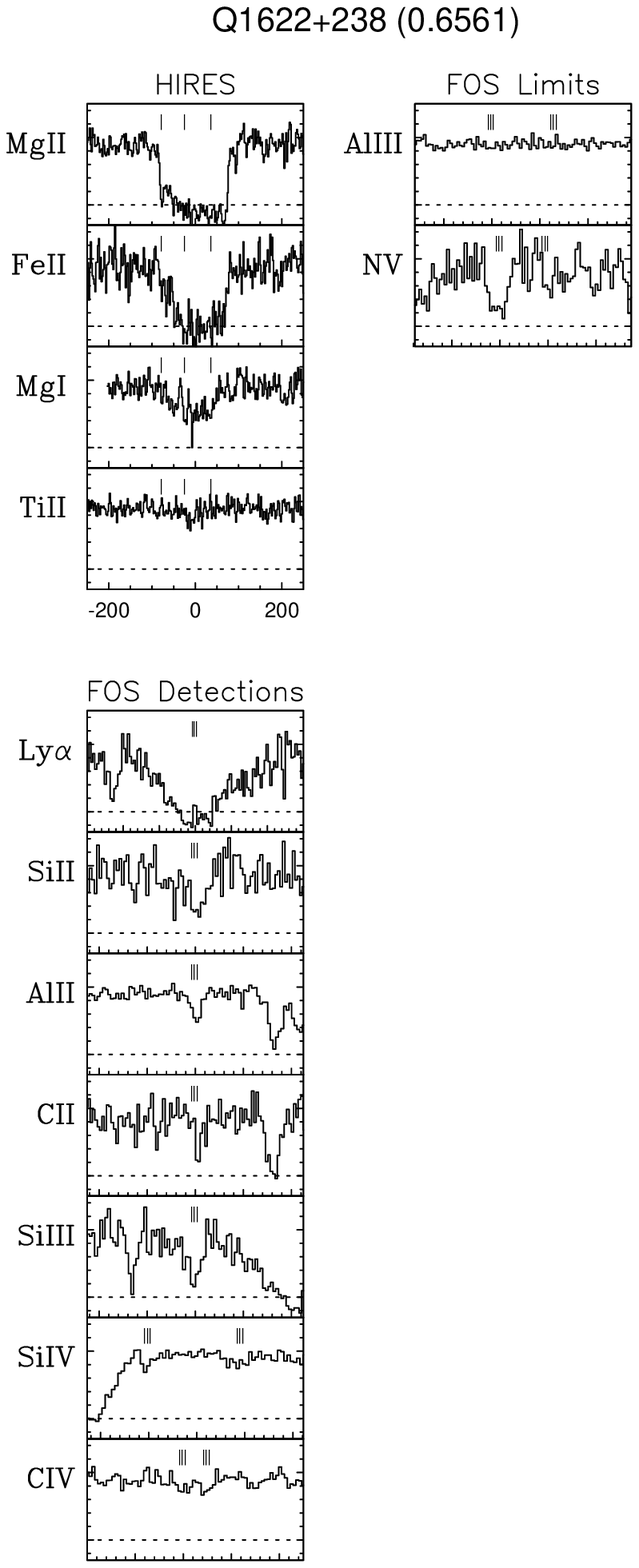, 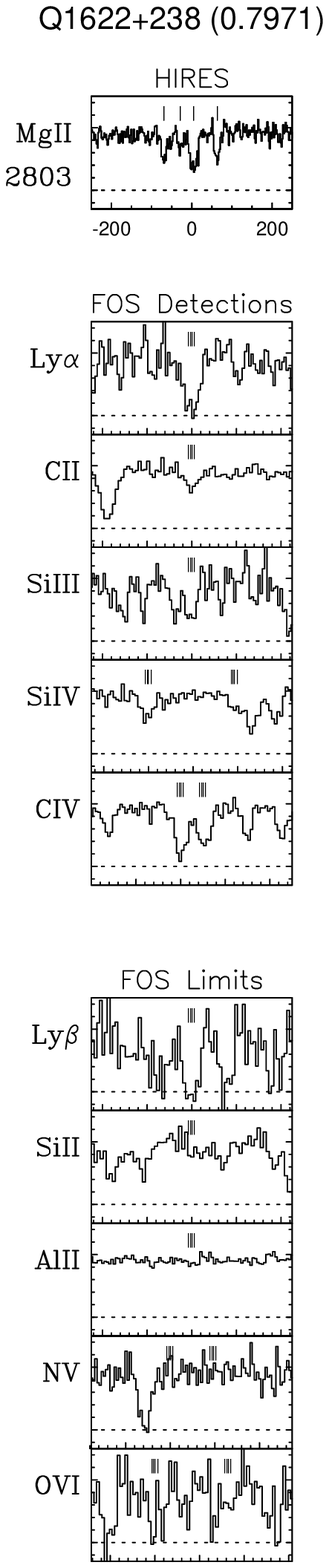, 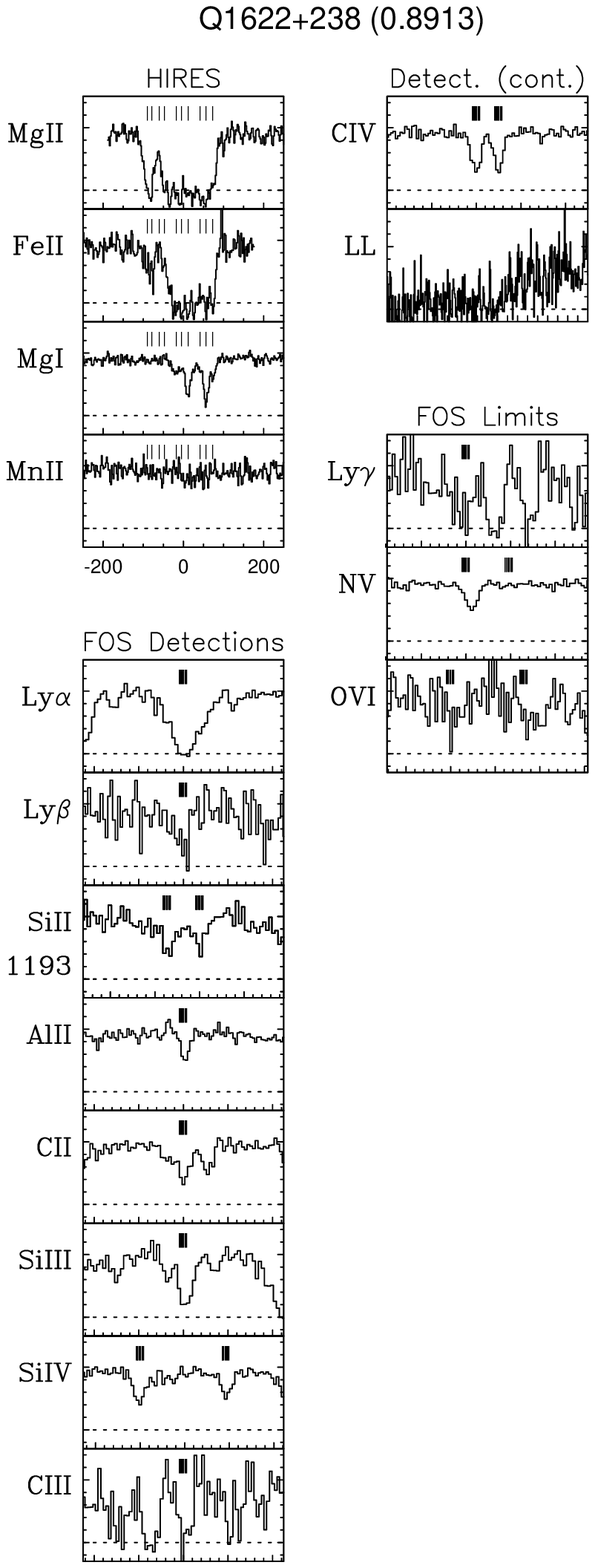, 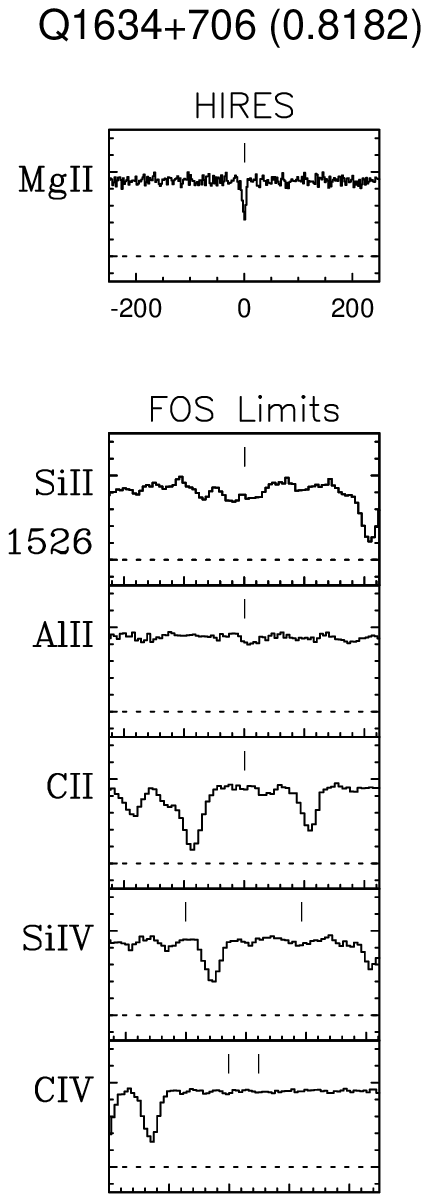, 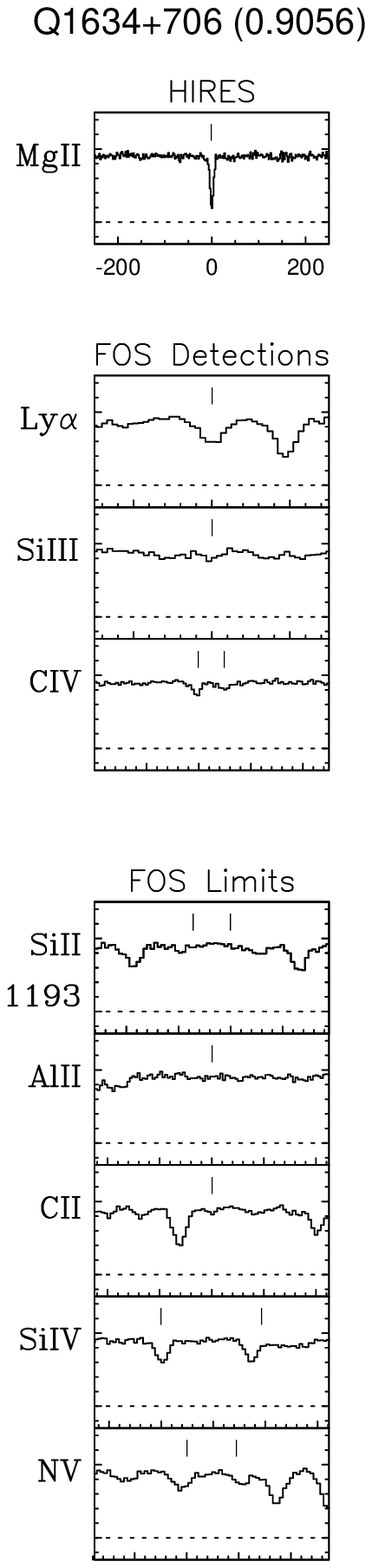, 
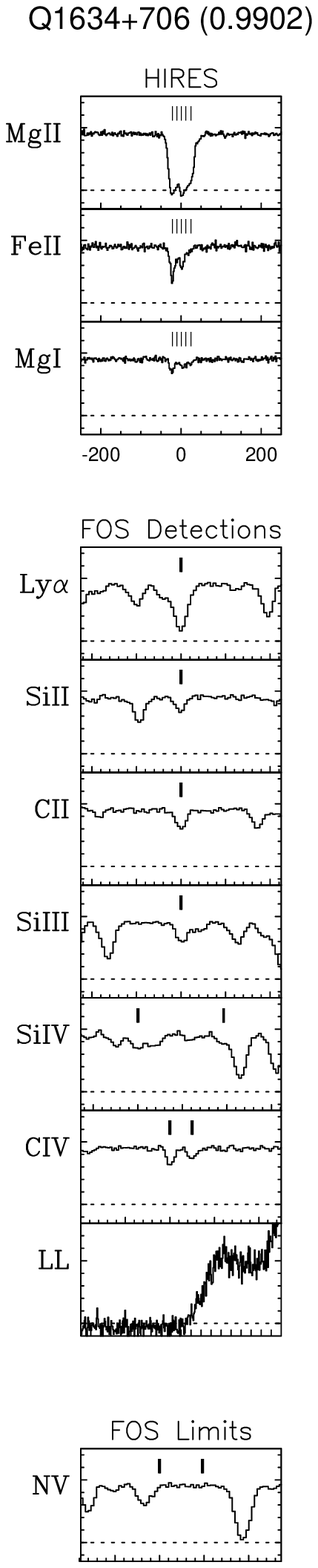, 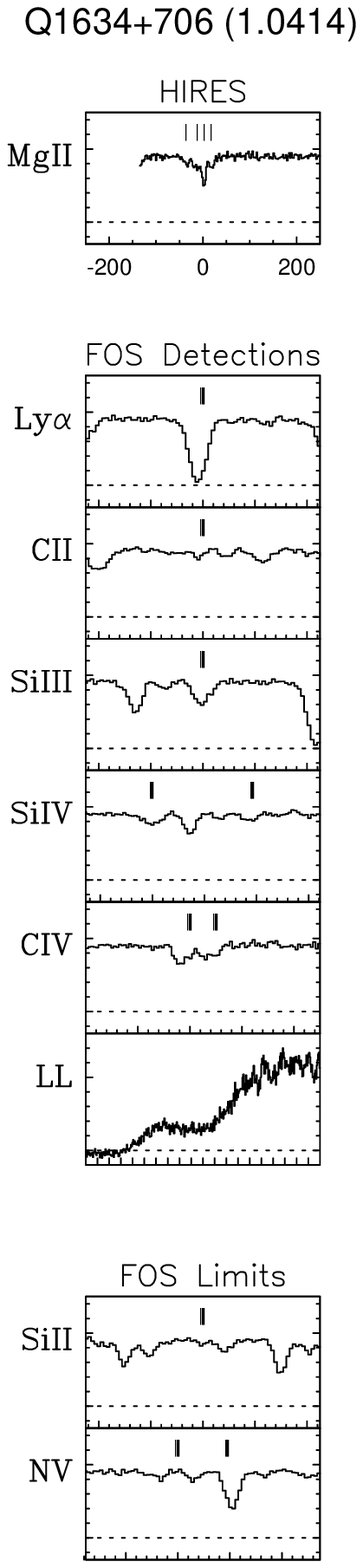, 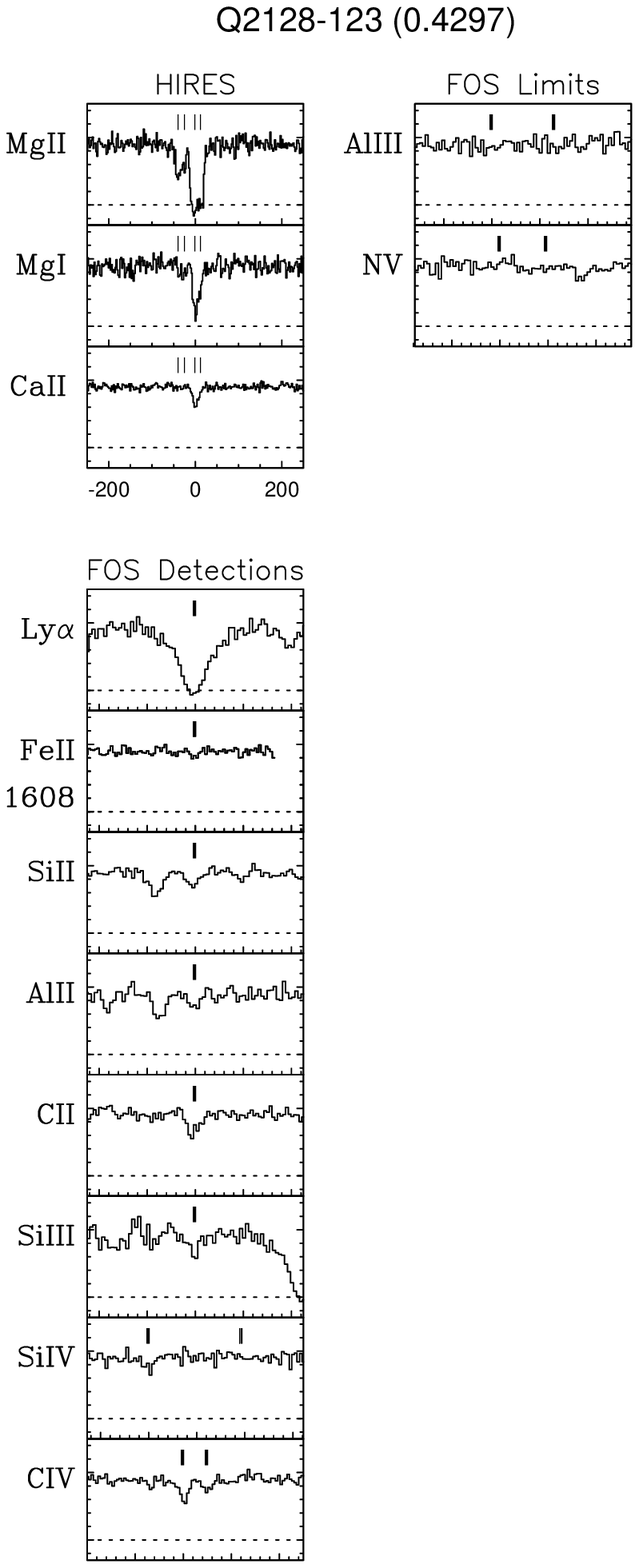, 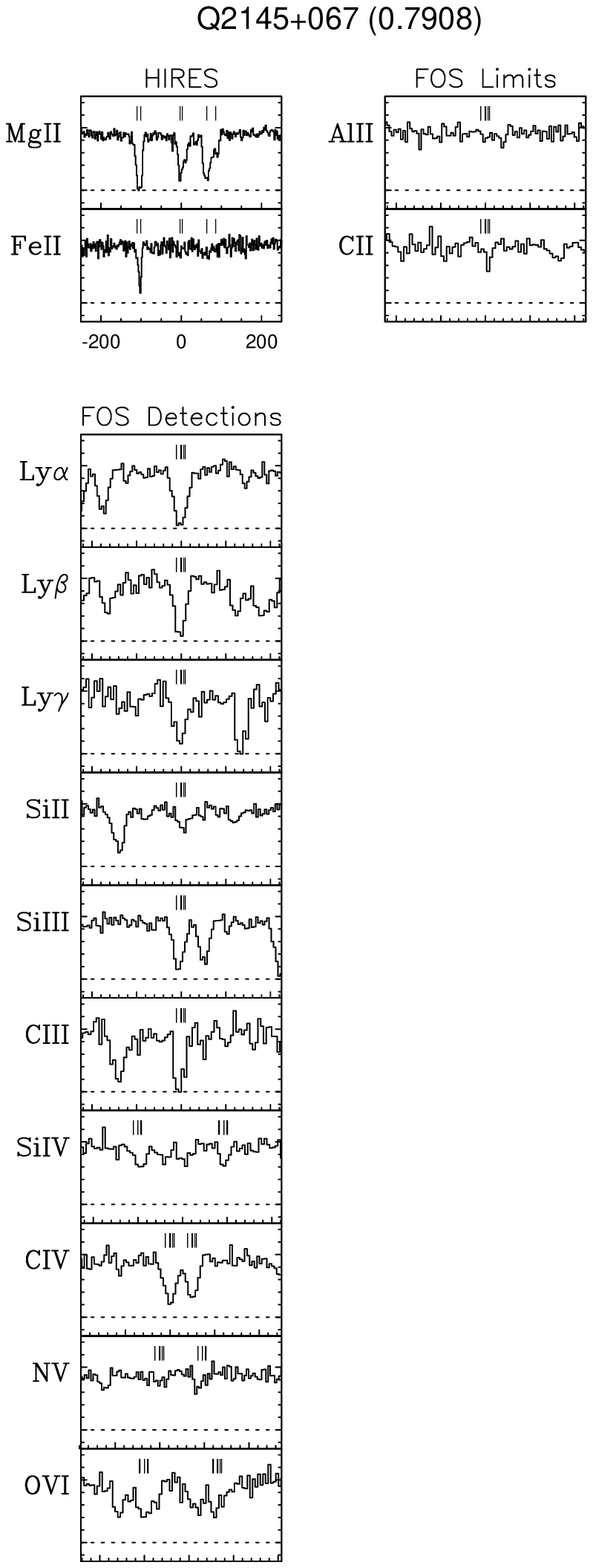]
{$a$--$ss$. For each system, the HIRES spectra of detected {\FeII},
{\MgI}, {\CaII}, and {\TiII}, and the FOS spectra of targetted
transitions, seperated by detections and non--detections.  HIRES
spectra ($\sim 6$~{\kms} resolution) span $500$~{\kms} and are
centered at the optical depth median integrated across the {\MgII}
profile.  FOS spectra ($\sim 230$~{\kms} resolution) are displayed in
a larger velocity window of $4500$~{\kms}, with small ticks on the
horizontal axis separated by $200$~{\kms}.  The vertical ticks above
the profile represent the different components of low ionization gas
derived from Voigt profile decomposition of the HIRES profiles (see
Paper~II.  The identical vertical ticks (derived from the HIRES data)
are superimposed on the velocity scale of the FOS data panels, with a
duplicate set of ticks for each member of the doublet
transitions. \label{fig:portraits}}
\end{figure*}

The results from the HIRES/Keck spectra are presented in
Table~\ref{tab:hiresresults}.
Tabulated are the quasar name, the absorber redshift, the {\MgII}
$\lambda 2796$ rest--frame equivalent width, $W_{r}({\MgII})$, the
{\FeII} $\lambda 2600$ rest--frame equivalent width, $W_{r}({\FeII})$,
and the {\MgII} $\lambda 2853$ rest--frame equivalent width, $W_{r}({\MgI})$.
Equivalent width limits are quoted at the $3\sigma$ level.
In a few systems, we have detected {\MnII}, {\CaII}, or
{\TiII}; their detection is mostly an arbitrary function of
wavelength coverage.
We have not listed them in Table~\ref{tab:hiresresults} (however, we
show these data in Figure~\ref{fig:portraits}).

In Table~\ref{tab:fosresults}, we present the results from the
FOS/{\it HST\/} spectra.
Tabulated are the quasar name, the absorber redshift, the {\Lya}, 
{\Lyb}, and {\Lyg} equivalent widths, status of the Lyman break,
and the equivalent widths of {\AlII}, {\AlIII}, {\SiII}, {\SiIII}, {\SiIV},
{\CII}, {\CIII}, {\CIV}, {\NV}, and {\OVI}.
All equivalent widths are rest frame.
The spectroscopic data are presented in
Figures~\ref{fig:portraits}$a$--$ss$.
For each system, three sets of panels are shown: the HIRES detections,
the FOS detections, and the FOS limits.
The spectra are all normalized by the continuum fits.
Both pre--COSTAR and COSTAR spectra are represented.
We note the pre--COSTAR spectra in the discussion of individual
systems (\S~\ref{sec:discusssystems}).

The panels with HIRES profiles show a velocity window of $-250$ to
$250$~{\kms} centered on the {\MgII} $\lambda 2796$ optical depth
mean.
Ticks above the continuum give the velocity positions of the Voigt
profiles components (the Voigt profile decomposition is described in 
Paper II\nocite{paper2}).
The panels with singlet FOS absorption lines show a velocity window of
$-1250$ to $1250$~{\kms} centered on the line.
The panels with doublet FOS absorption lines also show a velocity
window of $-1250$ to $1250$~{\kms}, but the zero point is arbitrarily
set half way between the doublet members in order to center the
doublet in the panel.
Ticks above the FOS spectra show the expected positions of the
absorption based upon the {\MgII} Voigt profile components.

For each system we show the flag transition or both members of a doublet
for the species presented in Tables~\ref{tab:hiresresults} and
\ref{tab:fosresults}.
If a weaker transition was used for a measurement, then it is labeled
in Figures~\ref{fig:portraits}$a$--$ss$ and a footnote is placed in  
Table~\ref{tab:hiresresults} or Table~\ref{tab:fosresults}.
For the HIRES spectra, we show the strongest {\TiII}, {\CaII},
and/or {\MnII}, even though these species are not listed in
Table~\ref{tab:hiresresults}.

\subsection{Discussion of the Individual Systems}
\label{sec:discusssystems}

\subsubsection{\rm Q~$0002+051$, UM 18~~~~($z_{\rm em}=1.90$)}

The FOS spectra of this quasar, which has four intervening {\MgII}
absorbers, were previously studied by
Jannuzi \etal (1998\nocite{cat3}) and Koratkar \etal
(1998\nocite{koratkar}).
Jannuzi \etal obtained only the G270H grating spectrum, with pre--COSTAR
optics, and using the $0.25{\arcsec} \times 2{\arcsec}$ slit.
Koratkar \etal obtained both the G190H and G270H grating spectra 
in spectropolarimetry mode ($1.0${\arcsec} aperture).
We have included only the Jannuzi \etal spectrum in our study,
except for a constraint on the Lyman limit break obtained from
the Koratkar \etal G190H spectrum.

\begin{description}

\item{$z=0.5915$} --- 
This system has no detectable {\FeII} or {\MgI} in the HIRES
spectrum.
In the FOS spectrum,
{\SiII} $\lambda 1526$ is blended with a {\Lyg} line at $z=1.5000$
and we thus conservatively fit the entire blend and quote an upper 
limit for $W_{r}({\SiII})$.
{\CIV} $\lambda 1548$ is blended with {\SiII} $\lambda 1260$ at
$z=0.9560$ and with {\Lyg} at $z=1.5359$.
The quoted upper limit for $W_{r}({\CIV})$ is taken to be
the $3\sigma$ equivalent width limit at the expected position of 
$\lambda 1550$.
{\AlII} $\lambda 1670$ is blended with {\Lyd} at $z=1.8016$
so we measure that line and record it as the limit on $W_{r}({\AlII})$.
The Lyman limit was not covered.

\item{$z=0.8514$} --- 
This system is rich in {\FeII} and {\MgI}  in the HIRES spectrum.
In the FOS spectrum, 
the quoted $W_{r}({\Lya})$ may be very slightly overestimated due to
the coincidence of {\SiIII} $\lambda 1206$ at $z=0.8665$.
We caution that the {\SiII} $\lambda 1526$ transition is
weak relative to our claimed $\lambda 1260$ detection, but
is consistent within the permitted scaling.
Jannuzi \etal (1998\nocite{cat3}) identified a {\Lya} line at
$z=1.3592$ at the location of {\CIV} $\lambda 1548$, but our
new knowledge of the presence of a {\MgII} system implies
our preferred {\CIV} $\lambda 1548$ identification.
The unphysical doublet ratio of {\CIVdblt} is due to a blend of
$\lambda 1550$ with {\Lyb} at $z=1.8016$, which is part of a clear 
Lyman series.
Both members of the {\SiIVdblt} doublet are blended, the former with
{\Lyb} at $z=1.5160$ and the latter with Galactic {\FeII} $\lambda
2586$.  A limit on $W_{r}({\SiIV})$ was recorded based on measuring 
the blend at the expected position of {\SiIV} $\lambda 1393$.
{\NV} $\lambda 1242$ is blended with {\Lyb} at $z=1.2472$.
There is a strong Lyman limit break with $\tau \simeq 1.4$.

\item{$z=0.8665$} --- 
This system has no detectable {\FeII} or {\MgI}  in the HIRES
spectrum.
In the FOS spectrum,
the blue wing of the {\Lya} line is blended with {\Lyd} at $z=1.3866$;
we deblended the lines with a Gaussian fit to obtain the quoted 
$W_{r}({\Lya})$.
Jannuzi \etal (1998\nocite{cat3}) favored the identification of
a {\Lya} line at $z=1.5653$ at the location of {\AlII} $\lambda 1670$.
Furthermore, the profile is extremely broad (greater than 500~{\kms}
in the rest--frame), suggesting a blend; we thus quote an upper limit
on $W_{r}({\AlII})$.
{\SiIII} $\lambda 1206$ is coincident with the strong {\Lya} line at
$z=0.8514$.
{\SiIV} $\lambda 1393$ is blended with Galactic {\FeII} $\lambda 2600$
so an upper limit on $W_{r}({\SiIV})$ was measured at the position
of $\lambda 1402$.
{\CIV} $\lambda 1548$ is blended with {\Lyb} at $z=1.8184$; we used
the $3\sigma$ limit on $\lambda 1550$ to compute the upper limit on 
$W_{r}({\CIV})$.
From a model of the Lyman limit, we determined that this system does
not contribute to the observed break from the $z=0.8514$ system.

\item{$z=0.9560$} --- 
This system has no detectable {\FeII} or {\MgI} in the HIRES
spectrum.  In the FOS spectrum, we claim a detection of 
{\SiIII} $\lambda 1206$, but note that it could be blended
with {\Lyb} at $z=1.3011$ (if {\Lya} is hidden in a complex
blend at $2797$~{\AA}).
{\SiII} $\lambda 1260$ is blended with 
{\Lyg} at $z=1.5359$, and other members of that series are cleanly
detected.  The other {\SiII} transitions are blended with other possible
lines so cannot be used to corroborate or refute a {\SiII} $\lambda 1260$
identification.  We quote a limit on {\SiII} from measurement of the blended
line at the position of $\lambda 1260$.  
{\SiIV} $\lambda 1393$ is blended in the red wing 
of {\Lyg} at $z=1.8016$, but a clean upper limit can be derived from
the position of {\SiIV} $\lambda 1402$.  The region where the Lyman limit
is expected exhibits a complex shape, however, there is no apparent
break.

\end{description}

\subsubsection{\rm Q~$0058+019$, PHL 938~~~~($z_{\rm em}=1.96$)} 

Only a fairly low signal--to--noise ratio, G190H FOS spectrum was
obtained by Rao \& Turnshek (1999\nocite{rao99}) using the $1.0${\arcsec}
aperture.  The signal--to--noise ratio is significantly reduced blueward of 
the Lyman limit break at $\sim 2250$~{\AA}.

\begin{description}

\item{$z=0.6127$} --- 
This system is a damped {\Lya} absorber and is seen to have {\FeII},
{\MgI} , {\MnII}, and {\TiII} absorption in the HIRES spectrum.
In the FOS spectrum, the {\SiIV} $\lambda 1393$ transition is
clearly detected, though the $\lambda 1402$ could have a contribution
from {\Lyeleven} from the strong Lyman series at $z=1.4638$.
The Lyman limit was not covered.

\item{$z=0.7252$} --- 
This system has {\MgI}, but no {\FeII}, in the HIRES spectrum.
In the FOS spectrum,
the Lyman series lines are somewhat ambiguous [the data are quite
noisy such that $W_{r}({\Lyb})$ is greater than $W_{r}({\Lya})$].
The {\OVI} $\lambda 1031$ transition is in the red wing of a strong
line, possibly {\Lyb} at $z=0.7337$; we determine 
an upper limit on $W_{r}({\OVI})$ at the position of $\lambda 1037$.
The Lyman limit was not covered.

\end{description}

\subsubsection{\rm PG~$0117+213$~~~~($z_{\rm em}=1.50$)}

A G270H FOS spectrum of this quasar was previously studied by
Jannuzi \etal (1998\nocite{cat3}) using the $0.3${\arcsec} aperture.
Koratkar \etal (1998\nocite{koratkar}) obtained spectra in the
spectropolarimetry mode using both the G190H and G270H gratings
($1.0${\arcsec} aperture).
We have limited our study to the G270H spectrum from Jannuzi \etal,
with the exception of analysis of Lyman limit breaks and of
the damped {\Lya} line in the G190H spectrum.
There are five intervening {\MgII} absorption systems.

\begin{description}

\item{$z=0.5764$} --- 
This system is a damped {\Lya} absorber (\cite{rao95}).
In the HIRES spectrum, {\MgI}, {\CaII} $\lambda 3969$ and several
{\TiII} transition were detected ({\CaII} $\lambda 3934$ was not
covered).
{\FeII}, expected to be very strong, was not covered in the HIRES 
spectrum.
In the FOS spectrum, {\FeII} $\lambda 1608$ was detected.
We quoted $W_{r}({\SiII})$ using $\lambda 1526$, since all other
{\SiII} transitions were available only in the spectropolarimetry
G190H spectrum.  {\AlII} $\lambda 1670$ is strong and could be
affected by a blend with a {\Lya} line; however we note that
$W_{r}({\AlII})$ is consistent with a black--bottom {\AlII} profile
with the same kinematic spread as {\MgII}.  
We point out that the strong damped {\Lya} line was measured in 
a G190H spectrum obtained in spectropolarimetry mode.
The Lyman limit was not covered.

\item{$z=0.7291$} --- 
This system has {\FeII} and {\MgI} in the HIRES spectrum.
We quoted $W_{r}({\SiII})$ using $\lambda 1526$, since all other
{\SiII} transitions were available only in the spectropolarimetry
G190H spectrum.
We identify {\CII} $\lambda 1334$ in this system.
The line was previously identified as {\Lyb} at $z=1.2501$ by Jannuzi
\etal (1998\nocite{cat3}), however, the ratio
$W_{r}({\Lyb})/W_{r}({\Lya})$ would be large for the small
$W_{r}({\Lya})$ in that system and there are no other
transitions to corroborate it. 
{\Lya} and the Lyman limit were only covered on the spectropolarimetry
G190H spectrum, and could not be measured.

\item{$z=1.0480$} --- 
This system has {\FeII} and very weak {\MgI}  detected
in the HIRES spectrum.  In the FOS spectrum, 
we have quoted a detection for {\SiIII} $\lambda 1206$, though we note
Jannuzi \etal (1998\nocite{cat3}) identify this absorption feature as
{\Lya} at $z=1.0326$.
We quote a {\CII} $\lambda 1334$ detection; though Jannuzi \etal have
identified this feature as {\FeII} $\lambda 1144$ at $z=1.3868$, 
their identification is not consistent with the absence of the
{\FeII} $\lambda 2383$ transition in the more sensitive HIRES spectrum.
The quoted upper limit for $W_{r}({\SiIV})$ is obtained from $\lambda
1402$ because $\lambda 1393$ is blended with Galactic {\MgI}.
$W_{r}({\CIV})$ is conservatively quoted as an upper limit because 
both the $\lambda 1548$ and $\lambda 1550$ are detected just below the
$3\sigma$ level.
There is a partial Lyman limit break with $\tau \simeq 0.9$ measured
in the G190H spectropolarimetry spectrum.

\item{$z=1.3250$} --- 
This system has {\FeII}, {\MgI}  and {\AlIII} $\lambda 1863$ in
the HIRES spectrum.
The {\CIV} doublet is taken from the ground--based spectrum of
SS92\nocite{ss92}, which has resolution $R\simeq860$.
In the FOS spectrum,
{\Lya} is coincident with {\SiIII} $\lambda 1206$ from the $z=1.3430$
{\MgII} system; the quoted $W_{r}({\Lya})$ may be larger than the true
{\Lya} absorption strength.
Gaussian deblending was used to measure $W_{r}({\Lyb})$ because the
blue wing of the {\Lyb} line is blended with Galactic {\FeII} $\lambda
2383$, but a clear asymmetry is apparent.
$W_{r}({\SiIV})$ $\lambda 1393$ could have a contribution from a
blend with {\AlII} $\lambda 1670$ at $z=0.9400$, but this contribution
should be small.  The {\SiIV} $\lambda 1402$ is not formally detected,
but this is not inconsistent with $\lambda 1393$ due to an uncertain
continuum fit near the spectrum edge; we claim a detection for {\SiIV}.
{\SiIII} $\lambda 1206$ is blended with Galactic {\MgII} $\lambda
2803$ in its blue wing and possibly with an unidentified line in its red
wing; the equivalent width of the full blend was taken as a limit on {\SiIII}.
The upper limit for $W_{r}({\NV})$ is obtained from the expected position
of $\lambda 1242$ because $\lambda 1238$ is blended with {\SiIII} $\lambda 1206$
at $z=1.3868$.
Both {\OVI} $\lambda 1031$ and $\lambda 1037$ are blended
with other lines, the former with {\Lyb} at $z=1.3389$, and the latter
with {\OVI} $\lambda 1031$, also from $z=1.3389$.
Absorption from {\OVI} might be present in this system, but it cannot be 
determined, so we quote an upper limit obtained by fitting the weaker
blended feature at the position of $\lambda 1037$.
The Lyman limit break shows structure suggestive of a ``double break''.
A model of the Lyman break revealed that both this system and the
$z=1.3430$ system contribute roughly equally, ach having $\tau \simeq
0.6$.

\item{$z=1.3430$} --- 
This system has {\FeII} and {\AlIIIdblt} in the HIRES spectrum.
The {\CIV} doublet is taken from the ground--based spectrum of
SS92\nocite{ss92}, which has resolution $R\simeq860$ at the position
of the doublet.
In the FOS spectrum,
{\NV} $\lambda 1238$ is blended with {\Lya} at $z=1.3868$.  
We quote a detection of {\NV} based upon $\lambda 1242$, which is 
blueward of {\Lya} at $z=1.3981$.
A significant feature was present at the location of {\SiIV} $\lambda
1393$ redward of the {\Lya} forest, but $\lambda 1402$ was not covered.
We tentatively claim a detection for the {\SiIV} doublet (consistent
with our policy on line identifications).
{\SiIII} $\lambda 1206$ is coincident with the strong {\Lya} line 
associated with the $z=1.3250$ {\MgII} absorber; we quote the equivalent
width of this line as an upper limit on $W_{r}({\SiIII})$.
There is a Lyman limit break with $\tau \simeq 1.3$ measured in the
G190H spectropolarimetry spectrum.  
The Lyman limit break shows structure suggestive of a ``double break''.
Modeling showed that both this system and the $z=1.3250$ system
contribute roughly equally to the break with each having $\tau \simeq
0.6$.

\end{description}

\subsubsection{\rm PKS~$0454-220$~~~~($z_{\rm em}=0.53$)}

G130H, G190H, and G270H FOS spectra of this quasar were obtained by
M.~Burbidge (see Cohen \etal (1991\nocite{cohen}).
Of the pre--COSTAR spectra included in our study, these are the only
ones obtained with a ``large'' aperture ($1${\arcsec}).
As such, the quoted equivalent widths may be systematically small;
we did not apply any correction factors.

\begin{description}

\item{$z=0.4744$} --- 
This is a near--DLA system (also see Rao \etal 1995\nocite{rao95});
the measured $W_{r}({\Lya}) = 5.5$~{\AA} places it on the logarithmic
part of the curve of growth with $N({\HI}) \sim 10^{19.5}$~{\cmsq}.
Due to a blend with {\SiIII} $\lambda 1206$ at $z=0.4833$ in the red 
wing of the {\Lya} line, $W_{r}({\Lya})$ could be slightly overestimated.
Detected in the HIRES spectrum are several {\FeII} transitions, 
{\MgI} , and {\CaII} $\lambda 3969$ ($\lambda 3934$ was not covered).
In the FOS spectrum,
the {\CIV} doublet is present on both the G190H and G270H gratings;
the equivalent widths of the $\lambda 1548$ transition measured
from the two spectra are consistent
($0.93\pm0.04$~{\AA} and $1.09\pm0.11$~{\AA}, respectively.)
The red wing of {\CIII} is blended with {\Lyg} at $z=0.4833$ and an
unidentified line, thus a limit is quoted based on a fit to the blend.
{\OVI} $\lambda 1031$ is blended with {\Lyb} at $z=0.4833$ and
$\lambda 1037$ is coincident with both the $\lambda 1031$ member of
{\OVI} at $z=0.4833$ and a dominating unidentified line.
We conservatively quote an upper limit on $W_{r}({\OVI})$ based on
the equivalent width of the blend at the expected position of
{\OVI} $\lambda 1031$.
Accurate measurement of the $N({\HI})$ optical depth from the Lyman
limit is compromised by geocoronal emission; given that this system is
a near--DLA, the flux below $\lambda 1345$~{\AA} should be zero.
Modeling showed that the break itself arises from the $z=0.4833$
system (see below).

\item{$z=0.4833$} --- 
In the HIRES spectrum, this system has detected {\FeII}, {\MgI}, and
{\CaII} $\lambda 3934$ ({\CaII} $\lambda 3969$ was not covered).
In the FOS spectrum,
{\CIV} was covered by both the G190H and G270H gratings, but the
profile shapes and equivalent widths are not consistent (the G190H 
appears compromised, but the G270H $\lambda 1548$ transition is large
relative to $\lambda 1550$); we have quoted the average of the
equivalent widths from the two gratings and present the G270H spectrum
in Figure~\ref{fig:portraits}.
Any {\SiIII} $\lambda 1206$ that is present in the spectrum is blended
in the red wing of the {\Lya} line at $z=0.4744$; we have quoted
$W_{r}({\Lya})$ for the upper limit on $W_{r}({\SiIII})$.
The upper limit on $W_{r}({\SiIV})$ was obtained from $\lambda 1402$,
since $\lambda 1393$ is blended with {\SiIV} $\lambda 1402$ at
$z=0.4744$.
{\OVI} $\lambda 1031$ is blended with {\OVI} $\lambda 1037$ at
$z=0.4744$ and a dominating unidentified line.
However, a clean upper limit on $W_{r}({\OVI})$ was obtained from
$\lambda 1037$.
Modeling showed that the Lyman limit break arises from this system,
however accurate measurement of the $N({\HI})$ optical depth is
compromised by geocoronal emission.

\end{description}

\subsubsection{\rm PKS~$0454+039$~~~~($z_{\rm em}=1.35$)}

G190H and G270H FOS spectra of this quasar (obtained by Bergeron
with the $1.0${\arcsec} square aperture)
were studied by Boiss\'{e} \etal (1998\nocite{boisse}), 
and a separate G270H spectrum ($1.0${\arcsec} round aperture)
was obtained by Bowen.
We independently reduced these data, and we present results below
from the Bergeron G190H and from a combined G270H spectrum from the 
Bowen and Bergeron observations.
Some slight differences between our analysis results and those of
Boiss\'{e} \etal are discussed on a system by system basis below.

\begin{description}

\item{$z=0.6428$} --- 
This system has {\FeII} in the HIRES spectrum.
In the FOS spectrum, 
{\Lya} is blended with {\Lyseven} at $z=1.1532$ and 
a {\Lya} line at $z=0.6448$, but it is distinct such that
accurate deblending is possible.
We claim a tentative detection of {\SiIV}, which appears to be
present in both the G190H and G270H spectra.
We note however, that the blue wing of $\lambda 1393$ is blended with
{\Lya} at $z=0.8801$ and that there may be a small contribution to
$\lambda 1402$ from {\NV} $\lambda 1242$ at $z=0.8596$, though the
latter is not formally detected.
Though Churchill \& Le~Brun (1998\nocite{cwc-lb98}) claimed a tentative
upper limit on {\CIV}, our reduction and continuum fit of the 
G190H spectrum yielded a detection above the $3\sigma$ level; 
we thus claim a detection for {\CIV}.
{\SiIII} $\lambda 1206$ is blended with {\Lynine} at $z=1.1532$;
we fit this feature to obtain an upper limit.
The Lyman limit was not covered and {\Lyb} is below the Lyman break 
from the $z=0.8596$ system where the flux is zero.

\item{$z=0.8596$} --- 
This is a well studied damped {\Lya} system (e.g.\ Steidel \etal
1995\nocite{sbbd95}; Lu \etal 1996b\nocite{lu_dla96}; Boiss\`{e} \etal
1998\nocite{boisse}).  
In the HIRES spectrum, a suite of five {\FeII} transitions are
detected, as are the {\MnII} $\lambda 2600$ triplet, {\MgI} and
{\TiII} $\lambda 3074$ and $\lambda 3243$.
In the FOS spectrum, numerous transitions are detected,
and the optical depth of the Lyman limit is $\tau \simeq 3.3$.

\item{$z=0.9315$} --- 
This system has {\FeII} in the HIRES spectrum.
Our measurement of the very weak {\Lya} line was consistent 
with that reported by Churchill \& Le Brun (1998\nocite{cwc-lb98}).
{\Lyb} is blended with the Lyman series at $z=1.1532$ and {\Lyg} is
blended with {\Lyd} at $z=0.9780$. 
{\SiIII} $\lambda 1206$ is in the blue wing of a {\Lya} line at
$z=0.9185$ so a non--restrictive limit was obtained by fitting this 
blend.
Though it appears that a clean {\CIV} doublet is present, $\lambda
1548$ is actually {\FeII} $\lambda 1608$ from the $z=0.8596$ DLA and
$\lambda 1550$ is {\MgII} $\lambda 2796$ from a dwarf galaxy at
$z=0.0714$.  A non--restrictive limit on $W_{r}({\CIV})$ is obtained by
fitting the feature coincident with the expected position of 
$\lambda 1548$.
{\OVI} $\lambda 1031$ is blended with the blue wing of {\Lyeight} at
$z=1.1532$ and $\lambda 1037$ is blended with {\Lysix}, also at
$z=1.1532$; we used $W_{r}({\Lyeight})$ to obtain an upper limit on
$W_{r}({\OVI})$.
There is no Lyman limit break.

\item{$z=1.1532$} --- 
In the HIRES spectrum, {\FeII}, {\MgI}  and the
{\AlIIIdblt} are present.  
The {\CIV} doublet is taken from the ground--based spectrum of 
SBS\nocite{sbs88}, which has resolution
$R\simeq2200$ at the observed wavelength of {\CIV}. 
In the FOS spectrum, 
the quoted $W_{r}({\SiIV})$ is taken from $\lambda 1402$ because
$\lambda 1393$ is blended with {\MgII} $\lambda 2796$ from the dwarf
galaxy at $z=0.0714$.
Boiss\`{e} \etal (1998\nocite{boisse}) reported detection of
{\NVdblt}, but the feature located where $\lambda 1238$ is expected
may be a {\Lya} line at $z=1.1941$, since $\lambda 1242$ would be in
the extreme blue wing of the possible {\Lya} line at $z=1.2025$.
We quote an upper limit on $W_{r}({\NV})$ from measuring the blend at
the position of $\lambda 1238$ .
Though {\OVI} may be present, $\lambda 1031$ is blended with {\Lyb}
at $z=1.1672$ and {\OVI} is in a complex region in the wing of the 
damped {\Lya} line at $z=0.8596$.
The optical depth of the Lyman limit is $\tau \simeq 1.1$.

\end{description}

\subsubsection{\rm PKS~$0823-223$~~~~($z_{\rm em} > 0.92$)}

This quasar, a luminous BL Lacertae object, has only a lower
limit on its emission redshift.
Only a G270H FOS spectrum was obtained by Rao \& Turnshek
(1999\nocite{rao99}) using the $1.0${\arcsec} aperture.

\begin{description}

\item{$z=0.7055$} --- 
This system has no detectable {\FeII} or {\MgI}  in the HIRES
spectrum.
{\Lya} was not covered in the FOS spectrum.
The upper limit on $W_{r}({\SiII})$ is obtained from the $\lambda
1526$ transition because the other {\SiII} transitions are not
covered.
{\CII} $\lambda 1334$, if present, is blended with strong {\SiII}
$\lambda 1190$ at $z=0.9110$, so this blend is fit to obtain an
upper limit on $W_{r}({\CIV})$.
For the upper limit on $W_{r}({\SiIV})$, we use $\lambda 1402$ because
$\lambda 1393$ is blended with Galactic {\FeII} $\lambda 2374$ and
$\lambda 2383$.
The Lyman limit was not covered.

\item{$z=0.9110$} --- 
This system was reported to be a ``double'' system by Veron--Cetty
\etal (1990\nocite{veron90}). 
Several {\FeII} transitions and {\MgI}  are present in
the HIRES spectrum.
The FOS spectrum is very rich in many, cleanly identified, transitions.
The $\lambda 1242$ member of the {\NV} doublet is blended with
Galactic {\FeII} $\lambda 2374$, but $\lambda 1238$ is unambiguously
detected.
The Lyman break was not covered. 

\end{description}


\subsubsection{\rm Q~$0958+551$, MRK 132~~~~($z_{\rm em}=1.76$)}

The G270H FOS spectrum, obtained with the $0.3${\arcsec} round aperture,
was studied by Jannuzi \etal (1998\nocite{cat3}).
The FOS spectrum covers only the {\Lya} forest down to a strong Lyman
limit break at 2200~{\AA} from a $z=1.7327$ system that wipes out all
flux below this wavelength.

\begin{description}

\item{$z=1.2113$} --- 
This system has no detectable {\FeII} or {\MgI}  in the HIRES
spectrum.
In the FOS spectrum, 
{\Lya} is blended with $\lambda 1550$ from a possible {\CIV}
doublet at $z=0.7330$.
There are no corroborating transitions for the {\CIV} redshift in the
HIRES spectrum, but there is a possible {\FeII} $\lambda 1608$ line in
the FOS spectrum.
We quote a non--restrictive upper limit on $W_{r}({\Lya})$.
{\SiIII}  $\lambda 1206$ is blended with {\CIII} $\lambda 977$ at
$z=1.7325$, the redshift of the strong Lyman limit break.
There is a strong absorption feature coincident with {\SiII} $\lambda
1260$ (could be {\FeII} $\lambda 1608$ at $z=0.7330$), but its
strength is not consistent with detection limits based to the
{\SiIIdblt}; we have used  $\lambda 1193$ to place an upper limit on
$W_{r}({\SiII})$.
The Lyman limit break was not covered.

\item{$z=1.2724$} --- 
This system has {\FeII} $\lambda 2383$ in the HIRES spectrum.
The {\CIV} doublet is taken from the ground--based spectrum of 
SBS\nocite{sbs88}, which has resolution
$R\simeq840$ at this wavelength.  In the FOS spectrum,
only {\Lya} is clearly detected above the $3\sigma$ threshold.
A possible detection of {\SiIV} is not clear; there could be 
coincident {\Lya} lines blended with both members of the doublet.
We conservatively quote an upper limit on $W_{r}({\SiIV})$.
The Lyman limit break was not covered.
{\NV} $\lambda 1242$ is possibly blended with {\OVI} $\lambda 1031$ at
$z=1.7327$, so $\lambda 1238$ provides an upper limit on
$W_{r}({\NV})$.

\end{description}

\subsubsection{\rm PG~$1206+459$~~~~($z_{\rm em}=1.16$)} 

Both the G190H and G270H FOS spectra of this quasar have been studied by
Jannuzi \etal (1998\nocite{cat3}).  They were obtained with pre--COSTAR
optics using the $0.25{\arcsec} \times 2{\arcsec}$ slit.
The metal line properties of the {\MgII} systems were studied in
detail by Churchill \& Charlton (1999\nocite{q1206}).

\begin{description}

\item{$z=0.9276$} --- 
This system was reported as a double system by Jannuzi \etal
(1998\nocite{cat3}) and by Churchill \& Charlton
(1999\nocite{q1206}), with redshifts $z=0.9254$ and $z=0.9276$.
In the HIRES spectrum, {\FeII} and {\MgI}  are present.
The FOS spectrum has the richest complement of transitions of
all the systems in our sample.
{\CII} $\lambda 1334$ is blended with a possible {\Lya} line at 
$z=1.1166$.
{\CII} $\lambda 1036$ is blended with {\OVI} $\lambda 1037$ and
with {\OVI} $\lambda 1031$ at $z=0.9343$.
The quoted $W_{r}({\CII})$ is based upon $\lambda 1334$ using a Gaussian
deblending, which likely yielded a slightly large value.
The {\CIV} profile required a double Gaussian fit and the
$\lambda 1550$ from $z=0.9254$ is blended with $\lambda 1548$
from $z=0.9276$.  We approximately measured $W_{r}({\CIV})$ 
(estimated uncertainty of $\sim 1$~{\AA}) as the 
sum of half of the equivalent width of this blend and the equivalent width
of the unblended $\lambda 1548$ from the $z=0.9254$ system.
{\OVI} also required a double Gaussian fit, but $W_{r}({\OVI})$ could be 
determined more accurately due to the larger separation of {\OVIdblt}.
{\CIII} $\lambda 977$ is clearly detected but is blended 
with {\Lya} at $z=0.5482$ and a smaller {\Lyg} line from the
$z=0.9343$ {\MgII} absorber; our upper limit is the equivalent
width of the blend.
Modeling of the Lyman limit showed that the break is entirely due to
the $z=0.9276$ system with $\tau \simeq 0.9$ [see Churchill \&
Charlton (1999\nocite{q1206})].

\item{$z=0.9343$} ---
This system has no detectable {\FeII} or {\MgI} in the HIRES spectrum.
$W_{r}({\OVI})$ was measured using $\lambda 1037$ because 
$\lambda 1031$ is blended with {\OVI} $\lambda 1037$ at $z=0.9276$.
In the FOS spectrum, {\Lyg} is blended with with {\Lya} at $z=0.5482$ 
and {\CIII} $\lambda 977$ at $z=0.9276$ such that only a non--restrictive
limit can be derived from a fit to the blend.
{\CII} $\lambda 1334$ is blended with strong {\Lya} at $z=1.1223$.
This system makes no contribution to the Lyman limit break
(see notes on the $z=0.9276$ system).

\end{description}

\subsubsection{\rm PG~$1241+176$~~~~($z_{\rm em}=1.27$)}

The pre--COSTAR G270H FOS spectrum of this quasar,
obtained with the $0.25{\arcsec} \times 2{\arcsec}$ slit, 
has been studied by Jannuzi \etal (1998\nocite{cat3}).

\begin{description}

\item{$z=0.5505$} --- 
In the HIRES spectrum, {\FeII}, {\MgI}  and {\CaII}
$\lambda 3969$ were detected ($\lambda 3934$ was not covered).
In the FOS spectrum, detected
{\SiII} $\lambda 1526$ was used to measure $W_{r}({\SiII})$ because
no other {\SiII} transitions are covered.
The Lyman limit was not covered.

\item{$z=0.5584$} --- 
This system has no detectable {\FeII} or {\MgI} in the HIRES spectrum.
In the FOS spectrum, {\SiII} $\lambda 1526$ was used to measure $W_{r}({\SiII})$ 
because no other {\SiII} transitions are covered.
{\CIV} $\lambda 1550$ was not detected at the
$3\sigma$ level, however, it is consistent with the weak, but
significant $\lambda 1548$ detection; we have quoted a detection for {\CIV}.
The Lyman limit was not covered.

\item{$z=0.8955$} --- 
This system has no detectable {\FeII} or {\MgI} in the HIRES spectrum.
In the FOS spectrum, only {\Lya} was unambiguously detected.
{\SiIII} $\lambda 1206$ is blended with {\OVI} $\lambda 1031$ at
$z=1.2154$; it is difficult to evaluate the relative contributions so
we quote an upper limit on $W_{r}({\SiIII})$.
The {\SiIV} doublet is somewhat ambiguous;
if real, it is offset redward from the predicted location.
The $\lambda 1393$ transition may be {\Lya} at $z=1.1742$ and the 
$\lambda 1402$ transition is blended with {\CII} $\lambda 1334$ at
$z=0.9927$, a system with other corroborating lines.
The limit on $W_{r}({\SiIV})$ was determined by fitting the blend
at the expected position of $\lambda 1393$.
The location of {\NV} $\lambda 1242$ is blended with an unknown
transition in the blue wing of {\OVI} $\lambda 1037$ at $z=1.2720$,
but a clean limit on $W_{r}({\NV})$ is determined at the position of 
$\lambda 1393$.  ({\CII} $\lambda 1036$ at $z=1.2720$ is ruled out as 
the blend based upon the absence of stronger {\CII} $\lambda 1334$).
The Lyman limit was not covered.

\end{description}

\subsubsection{\rm PG~$1248+401$~~~~($z_{\rm em}=1.03$)}

Both the G190H and G270H FOS spectra ($0.3${\arcsec} aperture)
of this quasar have been studied by Jannuzi \etal (1998\nocite{cat3}).

\begin{description}

\item{$z=0.7730$} --- 
This system has detected {\FeII} and {\MgI}  in the HIRES spectrum.
The FOS spectrum is very rich with clearly detected transitions.
As seen in {\Lya} and {\Lyb}, there is a higher redshift system
at $z=0.7760$ with no detectable {\MgII} in the HIRES spectrum.
There are no detected metal lines from that system that might
contaminate those of the $z=0.7730$ {\MgII} absorber.
$W_{r}({\SiII})$ was obtained by Gaussian deblending of $\lambda 1260$  
with {\SiIII} $\lambda 1206$ at $z=0.8546$.
We have tentatively claimed {\NV} as a detection, noting that $\lambda
1242$ is not formally detected, but is consistent with the expected
value from $\lambda 1238$.  {\NV} was also detected by the systematic 
procedures of Jannuzi \etal (1998\nocite{cat3}).
$W_{r}({\Lyg})$ is quoted as an upper limit, since its measured value
is inconsistent with $W_{r}({\Lya})$ and $W_{r}({\Lyb})$.
The Lyman limit is on the edge of the G190H spectrum where the
signal--to--noise ratio is hopeless.

\item{$z=0.8546$} ---
This system has several {\FeII} transitions but no {\MgI} in 
the HIRES spectrum.
In the FOS spectrum, 
this system is part of a group of ${\Lya}+{\Lyb}$ absorbers at
redshifts $0.8524$, $0.8585$, and $0.8614$.
This is one of the examples in the KP data set of extensive metal--line
systems occurring in overdensities in the distribution of {\Lya}
absorbers (\cite{cat2}; \cite{buellslg}).
The blend in the blue wing of {\SiIII} $\lambda 1206$ is {\SiII}
$\lambda 1260$ at $z=0.7730$.
$W_{r}({\SiIII})$ was obtained by Gaussian deblending.
We quote an upper limit on $W_{r}({\SiIV})$ because $\lambda 1393$ is
blended with Galactic {\FeII} $\lambda 2586$ and $\lambda 1402$  is
blended with Galactic {\FeII} $\lambda 2600$.
There is no Lyman limit break in a very clean portion of the spectrum.
\end{description}

\subsubsection{\rm Q~$1317+277$, TON 153~~~~($z_{\rm em}=1.02$)}

Both the G190H and G270H FOS spectra of this quasar have been studied by
Bahcall \etal (1996\nocite{cat2}).  These spectrum were obtained with
pre--COSTAR optics using the $0.25{\arcsec} \times 2{\arcsec}$ slit.
The G160L FOS spectrum, covering the Lyman limit, was presented by 
Bahcall \etal (1993\nocite{cat1}).

\begin{description}

\item{$z=0.6601$} --- 
In the HIRES spectrum, {\FeII}, {\MgI}  are present. 
In the FOS spectrum, 
there is an unknown blend with {\NV} $\lambda 1242$ but a clean upper
limit on $W_{r}({\NV})$ is taken from the position of $\lambda 1238$.
{\OVI} $\lambda 1031$ is blended with {\Lyb} at $z=0.6691$; we used
$\lambda 1037$ to place an upper limit on $W_{r}({\OVI})$.
There is a strong Lyman limit break, with $\tau \simeq 5.4$,
associated with this system.

\end{description}

\subsubsection{\rm PG~$1329+412$~~~~($z_{\rm em}=1.94$)}

The G190H grating spectrum ($1.0${\arcsec} aperture)
was obtained by K.~Lanzetta.
The G270H grating was obtained by Rao \& Turnshek
(1999\nocite{rao99}), also with the $1.0$ aperture.
This is a complex spectrum with at least 10 known metal--line
absorption redshifts, several of which give rise to a Lyman series.
In the G190H spectrum, there is a strong Lyman limit break at
$2083$~{\AA} from a $z = 1.2841$ system that wipes out all flux
below this wavelength.
The two strongest Lyman series systems at $z=1.8408$ and $z=1.8367$,
give rise to a partial Lyman limit break.
The former is associated with a {\CIV} absorber (SBS\nocite{sbs88}) and
has strong {\OVIdblt} absorption in the FOS spectrum.

\begin{description}

\item{$z=0.5008$} --- 
This system had no detectable {\FeII} nor {\MgI} in the
HIRES spectrum.
In the FOS spectrum, the upper limit on $W_{r}({\SiII})$ was obtained
using the $\lambda 1526$ transition since it was the only {\SiII} line
redward of the strong Lyman limit break at $z=1.2841$.
The {\SiIVdblt} is sitting on the shoulder of this break, and is thus
fairly noisy, with $\lambda 1402$ blended with a possible {\Lya} line
at $z=0.7308$; a non--restrictive upper limit is quoted from the noisy
region at the expected position of $\lambda 1393$.
The Lyman limit was not covered.

\item{$z=0.8933$} ---
The HIRES spectrum is relatively noisy so that some weak kinematic
outlying clouds could be missed.
Near {\MgII} $\lambda 2796$ there is an unidentified line at $v =
+80$~{\kms} (which cannot be {\MgII} because the $\lambda 2803$
transition is absent).
{\FeII} was detected, but {\MgI}  was not present.
In the FOS spectrum, only {\Lya} is detected.
{\AlII} $\lambda 1670$ is blended in the red wing of strong {\Lya} at
$z=1.6008$.
{\CII} $\lambda 1334$ is blended with {\Lyd}, also at $z=1.6008$.
The {\SiIV} doublet is on the shoulder of the ``double'', partial
Lyman limit break systems; $\lambda 1393$ is blended with {\Lysix} at
$z=1.8336$ and $\lambda 1402$ is blended with the blue wing of a
complex blend, including two {\Lye} lines at $z=1.8366$ and
$1.8408$, and {\Lyb} at $z=1.6008$. 
We quote an upper limit for $W_{r}({\SiIV})$.
The upper limit on $W_{r}({\CIV})$ is obtained using $\lambda 1550$
because $\lambda 1548$ is blended with {\OVI} $\lambda 1031$ at
$z=1.8408$.
The Lyman limit is wiped out by the break at $2083$~{\AA}.

\item{$z=0.9739$} --- 
This system has no detectable {\FeII} transitions nor {\MgI}  in
the HIRES spectrum.
In the FOS spectrum,
the {\Lya} line is the central member of a triple blend with {\Lyg} 
at $z=1.4716$ being the red--most member and an unidentified line 
(possibly {\Lya} at $z=0.9729$) to the blue.
The quoted $W_{r}({\Lya})$ was obtained by Gaussian deblending, but
because the profile is complicated the error is likely underestimated.
$W_{r}({\SiII})$ was also obtained using Gaussian deblending, due to
an unidentified line, possibly {\Lya} at $z=1.0499$.
The {\CII} $\lambda 1334$ transition is detected in a crowded
region of the spectrum, but is unblended between {\Lysix}
at $z=1.8366$ and {\Lyseven} at $z=1.8408$.
{\SiIV} $\lambda 1393$ is clearly detected, not being blended with any
Lyman series lines, however, $\lambda 1402$ is blended with {\Lya} at
$z=1.2830$, which is corroborated by a {\Lyb} line.
For {\CIV}, $\lambda 1550$ is stronger at the $4\sigma$ level,
suggesting a blend.
The upper limit on $W_{r}({\NV})$ is obtained from $\lambda 1242$
because $\lambda 1238$ is blended with {\NIII} $\lambda 989$ at
$z=1.4714$.
The Lyman limit is wiped out by the break at $2083$~{\AA}.

\item{$z=0.9984$} --- 
This system has several {\FeII} transitions, but {\MgI} was not
covered in the HIRES spectrum.
In the FOS spectrum, the {\Lya} is blended with an unidentified line;
we measured $W_{r}({\Lya})$ using Gaussian deblending, but note that
the value is uncertain.
{\CII} is blended with a strong {\Lya} line at $z=1.6008$.
{\SiIV} $\lambda 1402$ is blended with Galactic {\MgII} $\lambda
2803$.
The {\CIV} doublet, which brackets a possible {\Lya} line at
$z=1.5477$, is undetected.
There is also a possible {\Lya} line at $z=1.0371$ slightly to the
blue of {\NV} $\lambda 1238$.
The Lyman limit is wiped out by the break at $2083$~{\AA}.

\end{description}

\subsubsection{\rm PKS~$1354+195$~~~~($z_{\rm em}=0.72$)}

The G160L, G190H, and G270H FOS spectra of this quasar have been studied
by Jannuzi \etal (1998\nocite{cat3}) and Bergeron \etal 
(1994\nocite{bergeronKP}).  
These spectra were obtained pre--COSTAR, using the $0.25{\arcsec}
\times 2{\arcsec}$ slit.

\begin{description}

\item{$z=0.4566$} --- 
The HIRES spectrum is fairly noisy in the region of the {\MgII}
doublet and {\FeII} transitions, so it is possible that some kinematic
outlying clouds were missed.
{\FeII} $\lambda 2600$ and {\MgI} were detected.
In the FOS spectrum, the detections are all unambiguous.and the 
upper limits are not compromised by blends.
We note that there is a Lyman limit break with $\tau \simeq 0.9$ in the
G160L spectrum (Jannuzi \etal 1998\nocite{cat3}).

\item{$z=0.5215$} ---
This system has no detected {\FeII} transitions nor {\MgI}  in
the HIRES spectrum.
{\SiIII} $\lambda 1206$ is blended with {\SiII} $\lambda 1260$ at
$z=0.4566$, that latter being corroborated by the presence of a
proper--strength {\SiII} $\lambda 1526$ line.
The line at the expected wavelength of {\CII} $\lambda 1334$ is 
{\SiIV} $\lambda 1393$ at $z=0.4566$.
Based upon the G160L spectrum, there is no evidence for a Lyman limit
break at this redshift.

\end{description}

\subsubsection{\rm Q~$1622+238$, 3C~336~~~~($z_{\rm em}=0.93$)}

Both the G190H and G270H FOS spectra ($1.0${\arcsec} aperture)
of this quasar have been studied by Steidel \etal (1997\nocite{3c336}).
The background level for the G190H is not certain.
The spectrum has slightly negative flux in the core of the damped
{\Lya} line at $z=0.6561$ and below the Lyman limit break.  
Thus, the equivalent widths of lines measured in this grating may be
systematically small.
The HIRES spectrum is quite noisy, such that the detection threshold
is low for accompanying transitions.

\begin{description}

\item{$z=0.4720$} ---
In the HIRES spectrum, only the {\MgII} doublet is detected.
In the FOS spectrum,
the {\CIV} $\lambda 1550$ is blended with {\SiIII} $\lambda 1206$ at 
$z=0.8913$, but the $\lambda 1548$ appears to be clear.
The upper limit on $W_{r}({\SiII})$ was obtained from $\lambda 1526$
on the G270H grating because the stronger $\lambda 1260$ transition was
in a noisy region of the G190H spectrum.
{\CII} is ambiguous, being in a very noisy part of the spectrum and
possibly being blended with {\OVI} $\lambda 1037$ at $z=0.8913$;
we quote the $3 \sigma$ upper limit on $W_{r}({\CII})$.
The upper limit on $W_{r}({\SiIV})$ is obtained using $\lambda 1402$
because $\lambda 1393$ is blended with a strong line, which could be
{\Lya} at $z=0.6868$ with a contribution from {\NV} $\lambda 1238$ at
$z=0.6561$.
The Lyman limit break was not covered.

\item{$z=0.6561$} --- 
This system is a damped {\Lya} absorber (\cite{3c336}). 
In the HIRES spectrum, {\FeII} is strong and {\TiII} $\lambda 3385$ is
detected.
In the FOS spectrum, 
the upper limit on $W_{r}({\NV})$ is obtained using $\lambda 1242$
because $\lambda 1238$ is blended with a strong line, which could be
{\Lya} at $z=0.6868$ with a contribution from {\SiIV} $\lambda 1393$
at $z=0.4720$.
The Lyman limit break was not covered.

\item{$z=0.7971$} ---  
This system was first seen in {\CIV} absorption.
Only the {\MgII} $\lambda 2803$ transition was detected because 
$\lambda 2796$ was wiped out by the pen mark on the HIRES CCD.
Neither {\FeII} nor {\MgI}  was detected.
{\SiIV} $\lambda 1402$ is blended with {\CII} $\lambda 1334$, but
$\lambda 1393$ is cleanly measured.
A robust measurement of $W_{r}({\Lyb})$ was not possible due to the
signal to noise of the spectrum.
Also, {\Lyb} may suffer from blending so we have quoted an upper 
limit on $W_{r}({\Lyb})$.
Though {\SiIII} $\lambda 1206$ is in a busy part of the FOS spectrum,
there are no other candidates for the line; the line immediately to
the blue is {\FeII} $\lambda 1145$ at $z=0.8913$.
The {\OVI} was not formally detected in a noisy region of the
spectrum.
The region of the spectrum corresponding to the Lyman limit is very
noisy due to the break at $z=0.8913$. 

\item{$z=0.8913$} --- 
In this system, {\FeII}, {\MgI} , and {\MnII} $\lambda
2577$ and $\lambda 2594$ were detected in the HIRES spectrum.
In the FOS spectrum,
{\SiII} $\lambda 1260$ is blended with Galactic {\FeII} $\lambda
2383$, so the $\lambda 1193$ transition of the $\lambda \lambda 1190,
1193$ doublet is used to measure $W_{r}({\SiII})$.
$W_{r}({\CII})$ was measured using Gaussian deblending, where the blue
wing is {\SiIV} $\lambda 1402$ at $z=0.7971$.
{\NV} $\lambda 1238$ is blended with Galactic {\FeII} $\lambda 2344$;
we obtained the limit on $W_{r}({\NV})$ using $\lambda 1242$.
The {\OVI} doublet is ambiguous, being in a very noisy part of the
spectrum and possibly being blended with {\CII} $\lambda 1334$ at
$z=0.4720$; we quote a limit on $W_{r}({\OVI})$.
There is a Lyman limit break with a large optical depth whose accurate
measurement is difficult.

\end{description}

\subsubsection{\rm PG~$1634+706$~~~~($z_{\rm em}=1.34$)}

The G270H FOS spectrum of this quasar were studied by (Bahcall \etal
1996\nocite{cat2}).
This spectrum was taken with the pre--COSTAR optics using a
$0.25{\arcsec} \times 2{\arcsec}$ slit.
A pre--COSTAR spectropolarimetry G190H spectrum was obtained with the
wide, $4.3${\arcsec} by Impey \etal (1996\nocite{impey}).
We did not include the G190H spectrum in our study except to
investigate the Lyman limit breaks.

\begin{description}

\item{$z=0.8182$} --- 
This system has no detected {\FeII} transitions nor {\MgI}  in
the HIRES spectrum.
There were no detection in the FOS spectrum.
The upper limit on $W_{r}({\SiII})$ is taken from {\SiII} $\lambda
1526$, which provided the best limit despite being blended with
{\SiIV} $\lambda 1393$ at $z=0.9902$.  
The apparent, but slightly offset, {\AlII} $\lambda 1670$ is {\SiII}
$\lambda 1526$ at $z=0.9902$, as corroborated by {\SiII} $\lambda
1260$.
The upper limit on $W_{r}({\SiIV})$ is fairly clean, given the density
of lines near the $\lambda 1393$ transition; $\lambda 1402$ is likely
a blend, possibly with a {\Lya} line at $z=1.0984$.
There is no flux at the position of the Lyman limit due to the 
strong break from the $z=0.9902$ system.

\item{$z=0.9056$} --- 
This system has no detected {\FeII} transitions nor {\MgI}  in
the HIRES spectrum.
In the FOS spectrum,
the upper limit on $W_{r}({\SiII})$ is obtained using $\lambda 1193$
because $\lambda 1260$ is coincident with {\SiIII} $\lambda 1206$ at
$z=0.9902$.
{\CII} $\lambda 1334$ may be present, but the observed equivalent
width is $0.11\pm0.04$~{\AA}, which is below the suggested threshold
of $0.13$~{\AA} for a detection limit in order to avoid
flat fielding residuals in the reduced spectrum (Bahcall \etal
1996\nocite{cat2}).  
(The signal--to--noise ratio of this spectrum is higher than many of
the available flat fields for the epoch of these observations of
PG~$1634+706$.)
{\SiIV} $\lambda 1393$ is coincident with {\CII} $\lambda 1334$ at
$z=0.9902$; we obtained $W_{r}({\SiIV})$ using $\lambda 1402$, which
lies in the extended red wing of a possible {\Lya} line at $z=1.1979$.
The {\NV} $\lambda 1238$ line includes a contribution from 
{\SiII} $\lambda 1190$ at $z=0.9902$.  
This implies that the candidate $\lambda 1238$ line is unrealistically
strong relative to $\lambda 1242$, and also the separation of the
centroids of the two lines do not agree. 
We measure $W_{r}({\NV})$ for the candidate $\lambda 1238$ line, but
cautiously quote this value only as an upper limit.
There is no flux at the position of the Lyman limit due to the 
strong break from the $z=0.9902$ system.

\item{$z=0.9902$} --- 
In the HIRES spectrum, this system has several detected {\FeII}
transitions and detected {\MgI}.
In the FOS spectrum,
there is a {\Lya} line at $z=0.9785$ in the red wing of {\SiIII}
$\lambda 1206$; $W_{r}({\SiIII})$ was obtained using Gaussian
deblending.
Both members of the {\SiIV} doublet are blended, but easily modeled
with Gaussian fitting.
The $\lambda 1393$ transition is the central member of a three line
blend with {\Lya} at $z=1.2788$ and $z=1.2852$ and the $\lambda 1393$
transition is in the blue wing of Galactic {\MgII} $\lambda 2796$
(see Figure~\ref{fig:deblending}).
The feature just blueward of where {\NV} $\lambda 1238$ is expected is
{\SiIII} $\lambda 1206$ at $z=1.0414$; a $3\sigma$ limit was
determined at the expected position of $\lambda 1242$.
There is a strong Lyman limit break with $\tau > 6.0$ that wipes out
all flux blueward of $1815$~{\AA}.

\item{$z=1.0414$} --- 
This system has no detected {\FeII} transitions nor {\MgI}  in
the HIRES spectrum.  In the FOS spectrum,
the detections are without ambiguity, expect perhaps {\CII} $\lambda
1334$, which could be an artifact of flat fielding problems.
The relatively strong feature between {\SiIV} $\lambda 1393$ and
$\lambda 1402$ is Galactic {\MgI}.
{\NV} $\lambda 1242$ is blended with {\Lya} at $z=1.0881$, but
a $3\sigma$ limit was measured at the position of $\lambda 1238$.
There is a Lyman limit break with $\tau \simeq 1.4$.

\end{description}

\subsubsection{\rm PKS~$2128-123$, PHL~1598~~~~($z_{\rm em}=0.50$)}

Both the G190H and G270H FOS spectra of this quasar have been studied by
Jannuzi \etal (1998\nocite{cat3}).  The optics were pre--COSTAR and a
$0.25{\arcsec} \times 2{\arcsec}$ slit was used.

\begin{description}

\item{$z=0.4297$} --- 
This system is a near--DLA, with $W_{r}({\Lya}) = 2.92$~{\AA}.
In the HIRES spectrum, {\FeII} was not covered, whereas {\MgI}
and both members of the {\CaIIdblt} were detected.
In the FOS spectrum, there is no confusion with {\Lya} absorbers due
to the low redshift of the quasar.
We quote a detection for the {\SiIV} doublet, since $\lambda 1393$ is 
detected and $\lambda 1402$ is consistent with its expected strength
(recall that there is no {\Lya} forest confusion, nor are there any
other metal--line systems).
We obtained an upper limit (albeit not very constraining) on
$W_{r}({\FeII})$ using $\lambda 1145$. 
The Lyman limit was not covered.

\end{description}

\subsubsection{\rm PKS~$2145+067$~~~~($z_{\rm em}=1.00$)}

Both the G190H and G270H FOS spectra of this quasar have been studied by
Bahcall \etal (1993\nocite{cat1}) and by Bergeron \etal 
(1994\nocite{bergeronKP}).  These were obtained with pre--COSTAR
optics using the $0.25{\arcsec} \times 2{\arcsec}$ slit.

\begin{description}

\item{$z=0.7908$} --- 
In the HIRES spectrum, {\FeII} was detected, but {\MgI} was not.
From the FOS spectrum,
we quote a detection for {\SiII} $\lambda 1260$ based upon its
correct strength relative to detected {\SiII} $\lambda 1193$,
although we note Bahcall \etal (1996\nocite{cat2}) identified 
the line as {\Lya} at $z=0.8577$.
The line to the red of {\SiIII} $\lambda 1206$ is {\Lya} at
$z=0.7810$, which has a corroborating {\Lyb} line, thus 
this is not a doublet at another redshift, and a {\SiIII} detection 
is claimed.
{\NV} is detected, but we note that $\lambda 1242$ is probably blended
with another absorption feature.
{\CII} $\lambda 1334$ is not detected at $3\sigma$, which means that 
the strong feature to the blue of {\OVI} $\lambda 1037$ is not
{\CII} $\lambda 1036$; it is likely {\Lya} at $z=0.5266$.
The Lyman limit is located in an extremely noisy region of the G190H
spectrum and no measurement is possible.
See Bergeron \etal (1994\nocite{bergeronKP}) for an independent study
of the FOS data for this system.

\end{description}

\section{Overall Absorption Properties}
\label{sec:discussion}

For most systems, the measured absorption properties are not without
some ambiguity in one transition or another.  
We do not anticipate that highly detailed system by system
analyses can be made.
That is not to say that, in some cases where robust measurements of
extreme properties are measured, physical information cannot be
extracted [e.g.\ the absorbers toward PG~$1206+459$ (\cite{q1206})].
However, we are confident that our analysis of the data has yielded
measurements that provide a sound representation of the absorption
properties of the sample as a whole, given the consistency between 
our two independent methods for identifying absorption lines (see
\S~\ref{sec:data}).
Here, we present the general overall properties of the systems and
discuss a few of the more immediately obvious trends in the data.
We reserve further analysis for Paper II\nocite{paper2}.

In Figure~\ref{fig:ewall}, we present the rest--frame absorption
strengths of the species listed in Tables~\ref{tab:hiresresults} and 
\ref{tab:fosresults} vs.\ that of {\MgII}.
Only the ``flag transitions'' are plotted.
If a transition other than the ``flag transition'' was used to measure
an equivalent width or an upper limit on an equivalent width, the
value was not plotted on any of the figures.
However, this results in a negligible difference to the general appearance
of Figure~\ref{fig:ewall}.
For all data, the median errors are roughly the size of the data
points and those points with downward pointing arrows represent upper
limits.
Though there are very few {\CaIIdblt} data points (due to redshift
coverage), we have presented {\CaII} absorption strengths for
comparison with the numerous studies focused on the Galaxy, and on local
and low redshift galaxies (e.g.\ \cite{morton86}; \cite{robertson88};
\cite{bowen91a}, 1999b\nocite{bowen91b}; \cite{vallerga93};
\cite{ken93}; \cite{welty96}).

Spearman--Kendall non--parametric rank correlation tests, including
upper limits (\cite{isobe}; \cite{lavalley}), show that the absorption
strengths of all transitions with ionization potentials less than that
of {\CII} ($\sim 25$~eV) are correlated with $W_{r}({\MgII})$ at a
greater than 97\% confidence level (with the exception of {\CaII}).
We take this as additional confirmation that our measurements,
for the most part, provide an accurate representation of the overall
sample, since it is expected that these absorption strengths should be
correlated.
Note, however, that since most of the low ionization transitions are
saturated, their equivalent widths are a better measure of the overall
kinematic spread of the gas than of column densities (e.g.\
\cite{pb90}, 1994\nocite{pb94}).
An interesting trend with ionization level is that, as $W_{r}({\MgII})$
is increased above $1$~{\AA}, the low ionization strengths 
correspondingly increase whereas the higher ionization strengths have
considerable scatter and a relatively low mean.
Most of the latter data represent the damped {\Lya} systems, as can be
seen by their large $W_{r}({\Lya})$ values.
Accurate measurements of {\NV} and {\OVI} were fairly rare, but from the
few data points available, it would appear that $W_{r}({\NV})$ is
often less than $0.4$~{\AA}, and that {\OVI} may have a range of
absorption strengths similar to {\CIV}.

Based upon photoionization models (\cite{bs86}; SS92\nocite{ss92}), it
has been widely accepted that {\MgII} absorbers are optically thick 
at the Lyman limit in neutral hydrogen, i.e.\ they have neutral
hydrogen column densities of $N({\HI}) \geq 2 \times 10^{17}$~{\cmsq}
in one or more of the kinematic components.
In Figure~\ref{fig:vsLya}, we present the rest--frame equivalent
widths of {\MgII} and {\CIV} vs.\ that of {\Lya}.
Three data point types are shown.  
Those systems with a measured Lyman limit break are solid circles,
those with no break are open circles, and those for which the break
was not covered are open squares.
The majority of {\MgII} absorbers have {\Lya} strengths in the 
rough range $0.5$--$2$~{\AA}, with a sparsely populated tail extending
out to $\sim 10$~{\AA}.
It is not clear if there is a gap in the distribution of $W_{r}({\Lya})$;
this would occur if damped {\Lya} systems [by definition those with
$N({\HI}) \geq 2 \times 10^{20}$~{\cmsq}, which corresponds to
$W_{r}({\Lya}) \geq 9$~{\AA}] are more common than are ``{\HI} rich''
systems with $W_{r}({\Lya})$ in the intermediate range $\sim
2$--$8$~{\AA}..
Since these {\HI}--rich, but non--damped, systems have strong {\MgII}
absorption (most above $0.6$~{\AA}), their co--moving number density
probably evolves in close step with the strongest {\MgII} absorbers
(SS92\nocite{ss92}).

The {\Lya} transition is easily saturated and is thus not a good
measure of the neutral hydrogen column density.
The Lyman limit break can provide an estimate of $N({\HI})$ and is the
unambiguous signature of optically thick neutral hydrogen
(\cite{tytler82}).
The location of Lyman limit break was covered for 17 of the 45
absorbers.
All but two of the systems with $W_{r}({\Lya})>1$~{\AA} have breaks;
it is likely that nearly all metal--line systems in this equivalent
width range would also have breaks (also see \cite{cat3}).

Since the redshift number density of the weakest {\MgII} systems is
several times greater than that of systems with Lyman limit breaks
(\cite{weak}), it is expected that most {\MgII} absorbers with 
$W_{r}({\MgII}) < 0.3$~{\AA} will not have a break.
We find two weak systems with breaks, six without, and nine
undetermined.
We note that the majority of the systems have $W_{r}({\Lya}) \leq
1.0$~{\AA}, and none of these for which the location of the Lyman
limit was covered has a break.
This suggests, not surprisingly, that the weakest of the weak {\MgII}
absorbers are the ones lacking breaks.
Note that the presence of a Lyman break is independent of {\CIV}
strength in that the full range of observed $W_{r}({\CIV})$ exhibit
breaks.
Also, note that the {\CIV} strengths are relatively small in the
damped {\Lya} systems, implying that their {\CIV} kinematics are
probably similar to a more ``typical'' strong {\MgII} absorber.

In Figure~\ref{fig:vsCIV}, we present the rest--frame equivalent
widths of {\SiIV} and {\CII} vs.\ that of {\CIV}.
As with Figure~\ref{fig:ewall}, we plot ``flag transitions'' only.
In the plot of $W_{r}({\CII})$ vs.\ $W_{r}({\CIV})$, we illustrate the
combined effects of a range of kinematics and ionization conditions.
We also present $W_{r}({\SiIV})$ vs.\ $W_{r}({\CIV})$, which has
been central for studying the ionization conditions in the ISM and
Halo of the Galaxy (e.g.\ \cite{savage97}).
Future high resolution spectra of these particular transitions will be
essential for a detailed understanding of the kinematics and ionization
conditions of the ISM and halos of these intermediate redshift
galaxies.

For future detailed studies using higher resolution data, silicon has
the added virtue of sampling three ionization states.
Thus, we also present the rest--frame equivalent widths of {\SiIII}
vs.\ {\SiII}, and of {\SiIII} and {\SiII} vs.\ {\SiIV}, in
Figure~\ref{fig:vsSiIV}.
Again, we plot ``flag transitions'' only.
Since both silicon and magnesium are $\alpha$--process elements, and
{\SiII} and {\MgII} have nearly identical ionization potentials and
transitions with similar oscillator strengths, it is expected that
{\SiII} will tightly trace {\MgII}.
The {\SiIII} $\lambda 1206$ transition, however, is very strong and is
expected to be saturated, even at high resolution.

It is evident that the relative kinematics between the low and high
ionization phases vary from absorber to absorber as seen
in Figures~\ref{fig:vsCIV} and \ref{fig:vsSiIV}.
For equivalent widths greater than $\sim 0.4$~{\AA} there is a large
range of values (lack of correlations) for {\CII} and {\SiIV} vs.\
{\CIV} and for {\SiII} and {\SiIII} vs.\ {\SiIV}.

\section{Conclusion}
\label{sec:conclusions}

We have measured the absorption properties of 45 intermediate redshift
{\MgII} absorbers in FOS spectra from the {\it HST\/} archive and from
the database of the {\it HST\/} QSO Absorption Line Key Project.
The sample was selected from the 51 {\MgII} systems observed with
HIRES/Keck for which FOS spectra of the same quasars existed.
The {\MgII} profiles, and other transitions observed in the optical,
have been resolved at $\sim 6$~{\kms} resolution.
The UV FOS spectra have resolution $\sim 230$~{\kms}.
In this paper, we presented the data, the data analysis, and a brief
description of the properties of each system.
We present additional analysis of the data in a parallel companion
paper (Paper II\nocite{paper2}).

We have found evidence for a high ionization gaseous phase in
intermediate redshift {\MgII} absorbing galaxies.
Mostly, the high ionization species detected are {\CIV} and {\SiIV},
which are commonly seen in absorption in the Galaxy (e.g.\
\cite{savagearaa}; \cite{savage97}).
These data lead us to suggest that these galaxies have multiphase 
interstellar media and halos similar to those observed locally
(\cite{dahlem}; also see \cite{letter}).

What is the kinematic spread of the {\CIV} and what is its line of
sight velocity structure?
How is this structure related to that seen in {\MgII}?
High resolution spectroscopic observation are sorely needed for
sorting out the physical nature of this high ionization material and
its relation to the kinematically complex low ionization gas.
Only if the high ionization profiles are resolved at resolutions
comparable to the HIRES/Keck data, can the {\it relative\/} kinematics
of the low and high ionization gas can be quantified.

\acknowledgements
Support for this work was provided by the NSF (AST--9617185), and NASA
(NAG 5--6399 and AR--07983.01--96A) the latter from the Space
Telescope Science Institute, which is operated by AURA, Inc.,
under NASA contract NAS5--26555.
BTJ acknowledges support from NOAO, which is operated by AURA,
Inc., under cooperative agreement with the NSF.
We thank Blair Savage for providing a table of Galactic absorption
properties in electronic form.


\newpage

\begin{table}
\tablenum{1}
\label{tab:hiresobsjournal}
 
\begin{center}
\begin{tabular}{lcrlrc}
\multicolumn{6}{c}{TABLE 1} \\
\multicolumn{6}{c}{\sc Journal of HIRES/Keck Observations} \\
\hline \hline
QSO & V [mag] & $z_{\rm em}$ & Date [UT] & Exp [s] & $\lambda $ Range [\AA] \\
\hline 
$0002+051$ & 16.2 & $1.90$ & 1994 Jul 5  &    2700 & 3655.7--6079.0 \\
$0058+019$ & 17.2 & $1.96$ & 1996 Jul 18 &    3000 & 3766.2--5791.3 \\
$0117+213$ & 16.1 & $1.50$ & 1995 Jan 23 &    5400 & 4317.7--6775.1 \\
$0454-220$ & 15.5 & $0.53$ & 1995 Jan 22 &    5400 & 3765.8--6198.9 \\
$0454+039$ & 16.5 & $1.35$ & 1995 Jan 22 &    4500 & 3765.8--6198.9 \\
$0823-223$ & 15.7 & $>0.92$ & 1995 Jan 24 & 3600 & 3977.8--6411.8 \\
$0958+551$ & 16.0 & $1.76$ & 1995 Jan 23 &    3600 & 5400.0--7830.0 \\
$1206+459$ & 16.1 & $1.16$ & 1995 Jan 23 &    3600 & 3810.5--6304.9 \\
$1241+176$ & 15.4 & $1.27$ & 1995 Jan 22 &    2400 & 3765.8--6189.9 \\
$1248+401$ & 16.3 & $1.03$ & 1995 Jan 22 &    4200 & 3765.8--6189.9 \\
$1317+277$ & 16.0 & $1.02$ & 1995 Jan 23 &    3600 & 3810.5--6304.9 \\
$1329+412$ & 16.3 & $1.94$ & 1996 Jul 18 &    6300 & 3766.2--5791.3 \\
$1354+195$ & 16.0 & $0.72$ & 1995 Jan 22 &    3600 & 3765.8--6189.9 \\
$1622+238$ & 18.3 & $0.93$ & 1994 Jul 4,5 & 16,200 & 3726.9--6191.0 \\
$1634+706$ & 14.9 & $1.34$ & 1994 Jul 4,5 &   2700 & 3723.3--6185.7 \\
$2128-123$ & 15.9 & $0.50$ & 1996 Jul 19 &    3900 & 3766.2--5791.3 \\
$2145+067$ & 16.3 & $1.00$ & 1996 Jul 18 &    4500 & 3766.2--5791.3 \\
\hline
\end{tabular}
\end{center}
\end{table}

  
\newpage

\begin{table}
\tablenum{2}
\label{tab:fosobsjournal}

\begin{center}
\begin{tabular}{lllll}
\multicolumn{5}{c}{TABLE 2} \\
\multicolumn{5}{c}{\sc FOS Archival Data Used for this Study} \\
\hline \hline
QSO & Alias & \multicolumn{3}{c}{Program ID and PI} \\
\cline{3-5}
 & &   G130H  &  G190H   &   G270H \\
 & & (1150--1600~{\AA}) & (1600--2300~{\AA}) & (2225--3275~{\AA}) \\
\hline 
$0002+051$ & UM~$18$    & \nodata &  6705(Koratkar)    & 4581(KP) \\
$0058+019$ & PHL~$938$  & \nodata &  6577(Rao)         &  \nodata      \\
$0117+213$ & PG         & \nodata &  6109(Koratkar)$^{*}$ & 5664(KP)  \\
$0454-220$ & PKS        & 1026(Burbidge) &  1026(Burbidge) &  1026(Burbidge) \\
$0454+039$ & PKS        & \nodata &  5351(Bergeron)    & 5351(Bergeron)+5451(Bowen) \\
$0823-223$ & PKS        & \nodata &  \nodata           & 6577(Rao)      \\
$0958+551$ & MRK~$132$ & \nodata &  \nodata           & 5664(KP)  \\
$1206+459$ & PG         & \nodata &  2424(KP)    & 2424(KP)  \\
$1241+176$ & PG         & \nodata &  \nodata           & 4112(KP)  \\
$1248+401$ & PG         & \nodata &  5664(KP)    & 5664(KP)  \\
$1317+277$ & TON~$153$  & \nodata &  2424(KP)    & 2424(KP)  \\
$1329+412$ & PG         & \nodata &  5948(Lanzetta)    & 6577(Rao)      \\
$1354+195$ & PKS        & \nodata &  2424(KP)    & 2424(KP)  \\
$1622+238$ & $3$C~$336$ & \nodata &  5304(Steidel)     & 5304(Steidel)  \\
$1634+706$ & PG         & \nodata &  3732(Impey)$^{*}$ & 3221(KP)  \\
$2128-123$ & PKS, PHL~1598  & \nodata &  4581(KP)    & 4581(KP)  \\
$2145+067$ & PKS        & \nodata &  2424(KP)    & 2424(KP)  \\
\hline
\multicolumn{5}{l}{$^{*}$ Spectropolarimtrey mode; limited use (see text).} \\
\multicolumn{5}{l}{Note:--- KP indicates the observation was either
part of the QSO Absorption Line Key Project (PI J. Bahcall)} \\
\multicolumn{5}{l}{observations or from the GTO observations of J. Bahcall.} \\
\end{tabular}
\end{center}
\end{table}

  
\newpage

\begingroup
\small
\begin{table}
\tablenum{3}
\label{tab:hiresresults}
\begin{center}
\begin{tabular}{cllll}
\multicolumn{5}{c}{TABLE 3}\\
\multicolumn{5}{c}{\sc Database and HIRES/Keck Results$^{a}$}\\
\hline \hline
QSO & $z_{abs}$ & $W_{r}({\MgII})$ [{\AA}] & $W_{r}({\FeII})$ [{\AA}] & $W_{r}({\MgI})$ [{\AA}] \\
\hline
$0002+051$ &  0.5915 &   $ 0.103\pm0.008$     & $< 0.012$             & $< 0.010$ \nl
           &  0.8514 &   $ 1.086\pm0.016$     & $ 0.419\pm0.022$
& $ 0.166\pm0.027$ \nl
           &  0.8665 &   $ 0.023\pm0.008$     & $< 0.010$             & $< 0.007$ \nl
           &  0.9560 &   $ 0.052\pm0.007$     & $< 0.005$             & $< 0.005$ \nl
$0058+019$ &  0.6127 &   $ 1.625\pm0.013$     & $ 1.274\pm0.037$      & $ 0.300\pm0.062$ \nl
           &  0.7252 &   $ 0.253\pm0.012$     & $< 0.034$             & $ 0.041\pm0.026$ \nl
$0117+213$ &  0.5764 &   $ 0.906\pm0.100$     & $ 0.270\pm0.050$$^{b}$ & $0.155\pm0.023$ \nl
           &  0.7291 &   $ 0.238\pm0.009$     & $ 0.075\pm0.016$      & $ 0.013\pm0.009$ \nl
           &  1.0480 &   $ 0.415\pm0.009$     & $ 0.065\pm0.010$      & $ 0.027\pm0.011$ \nl
           &  1.3250 &   $ 0.291\pm0.011$     & $ 0.026\pm0.009$      & $ 0.005\pm0.002$ \nl
           &  1.3430 & $ 0.153\pm0.008$$^{b}$ & $ 0.029\pm0.004$$^{b}$ &  $< 0.010$ \nl
$0454-220$ &  0.4744 &   $ 1.382\pm0.009$     & $ 0.975\pm0.029$      & $ 0.333\pm0.014$ \nl
           &  0.4833 &   $ 0.431\pm0.007$     & $ 0.162\pm0.043$      & $ 0.070\pm0.010$ \nl
$0454+039$ &  0.6428 &   $ 0.118\pm0.008$     & $ 0.037\pm0.014$      & $< 0.005$ \nl
           &  0.8596 &   $ 1.445\pm0.014$     & $ 1.232\pm0.014$      & $ 0.306\pm0.022$ \nl
           &  0.9315 &   $ 0.042\pm0.005$     & $ 0.030\pm0.008$$^{b}$ &      $< 0.005$ \nl
           &  1.1532 &   $ 0.432\pm0.012$     & $ 0.084\pm0.015$      & $ 0.026\pm0.012$ \nl
$0823-223$ &  0.7055 &   $ 0.092\pm0.007$     & $< 0.008$             & $< 0.005$ \nl  
           &  0.9110 &   $ 1.276\pm0.016$     & $ 0.416\pm0.026$      & $ 0.224\pm0.031$ \nl
$0958+551$ &  1.2113 &   $ 0.060\pm0.007$     & $< 0.006$             & $< 0.010$ \nl
           &  1.2724 &   $ 0.081\pm0.007$     & $ 0.017\pm0.004$      & $ 0.007\pm0.002$ \nl
$1206+459$ &  0.9276 &   $ 0.878\pm0.016$     & $ 0.077\pm0.020$      & $ 0.042\pm0.020$ \nl
           &  0.9343 &   $ 0.049\pm0.005$     & $< 0.004$             & $< 0.004$ \nl
$1241+176$ &  0.5505 &   $ 0.481\pm0.019$     & $ 0.236\pm0.048$      & $ 0.098\pm0.030$ \nl
           &  0.5584 &   $ 0.135\pm0.014$     & $< 0.012$             & $< 0.008$ \nl
           &  0.8955 &   $ 0.018\pm0.005$     & $< 0.005$             & $< 0.004$ \nl
$1248+401$ &  0.7730 &   $ 0.694\pm0.009$     & $ 0.247\pm0.020$      & $ 0.065\pm0.022$ \nl
           &  0.8546 &   $ 0.235\pm0.014$     & $ 0.031\pm0.007$      & $< 0.019$ \nl
$1317+277$ &  0.6601 &   $ 0.338\pm0.011$     & $ 0.126\pm0.016$      & $ 0.026\pm0.009$ \nl
$1329+412$ &  0.5008 &   $ 0.258\pm0.035$     & $< 0.100$             & $< 0.038$ \nl
           &  0.8933 &   $ 0.400\pm0.023$     & $ 0.080\pm0.035$      & $< 0.055$ \nl
           &  0.9739 &   $ 0.181\pm0.035$     & $< 0.028$             & $< 0.024$ \nl
           &  0.9984 &   $ 0.142\pm0.010$     & $ 0.058\pm0.017$      & \nodata \nl
$1354+195$ &  0.4566 &   $ 0.751\pm0.023$     & $ 0.149\pm0.088$      & $ 0.038\pm0.028$ \nl
           &  0.5215 &   $ 0.030\pm0.007$     & $< 0.012$             & $< 0.007$ \nl
$1622+238$ &  0.4720 &   $ 0.681\pm0.048$     & $< 0.118$             & $< 0.057$ \nl
           &  0.6561 &   $ 1.449\pm0.029$     & $ 1.015\pm0.050$      & $ 0.291\pm0.027$ \nl
           &  0.7971 &   $ 0.274\pm0.029$     & $< 0.092$             & $< 0.254$ \nl
           &  0.8913 &   $ 1.534\pm0.025$     & $ 1.080\pm0.421$      & $ 0.307\pm0.033$ \nl
$1634+706$ &  0.8182 &   $ 0.030\pm0.018$     & $< 0.008$             & $< 0.004$ \nl
           &  0.9056 &   $ 0.064\pm0.004$     & $< 0.005$             & $< 0.004$ \nl
           &  0.9902 &   $ 0.558\pm0.005$     & $ 0.127\pm0.011$      & $ 0.058\pm0.010$ \nl
           &  1.0414 &   $ 0.097\pm0.008$     & $< 0.038$             & $< 0.003$ \nl
$2128-123$ &  0.4297 &   $ 0.406\pm0.014$     & $< 0.260$$^{b}$       & $ 0.162\pm0.026$ \nl
$2145+067$ &  0.7908 &   $ 0.483\pm0.015$     & $ 0.041\pm0.014$      & $< 0.038$ \nl
\hline
\multicolumn{5}{l}{$^{a}$ All equivalent widths are rest--frame and
their limits are $3\sigma$.}\\
\multicolumn{5}{l}{$^{b}$ Not ``flag'' transition; see system notes (\S3)}\\
\end{tabular}
\end{center}
\end{table}
\endgroup

  
\newpage
 
\begingroup
\tiny
\begin{table}
\tablenum{4}
\label{tab:fosresults}
\rotate[l]{\makebox[1.025\textheight][l]{\vbox{
\begin{center}
\vglue 0.1in
\begin{tabular}{lllllcllllllllll}
\multicolumn{16}{c}{\normalsize TABLE 4}\\
\multicolumn{16}{c}{\normalsize\sc FOS/{\it HST\/} Results$^{a}$}\\
\hline \hline
QSO & $z_{abs}$ & $W_{r}({\Lya})$ & $W_{r}({\Lyb})$ & $W_{r}({\Lyg})$
& Ly lim & $W_{r}({\AlII})$ & $W_{r}({\AlIII})$ & $W_{r}({\SiII})$ & $W_{r}({\SiIII})$ &
$W_{r}({\SiIV})$ & $W_{r}({\CII})$ & $W_{r}({\CIII})$ & $W_{r}({\CIV})$
& $W_{r}({\NV})$ & $W_{r}({\OVI})$\\
\hline
$0002+051$ &  0.5915 &            \nodata &            \nodata &            \nodata & \nodata &           $< 0.70$ &           $< 0.11$ &     $< 0.30$$^{b}$ &            \nodata &            \nodata &            \nodata &            \nodata &     $< 0.23$$^{b}$ &            \nodata &            \nodata\nl
           &  0.8514 &     $ 2.47\pm0.08$ &            \nodata &            \nodata &     $+$ &     $ 0.25\pm0.05$ &            \nodata &     $ 0.66\pm0.09$ &     $ 1.15\pm0.11$ &           $< 0.82$ &     $ 0.54\pm0.06$ &            \nodata &     $ 1.26\pm0.06$ &           $< 0.15$ &            \nodata\nl
           &  0.8665 &     $ 0.81\pm0.10$ &            \nodata &            \nodata &     $-$ &           $ <0.66$ &            \nodata &           $< 0.12$ &           $< 2.45$ &     $< 0.11$$^{b}$ &           $< 0.12$ &            \nodata &     $< 0.11$$^{b}$ &           $< 0.18$ &            \nodata\nl
           &  0.9560 &     $ 0.85\pm0.07$ &            \nodata &            \nodata &     $-$ &           $< 0.15$ &            \nodata &           $< 0.61$ &     $ 0.32\pm0.07$ &     $< 0.10$$^{b}$ &           $< 0.10$ &            \nodata &     $ 0.52\pm0.04$ &           $< 0.12$ &            \nodata\nl
$0058+019$ &  0.6127 &     $ 6.77\pm0.40$ &     $ 1.22\pm0.45$ &            \nodata & \nodata &            \nodata &            \nodata &     $ 0.76\pm0.14$ &     $ 0.64\pm0.18$ &     $ 0.51\pm0.11$ &     $ 0.74\pm0.12$ &            \nodata &            \nodata &           $< 0.36$ &           $< 0.71$\nl
           &  0.7252 &     $ 0.46\pm0.13$ &     $ 0.75\pm0.19$ &            \nodata & \nodata &            \nodata &            \nodata &           $< 0.30$ &           $< 0.27$ &            \nodata &           $< 0.18$ &           $< 0.85$ &            \nodata &           $< 0.30$ &     $< 0.40$$^{b}$\nl
$0117+213$ &  0.5764 &     $11.15\pm1.09$ &            \nodata &            \nodata & \nodata &     $ 0.96\pm0.04$ &            $< 0.11$& $ 0.33\pm0.05$$^{b}$ &          \nodata &            \nodata &            \nodata &            \nodata &     $ 0.58\pm0.06$ &            \nodata &            \nodata\nl
           &  0.7291 &            \nodata &            \nodata &            \nodata & \nodata &           $< 0.10$ &     $< 0.12$$^{a}$ &     $< 0.10$$^{b}$ &            \nodata &           $< 0.12$ &     $ 0.23\pm0.06$ &            \nodata &           $< 0.10$ &            \nodata &            \nodata\nl
           &  1.0480 &     $ 1.93\pm0.04$ &            \nodata &            \nodata &     $+$ &            \nodata &            \nodata &           $< 0.08$ &     $ 0.44\pm0.04$ &     $< 0.09$$^{b}$ &     $ 0.31\pm0.07$ &            \nodata &           $< 0.08$ &           $< 0.07$ &            \nodata\nl
           &  1.3250 &     $ 1.51\pm0.03$ &     $ 0.30\pm0.04$ &            \nodata &     $+$ &            \nodata &            \nodata &           $< 0.07$ &           $< 0.30$ &     $ 0.33\pm0.04$ &     $ 0.19\pm0.03$ &     $ 0.65\pm0.05$ &     $ 0.89\pm0.03$ &     $< 0.08$$^{b}$ &     $< 0.19$$^{b}$\nl
           &  1.3430 &     $ 1.19\pm0.04$ &     $ 0.70\pm0.03$ &     $ 0.64\pm0.05$ &     $+$ &            \nodata &            \nodata &     $ 0.21\pm0.03$ &           $< 1.49$ &     $ 0.26\pm0.04$ &     $ 0.17\pm0.03$ &     $ 0.57\pm0.04$ &     $ 0.67\pm0.02$ & $ 0.17\pm0.03$$^{b}$ &     $ 0.18\pm0.03$\nl
$0454-220$ &  0.4744 &     $ 3.74\pm0.05$ &     $ 0.96\pm0.07$ &     $ 0.86\pm0.08$ & \nodata &     $ 0.61\pm0.05$ &     $ 0.16\pm0.05$ &     $ 0.51\pm0.02$ &     $ 0.67\pm0.04$ &     $ 0.45\pm0.03$ &     $ 0.76\pm0.03$ &           $< 1.59$ &     $ 0.63\pm0.03$ &           $< 0.06$ &           $< 0.75$\nl
           &  0.4833 &     $ 1.56\pm0.03$ &     $ 0.74\pm0.07$ &           $< 1.58$ &     $+$ &     $ 0.13\pm0.06$ &           $< 0.11$ &     $ 0.12\pm0.02$ &           $< 3.60$ &     $< 0.07$$^{b}$ &     $ 0.53\pm0.04$ &     $ 0.43\pm0.07$ &     $ 0.38\pm0.11$ &           $< 0.14$ &           $< 0.84$\nl
$0454+039$ &  0.6428 &     $ 0.70\pm0.05$ &            \nodata &            \nodata & \nodata &           $< 0.10$ &           $< 0.11$&           $< 0.13$ &           $< 0.49$ &     $ 0.29\pm0.10$ &           $< 0.11$ &            \nodata &     $ 0.46\pm0.05$ &           $< 0.12$ &            \nodata\nl
           &  0.8596 &     $10.70\pm0.26$ &     $ 2.36\pm0.13$ &     $ 2.26\pm0.04$ &     $+$ &     $ 0.69\pm0.04$ &            \nodata &     $ 0.89\pm0.05$ &     $ 0.75\pm0.03$ &     $ 0.70\pm0.04$ &     $ 0.77\pm0.04$ &     $ 1.39\pm0.11$ &     $ 0.66\pm0.03$ &     $< 0.17$$^{b}$ &     $ 0.58\pm0.11$\nl
           &  0.9315 &     $ 0.17\pm0.04$ &           $< 0.41$ &           $< 0.95$ &     $-$ &           $< 0.09$ &            \nodata &           $< 0.10$ &           $< 0.65$ &           $< 0.09$ &     $ 0.18\pm0.03$ &     $ <0.17$ &           $< 0.46$ &           $< 0.08$ &           $< 0.49$\nl
           &  1.1532 &     $ 1.61\pm0.03$ &     $ 1.63\pm0.03$ &     $ 1.26\pm0.03$ &     $+$ &            \nodata &            \nodata &     $ 0.34\pm0.03$ &     $ 1.09\pm0.05$ &$ 0.27\pm0.05$$^{b}$ &     $ 0.38\pm0.03$ &     $ 0.92\pm0.03$ &     $ 0.94\pm0.06$ &          $< 0.46$ &           $< 1.18$\nl
$0823-223$ &  0.7055 &            \nodata &            \nodata &            \nodata & \nodata &           $< 0.13$ &     $< 0.12$$^{a}$ &     $< 0.13$$^{b}$ &            \nodata &     $< 0.16$$^{b}$ &           $< 1.37$ &            \nodata &           $< 0.18$ &            \nodata &            \nodata\nl
           &  0.9110 &     $ 2.68\pm0.07$ &            \nodata &            \nodata & \nodata &     $ 0.31\pm0.05$ &            \nodata &     $ 0.91\pm0.07$ &     $ 1.34\pm0.06$ &     $ 0.44\pm0.06$ &     $ 1.50\pm0.06$ &            \nodata &     $ 1.34\pm0.06$ &     $ 0.18\pm0.06$ &            \nodata\nl
$0958+551$ &  1.2113 &           $< 0.92$ &            \nodata &            \nodata & \nodata &            \nodata &            \nodata &     $< 0.22$$^{b}$ &           $< 2.15$ &           $< 0.24$ &           $< 0.24$ &            \nodata &            \nodata &           $< 0.23$ &            \nodata\nl
           &  1.2724 &     $ 0.75\pm0.15$ &            \nodata &            \nodata & \nodata &            \nodata &            \nodata &           $< 0.23$ &           $< 0.22$ &           $< 0.24$ &           $< 0.24$ &            \nodata &     $ 0.44\pm0.03$ &           $< 0.22$ &            \nodata\nl
$1206+459$ &  0.9276 &     $ 2.47\pm0.07$ &     $ 1.70\pm0.05$ &     $ 1.41\pm0.06$ &     $+$ &           $< 0.15$ &            \nodata &     $ 0.61\pm0.07$ &     $ 1.06\pm0.09$ &     $ 0.64\pm0.06$ &     $ 0.84\pm0.05$ &           $< 2.22$ &     $ 1.84\pm0.52$ &     $ 1.00\pm0.16$ &     $ 1.32\pm0.06$\nl
           &  0.9343 &     $ 0.47\pm0.07$ &     $ 0.30\pm0.04$ &           $< 2.29$ &     $-$ &           $< 0.15$ &            \nodata &           $< 0.13$ &           $< 0.15$ &           $< 0.10$ &           $< 0.84$ &     $ 0.24\pm0.04$ &     $ 0.25\pm0.05$ &     $< 0.12$$^{b}$ & $ 0.11\pm0.04$$^{b}$\nl
$1241+176$ &  0.5505 &            \nodata &            \nodata &            \nodata & \nodata &     $ 0.22\pm0.07$ &           $< 0.10$ & $ 0.37\pm0.06$$^{b}$ &          \nodata &            \nodata &            \nodata &            \nodata &     $ 0.83\pm0.07$ &            \nodata &            \nodata\nl
           &  0.5584 &            \nodata &            \nodata &            \nodata & \nodata &           $< 0.11$ &           $< 0.12$&     $< 0.37$$^{b}$ &            \nodata &            \nodata &            \nodata &            \nodata &     $ 0.21\pm0.06$ &            \nodata &            \nodata\nl
           &  0.8955 &     $ 0.45\pm0.05$ &            \nodata &            \nodata & \nodata &           $< 0.10$ &            \nodata &           $< 0.10$ &           $< 0.40$ &           $< 0.27$ &           $< 0.10$ &            \nodata &           $< 0.10$ &           $< 0.08$ &            \nodata\nl
$1248+401$ &  0.7730 &     $ 1.45\pm0.04$ &     $ 0.82\pm0.06$ &           $< 1.25$ & \nodata &     $ 0.16\pm0.06$ &            \nodata &     $ 0.29\pm0.03$ &     $ 0.65\pm0.03$ &     $ 0.30\pm0.03$ &     $ 0.25\pm0.06$ &     $ 0.37\pm0.08$ &     $ 0.65\pm0.06$ &     $ 0.12\pm0.03$ &     $ 0.23\pm0.05$\nl
           &  0.8546 &     $ 1.46\pm0.29$ &     $ 1.06\pm0.04$ &     $ 0.69\pm0.05$ &     $-$ &           $< 0.12$ &            \nodata &           $< 0.09$ &     $ 1.04\pm0.04$ &     $< 0.13$$^{b}$ &     $ 0.23\pm0.03$ &     $ 0.54\pm0.04$ &     $ 0.68\pm0.06$ &     $ 0.40\pm0.12$ &     $ 0.34\pm0.05$\nl
$1317+277$ &  0.6601 &     $ 1.48\pm0.03$ &     $ 0.84\pm0.05$ &            \nodata &     $+$ &           $< 0.14$ &           $< 0.14$ &     $ 0.43\pm0.03$ &     $ 0.42\pm0.04$ &           $< 0.15$ &     $ 0.22\pm0.04$ &     $ 0.66\pm0.14$ &           $< 0.14$ &           $< 0.07$ &     $< 0.13$$^{b}$\nl
$1329+412$ &  0.5008 &            \nodata &            \nodata &            \nodata & \nodata &           $< 0.22$ &           $< 0.16$&     $< 0.41$$^{b}$ &            \nodata &           $< 0.96$ &            \nodata &            \nodata &           $< 0.40$ &            \nodata &            \nodata\nl
           &  0.8933 &     $ 1.15\pm0.16$ &            \nodata &            \nodata & \nodata &           $< 3.17$ &            \nodata &           $< 0.25$ &           $< 0.34$ &           $< 0.47$ &           $< 1.81$ &            \nodata &     $< 0.12$$^{b}$ &           $< 0.32$ &            \nodata\nl
           &  0.9739 &     $ 1.15\pm0.23$ &            \nodata &            \nodata & \nodata &            \nodata &            \nodata &     $ 0.17\pm0.08$ &     $ 0.53\pm0.10$ &     $ 0.66\pm0.05$ &     $ 0.23\pm0.06$ &            \nodata &     $ 0.87\pm0.05$ &     $< 0.17$$^{b}$ &            \nodata\nl
           &  0.9984 &     $ 0.31\pm0.20$ &            \nodata &            \nodata & \nodata &            \nodata &            \nodata &           $< 0.16$ &           $< 0.19$ &           $< 0.12$ &           $< 0.73$ &            \nodata &           $< 0.11$ &           $< 0.18$ &            \nodata\nl
$1354+195$ &  0.4566 &     $ 1.72\pm0.08$ &            \nodata &            \nodata &     $+$ &           $< 0.24$ &           $< 0.17$ &     $ 0.39\pm0.07$ &     $ 0.42\pm0.08$ &     $ 0.56\pm0.08$ &     $ 0.78\pm0.10$ &            \nodata &     $ 0.91\pm0.04$ &           $< 0.17$ &            \nodata\nl
           &  0.5215 &     $ 1.08\pm0.08$ &            \nodata &            \nodata &     $-$ &           $< 0.22$ &           $< 0.19$ &           $< 0.18$ &           $< 0.37$ &           $< 0.09$ &           $< 0.53$ &            \nodata &           $< 0.24$ &           $< 0.17$ &            \nodata\nl
$1622+238$ &  0.4720 &     $ 1.27\pm0.27$ &            \nodata &            \nodata & \nodata &           $< 0.20$ &            \nodata &     $< 0.28$$^{b}$ &           $< 0.67$ &     $< 0.48$$^{b}$ &           $< 0.56$ &            \nodata &     $ 0.46\pm0.14$ &           $< 0.62$ &            \nodata\nl
           &  0.6561 &     $ 9.42\pm0.34$ &            \nodata &            \nodata & \nodata &     $ 0.53\pm0.07$ &           $< 0.16$ &     $ 1.00\pm0.16$ &     $ 1.06\pm0.13$ &     $ 0.16\pm0.07$ &     $ 0.64\pm0.13$ &            \nodata &     $ 0.29\pm0.09$ &     $< 0.43$$^{b}$ &            \nodata\nl
           &  0.7971 &     $ 1.38\pm0.14$ &           $< 1.63$ &            \nodata & \nodata &           $< 0.10$ &            \nodata &           $< 0.23$ &     $ 0.61\pm0.16$ &     $ 0.48\pm0.07$ &     $ 0.40\pm0.06$ &            \nodata &     $ 1.25\pm0.06$ &           $< 0.29$ &           $< 0.65$\nl
           &  0.8913 &     $ 2.98\pm0.10$ &     $ 1.08\pm0.20$ &           $< 0.58$ &     $+$ &     $ 0.50\pm0.05$ &            \nodata & $ 0.44\pm0.08$$^{b}$ &   $ 0.96\pm0.07$ &     $ 0.70\pm0.07$ &     $ 0.71\pm0.09$ &     $ 1.21\pm0.20$ &     $ 0.82\pm0.05$ &     $< 0.08$$^{b}$ &           $< 0.44$\nl
$1634+706$ &  0.8182 &            \nodata &            \nodata &            \nodata & \nodata &           $< 0.16$ &            \nodata &     $< 0.35$$^{b}$ &            \nodata &           $< 0.07$ &           $< 0.07$ &            \nodata &           $< 0.07$ &            \nodata &            \nodata\nl
           &  0.9056 &     $ 0.49\pm0.03$ &            \nodata &            \nodata & \nodata &           $< 0.07$ &            \nodata &     $< 0.07$$^{b}$ &     $ 0.11\pm0.02$ &     $< 0.12$$^{b}$ &           $< 0.07$ &            \nodata &     $ 0.18\pm0.02$ &           $< 0.38$ &            \nodata\nl
           &  0.9902 &     $ 1.09\pm0.03$ &            \nodata &            \nodata &     $+$ &            \nodata &            \nodata &     $ 0.25\pm0.02$ &     $ 0.39\pm0.03$ &     $ 0.32\pm0.03$ &     $ 0.38\pm0.02$ &            \nodata &     $ 0.32\pm0.02$ &           $< 0.07$ &            \nodata\nl
           &  1.0414 &     $ 1.42\pm0.01$ &            \nodata &            \nodata &     $+$ &            \nodata &            \nodata &           $< 0.06$ &     $ 0.43\pm0.03$ &     $ 0.26\pm0.01$ &     $ 0.10\pm0.01$ &            \nodata &     $ 0.40\pm0.02$ &           $< 0.06$ &            \nodata\nl
$2128-123$ &  0.4297 &     $ 2.92\pm0.08$ &            \nodata &            \nodata & \nodata &     $ 0.31\pm0.09$ &           $< 0.33$ &     $ 0.26\pm0.04$ &     $ 0.27\pm0.08$ &     $ 0.20\pm0.05$ &     $ 0.37\pm0.05$ &            \nodata &     $ 0.40\pm0.04$ &           $< 0.17$ &            \nodata\nl
$2145+067$ &  0.7908 &     $ 1.22\pm0.04$ &     $ 0.79\pm0.07$ &     $ 0.77\pm0.09$ & \nodata &           $< 0.20$ &            \nodata &     $ 0.40\pm0.06$ &     $ 0.84\pm0.04$ &     $ 0.41\pm0.07$ &           $< 0.27$ &     $ 0.84\pm0.07$ &     $ 1.13\pm0.08$ &     $ 0.30\pm0.08$ &     $ 0.83\pm0.11$\nl
\hline
\multicolumn{16}{l}{$^{a}$ All equivalent widths are rest--frame and
their limits are $3\sigma$.}\\
\multicolumn{16}{l}{$^{b}$ Not ``flag'' transition; see system notes (\S3)}\\
\end{tabular}
\end{center}
}}}
\end{table}
\endgroup
 


\clearpage

\newpage

\newpage

\newpage

\clearpage

\clearpage
\begin{figure*}[t]
\figurenum{5$d$--$f$}
\plotfiddle{fig5d.ps}{8.in}{0}{100}{100}{-338}{-135}
\plotfiddle{fig5e.ps}{0.in}{0.}{100.}{100.}{-212}{-113}
\plotfiddle{fig5f.ps}{0.in}{0.}{100.}{100.}{40}{-91}
\end{figure*}

\clearpage
\begin{figure*}[t]
\figurenum{5$g$--$i$}
\plotfiddle{fig5g.ps}{8.in}{0}{100}{100}{-338}{-135}
\plotfiddle{fig5h.ps}{0.in}{0.}{100.}{100.}{-212}{-113}
\plotfiddle{fig5i.ps}{0.in}{0.}{100.}{100.}{-86}{-91}
\end{figure*}

\clearpage
\begin{figure*}[t]
\figurenum{5$j$--$k$}
\plotfiddle{fig5j.ps}{8.in}{0}{100}{100}{-338}{-135}
\plotfiddle{fig5k.ps}{0.in}{0.}{100.}{100.}{-86}{-113}
\end{figure*}

\clearpage
\begin{figure*}[t]
\figurenum{5$l$--$m$}
\plotfiddle{fig5l.ps}{8.in}{0}{100}{100}{-338}{-135}
\plotfiddle{fig5m.ps}{0.in}{0.}{100.}{100.}{-86}{-113}
\end{figure*}

\clearpage
\begin{figure*}[t]
\figurenum{5$n$--$o$}
\plotfiddle{fig5n.ps}{8.in}{0}{100}{100}{-338}{-135}
\plotfiddle{fig5o.ps}{0.in}{0.}{100.}{100.}{-212}{-113}
\end{figure*}

\clearpage
\begin{figure*}[t]
\figurenum{5$p$--$q$}
\plotfiddle{fig5p.ps}{8.in}{0}{100}{100}{-338}{-135}
\plotfiddle{fig5q.ps}{0.in}{0.}{100.}{100.}{-86}{-113}
\end{figure*}

\clearpage
\begin{figure*}[t]
\figurenum{5$r$--$u$}
\plotfiddle{fig5r.ps}{8.in}{0}{100}{100}{-338}{-135}
\plotfiddle{fig5s.ps}{0.in}{0.}{100.}{100.}{-212}{-113}
\plotfiddle{fig5t.ps}{0.in}{0.}{100.}{100.}{-86}{-91}
\plotfiddle{fig5u.ps}{0.in}{0.}{100.}{100.}{40}{-69}
\end{figure*}

\clearpage
\begin{figure*}[t]
\figurenum{5$v$--$w$}
\plotfiddle{fig5v.ps}{8.in}{0}{100}{100}{-338}{-135}
\plotfiddle{fig5w.ps}{0.in}{0.}{100.}{100.}{-86}{-113}
\end{figure*}

\clearpage
\begin{figure*}[t]
\figurenum{5$x$--$z$}
\plotfiddle{fig5x.ps}{8.in}{0}{100}{100}{-338}{-135}
\plotfiddle{fig5y.ps}{0.in}{0.}{100.}{100.}{-212}{-113}
\plotfiddle{fig5z.ps}{0.in}{0.}{100.}{100.}{-86}{-91}
\end{figure*}

\clearpage
\begin{figure*}[t]
\figurenum{5$aa$--$bb$}
\plotfiddle{fig5aa.ps}{8.in}{0}{100}{100}{-338}{-135}
\plotfiddle{fig5bb.ps}{0.in}{0.}{100.}{100.}{-86}{-113}
\end{figure*}

\clearpage
\begin{figure*}[t]
\figurenum{5$cc$--$ee$}
\plotfiddle{fig5cc.ps}{8.in}{0}{100}{100}{-338}{-135}
\plotfiddle{fig5dd.ps}{0.in}{0}{100.}{100.}{-86}{-113}
\plotfiddle{fig5ee.ps}{0.in}{0}{100.}{100.}{40}{-91}
\end{figure*}

\clearpage
\begin{figure*}[t]
\figurenum{5$ff$--$hh$}
\plotfiddle{fig5ff.ps}{8.in}{0}{100}{100}{-338}{-135}
\plotfiddle{fig5gg.ps}{0.in}{0.}{100.}{100.}{-212}{-113}
\plotfiddle{fig5hh.ps}{0.in}{0.}{100.}{100.}{-86}{-91}
\end{figure*}

\clearpage
\begin{figure*}[t]
\figurenum{5$ii$--$kk$}
\plotfiddle{fig5ii.ps}{8.in}{0}{100}{100}{-338}{-135}
\plotfiddle{fig5jj.ps}{0.in}{0.}{100.}{100.}{-212}{-113}
\plotfiddle{fig5kk.ps}{0.in}{0.}{100.}{100.}{-86}{-91}
\end{figure*}

\clearpage
\begin{figure*}[t]
\figurenum{5$ll$--$nn$}
\plotfiddle{fig5ll.ps}{8.in}{0}{100}{100}{-338}{-135}
\plotfiddle{fig5mm.ps}{0.in}{0.}{100.}{100.}{-212}{-113}
\plotfiddle{fig5nn.ps}{0.in}{0.}{100.}{100.}{40}{-91}
\end{figure*}

\clearpage
\begin{figure*}[t]
\figurenum{5$oo$--$qq$}
\plotfiddle{fig5oo.ps}{8.in}{0}{100}{100}{-338}{-135}
\plotfiddle{fig5pp.ps}{0.in}{0.}{100.}{100.}{-212}{-113}
\plotfiddle{fig5qq.ps}{0.in}{0.}{100.}{100.}{-86}{-91}
\end{figure*}

\clearpage
\begin{figure*}[t]
\figurenum{5$rr$--$ss$}
\plotfiddle{fig5rr.ps}{8.in}{0}{100}{100}{-338}{-135}
\plotfiddle{fig5ss.ps}{0.in}{0.}{100.}{100.}{-86}{-113}
\end{figure*}

\newpage
\begin{figure*}[t]
\plotfiddle{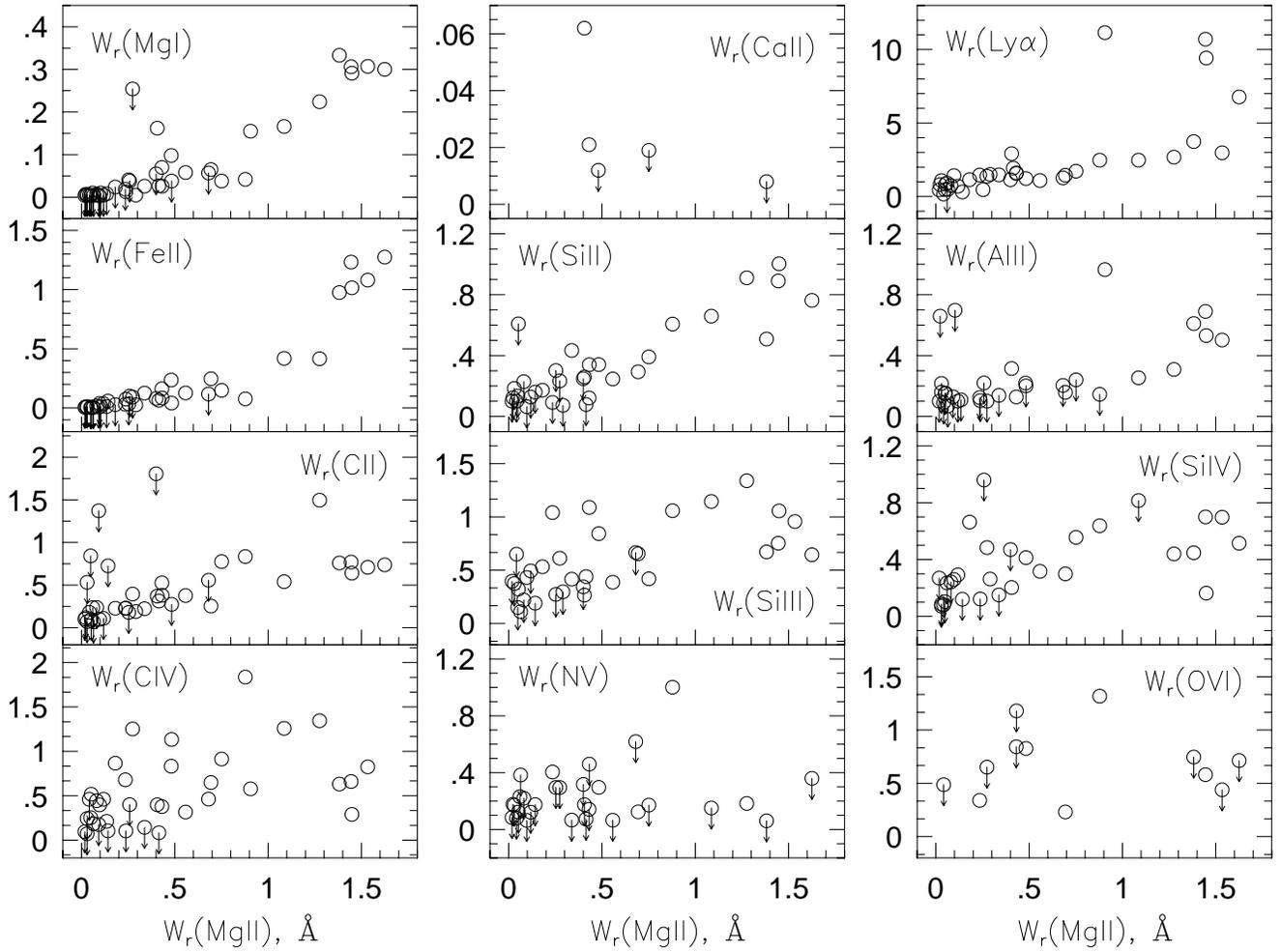}{5.in}{0}{77}{77}{-310}{0}
\caption[fig6.ps]
{The rest--frame equivalent widths of the elemental species listed in
Tables~\ref{tab:hiresresults} and \ref{tab:fosresults} vs.\ that of
{\MgII} $\lambda 2796$. Also included is {\CaII}. The panels are in
order of increasing ionization potential from the upper left to the
lower right. Downward pointing arrows denote upper limits. Only ``flag
transitions'' are plotted.  If a measurement listed in
Table~\ref{tab:fosresults} was based upon a transition other than the
``flag transition'', it was not included. \label{fig:ewall}}
\end{figure*}

\newpage
\begin{figure*}[t]
\plotone{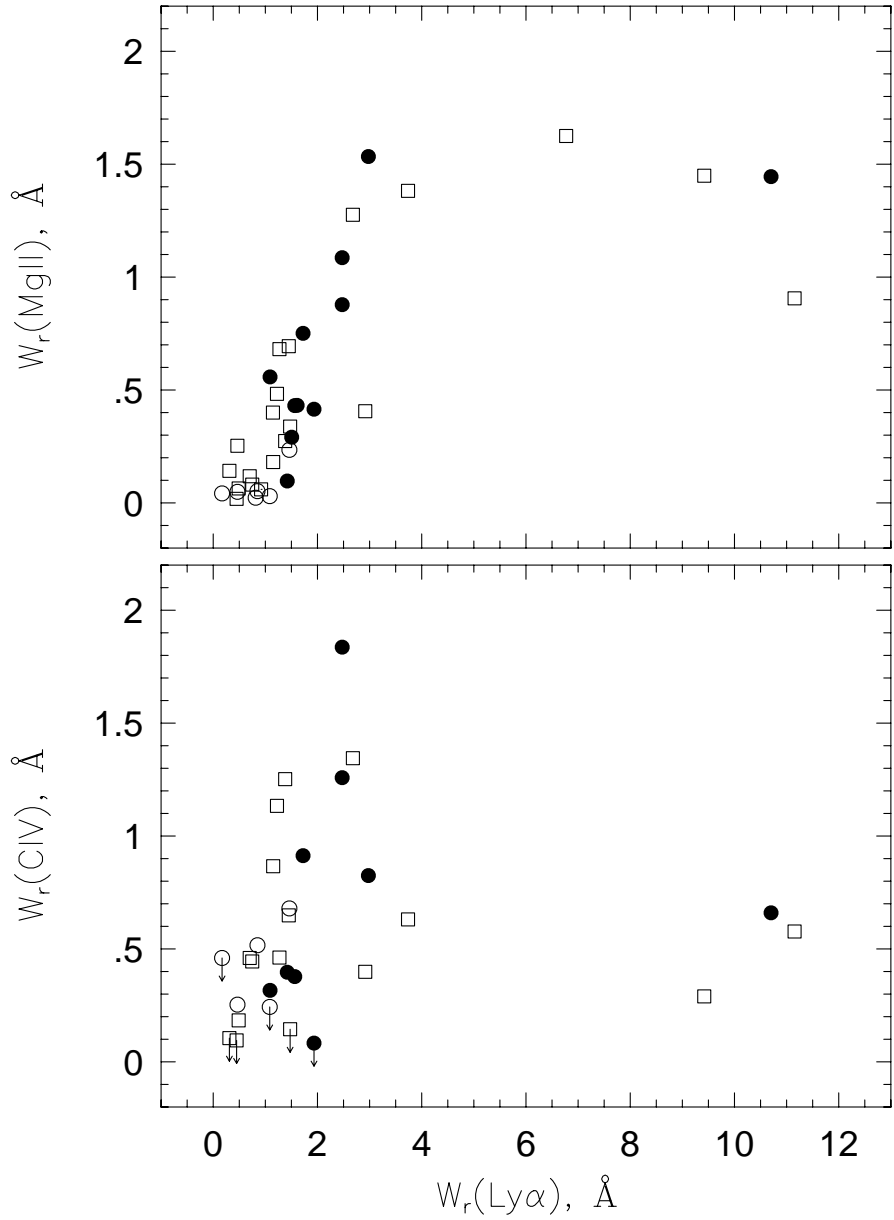}
\caption[fig7.ps]
{The rest--frame equivalent widths of {\MgII} and {\CIV} vs.\ that of
{\Lya}. The data point types are as follows: solid circles have
measured Lyman limit breaks, open circles have no break measured, and
open squares  are systems for which the break was not covered.
The median error of the data is roughly the size of the data
points. Downward pointing arrows denote upper
limits. \label{fig:vsLya}}
\end{figure*}

\newpage
\begin{figure*}[t]
\plotone{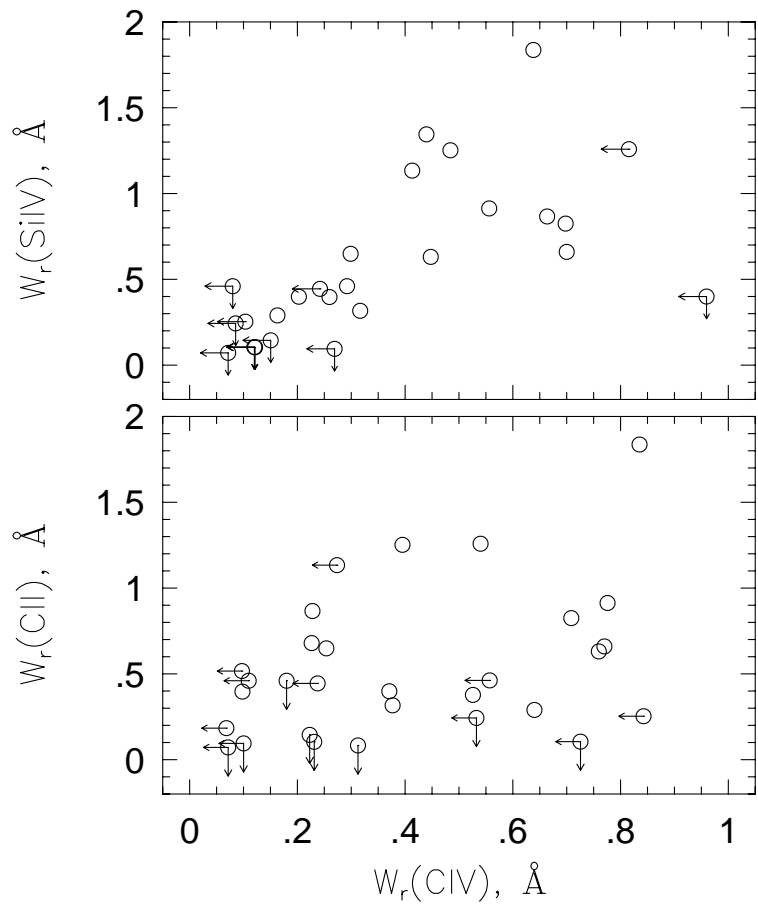}
\caption[fig8.ps]
{The rest--frame equivalent widths of {\SiIV} and  {\CII} vs.\ that of
{\CIV}. The median error of the data is roughly the size of the data points.
Downward pointing arrows denote upper limits on {\SiIV} and {\CII},
respectively, and leftward pointing arrows denote upper limits on 
{\CIV}.  Systems in which the ``flag transition'' was not measured
are not included on this plot. \label{fig:vsCIV}}
\end{figure*}

\newpage
\begin{figure*}[t]
\plotone{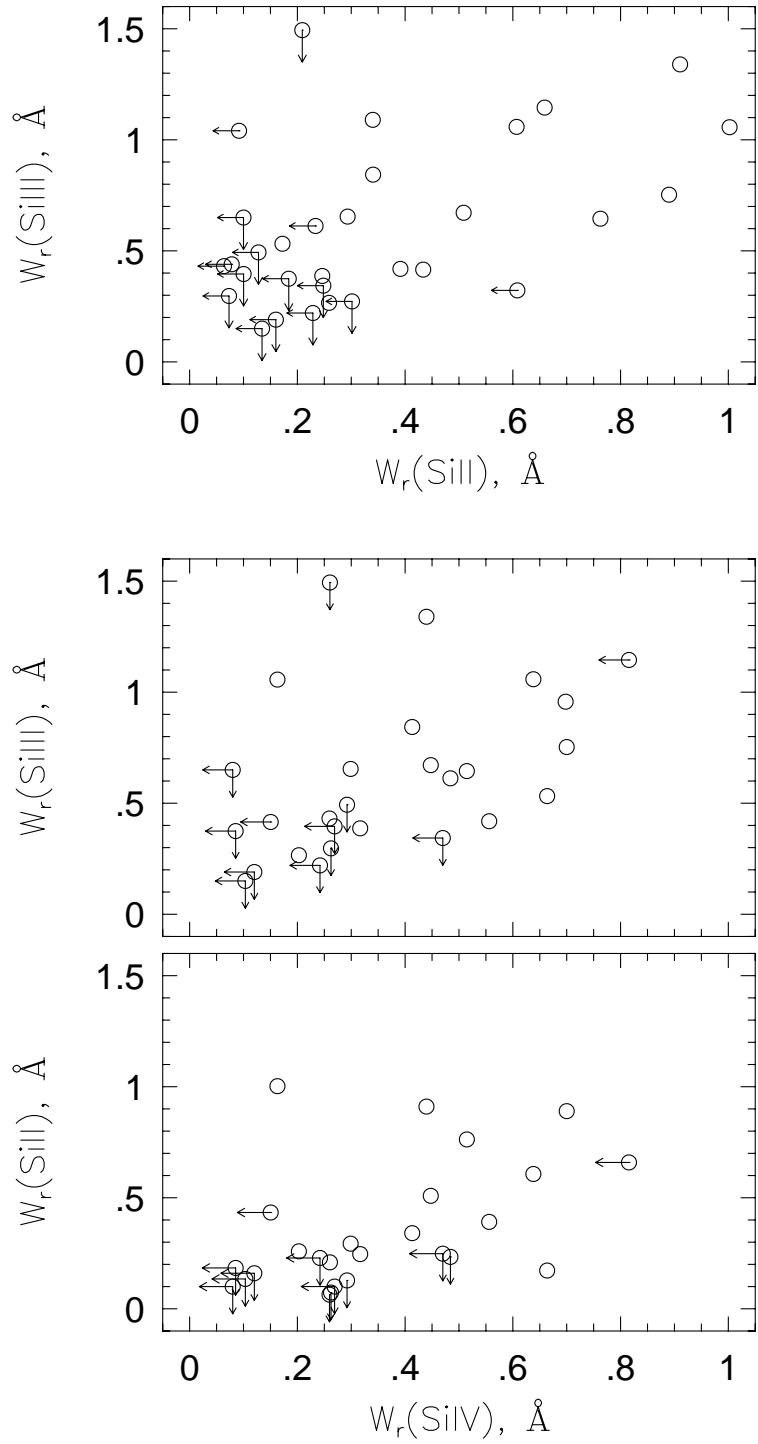}
\caption[fig9.ps]
{The rest--frame equivalent widths of {\SiIII} vs.\ {\SiII} and of
{\SiIII} and {\SiII} vs.\ that of {\SiIV}. The median error of the
data is roughly the size of the data points. Downward pointing arrows
denote upper limits on {\SiIII} and {\SiII}, respectively, and
leftward pointing arrows denote upper limits on  {\SiII} and {\SiIV}.
As with previous figures, only ``flag transition'' measurements are
included. \label{fig:vsSiIV}}
\end{figure*}

\end{document}